\documentclass[sn-basic,pdflatex]{sn-jnl}
\usepackage{amsmath}    
\usepackage{amssymb}    
\usepackage{amsthm}    	
\usepackage{algorithm}
\usepackage{algpseudocode}

\usepackage[draft]{minted}

\usepackage{qsharp}
\usepackage{subcaption}
\usepackage{standalone}

\usepackage{graphicx} 
\usepackage{hyperref}

\usepackage{textcomp}%
\usepackage{manyfoot}%
\usepackage{booktabs}%

\usepackage{algorithm}
\usepackage{algpseudocode}

\usepackage{amsfonts}
\DeclareMathOperator*{\argmax}{argmax}  
\DeclareMathOperator*{\tr}{tr}  
\newcommand\WMC{\text{WMC}}
\newcommand\MPE{\text{MPE}}
\newcommand\MAP{\text{MAP}}
\newcommand\WG{\text{WG}}

\newtheorem{lemma}{Lemma}
\newtheorem{theorem}{Theorem}

\newtheorem{example}{Example}

\usepackage{braket}
\usepackage{tikz}
\usetikzlibrary{decorations.pathreplacing,arrows,quotes,angles}
\newcommand\C{{\mathbf C}}
\newcommand\CI{\mathrm{CI}}
\newcommand\psiangle{18}
\begin{document}
\title[QWCS, QWMC]{Quantum Algorithms for Weighted Constrained Sampling and Weighted Model Counting}

\author*{\fnm{Fabrizio} \sur{Riguzzi}}\email{fabrizio.riguzzi@unife.it}

\affil{\orgdiv{Department of Mathematics and Computer Science}, \orgname{University of Ferrara}, \orgaddress{\street{Via Machiavelli 30}, \city{Ferrara}, \postcode{44121},  \country{Italy}}}

\abstract{
We consider the problems of weighted constrained sampling and weighted model counting, where we are given a propositional formula 
and a weight for each world. The first problem consists of sampling worlds with a probability proportional to their weight given that the formula is satisfied.
The latter is the problem of computing the sum of the weights of the models of the formula. Both have applications in many  fields such as probabilistic reasoning,
graphical models, statistical physics, statistics and hardware verification.

In this article, we propose QWCS and QWMC, quantum algorithms for performing  weighted constrained sampling and weighted model counting, respectively.
Both are  based on the quantum search/quantum model counting algorithms that are modified to take into account the weights. 

In the black box model of computation, where we can only query an oracle for evaluating the Boolean function given an assignment,
QWCS  requires $O(2^{\frac{n}{2}}+1/\sqrt{\WMC})$ oracle calls, where where $n$ is the number of Boolean variables and  $\WMC$ is the normalized between 0 and 1
weighted model count of the formula, while a classical algorithm has a complexity of $\Omega(1/\WMC)$.
QWMC takes $\Theta(2^{\frac{n}{2}})$ oracle calss,  while classically the best complexity is  $\Theta(2^n)$, thus achieving a quadratic speedup. 
}


\maketitle
\section{Introduction}
Given a Boolean formula and a functions assigning weights to assignments of values to the Boolean variable, we consider the problems of Weighted Constrained Sampling (WCS) and Weighted Model Counting (WMC). The first, also called distribution-aware sampling \citep{Chekraborty-Fremont}, involves sampling assignments to the Boolean variables  with a probability proportional to their weight given that the formula is satisfied.
The latter   \citep{DBLP:conf/aaai/SangBK05}  consists in computing the sum of the weights of the models of the formula, i.e. the weighted model count.

WCS has important applications in a variety of domanis, including statistical physics \citep{Jerrum1996TheMC}, statistics \citep{Madras-Piccioni},  hardware verification \citep{Naveh}, and probabilistic reasoning, where it can be used to solve the problem of Most Probable Explanation (MPE) and Maximum A Posteriori (MAP).
MPE \citep{sang2007dynamic}  involves finding an assignment to all variables that satisfies a Boolean formula and has the
maximum weight.
The related MAP problem means finding an assignment of a subset of the variables such that 
the sum of the weights of the models of the formula that agree on the assignment is maximum.

WMC was successfully applied, among others, to the problem of performing inference in graphical models \citep{DBLP:journals/ai/ChaviraD08,DBLP:conf/aaai/SangBK05}. 
In particular, other graphical model inference algorithms \citep{lauritzen1988local,DBLP:journals/jair/ZhangP96,dechter1999bucket,darwiche2001recursive}
take time $\Theta(n2^w)$,  where $n$ is the number of variables
and $w$ is the treewidth  \citep{bodlaender1993tourist} of the network, a measure of the complexity of the network.
WMC instead takes time $O(n2^w)$, i.e., exponential in the treewidth
in the worst case \citep{DBLP:journals/ai/ChaviraD08}. 
This is possible because WMC exploits the structure of graphical models in the form of  context-specific independence and determinism.

In this paper we propose to use quantum  computing for  performing WCS and WMC.
We call QWCS and  QWMC the resulting algorithms. The first is based on quantum search using Grover's algorithm \citep{Grover:1996:FQM:237814.237866,grover1996fast,grover1997quantum}.
The latter on quantum model counting \citep{boyer1998tight,DBLP:conf/icalp/BrassardHT98}. We modified these algorithms to take into account weights. 
In particular, the proposed algorithms modify the algorithms for unweighted search and counting by replacing  the Hadamard gates with rotation gates, with the rotations depending on the weights. 

QWCS and QWMC work under a black box computation model where we don't know 
anything about the propositional formula, we only have the possibility of querying an oracle giving the
value of the formula for an assignment
of the propositional variables.
QWCS take $O(2^{\frac{n}{2}}+1/\sqrt{\WMC})$ oracle calls to solve WCS with a probability of at least $\sqrt{\frac{1}{2}}\approx 0.707$ , where $\WMC$ is the
weighted model count  normalized  between 0 and 1 (cf. Theorem \ref{qwcs-compl}) and $n$ is the number of variables, while any classical algorithm takes $\Omega(1/\WMC)$ oracle calls.

QWMC  takes  $\Theta(2^{\frac{n}{2}})$ oracle calls (cf. Theorem \ref{qwmc-compl}),  to bound the error to $2^{-\frac{n+1}{2}}$ with probability $\frac{11}{12}$, while any classical algorithm takes  $\Theta(2^n)$ oracle calls, thus achieving a quadratic speedup. 

Existing approaches for WMC (knowledge compilation \citep{DBLP:journals/jair/DarwicheM02,lagniez2017improved,huang2006solving}, backtracking search \citep{bacchus2009solving}, reduction
to unweighted counting \citep{DBLP:conf/ijcai/ChakrabortyFMV15},
see also \citep{fichte2021model} for a recent competition among solvers) assume a white box computation model where the formula is known and can be manipulated.

QWMC may be useful for probabilistic inference from models with high treewidth: supposing the cost of implementing the oracle is linear in $n$, if the treewidth is larger than half  
the number of variables, then QWMC performs better than classical inference algorithms in the worst case, because they take time $O(n2^w)$ while
QWMC takes time $\Theta(n2^{\frac{n}{2}})$.

QWMC can also be used as a subroutine for probabilistic inference systems over graphical models.
For example, in the junction tree algorithm \citep{DBLP:conf/uai/ShenoyS88,lauritzen1988local}, it can be used after the probabilities are propagated in the tree to compute the marginals of
the variables in tree nodes.

This article extends \citep{Rig20-ECAI-IC} by adding the QWCS algorithm. Moreover, it fixes an error  in the QWMC algorithm.

The paper is organized as follows.   Section \ref{wmc} presents the WCS and WMC problems. 
Section  \ref{qs}  describes quantum search and Section \ref{comparison} compares it with classical algorithms in terms of cost.
Section \ref{qmap} presents QWCS, whose complexity is discussed in Section \ref{qmap-class} and compared
with classical algorithms.
The algorithm for  
quantum counting is discussed in Section \ref{qmc}.
Section \ref{compl} compares the complexity of the algorithm  to that of classical algorithms. 
Section \ref{qwmc} illustrates the QWMC algorithm and 
Section \ref{wcompl} compares it with classical algorithms.
Related work is described in Section \ref{related}.
Finally, Section \ref{discussion} presents a discussion of the work
and Section \ref{conc} concludes the paper.

The supplementary material includes a brief introduction to the quantum computing concepts used in the paper, the code for the quantum algorithms in Q\# and Qiskit, and the result of a run of the algorithms applied to the 
problem of WMC, MPE and MAP.
\section{Weighted Constrained Sampling and Weighted Model Counting}
\label{wmc}
Let $X$ be a vector of $n$ Boolean variables $[X_1,\ldots,X_n]$ and let $x$ be an assignment of values to $X$, i.e., a vector of $n$ Boolean values $[x_1,\ldots,x_n]$.
We call $x$ a \emph{world}, a \emph{configuration} or an \emph{assignment}.
Consider a propositional logic formula $\phi$ over   $X$ built from the standard propositional connectives $[\neg,\wedge,\vee,\rightarrow,\leftrightarrow,\oplus]$ (not, and, or, imply, iff, xor). If an assignment $x$ of variables $X$ makes formula $\phi$ evaluate to true, we write $x\models \phi$ and
we say that $x$ \emph{satisfies} $\phi$ or that $x$ is a \emph{model} of $\phi$. Let us call $M$ the number of models.
We can also see $\phi$ as a function from $\mathbb{B}^n$ to $\mathbb{B}$, where $\mathbb{B}=\{0,1\}$, and express that $x$ makes $\phi$ evaluate to true by
 $\phi(x)=1$.

The Satisfiability problem (SAT) is the problem of deciding whether a formula $\phi$ has a model, i.e., whether $M>0$. 
The Functional Satisfiability Problem (FSAT) is defined as:  given a formula $\phi$, return an assignment  $x$ that is a model of $\phi$
or answer NO if no such assignment exists. 
The problem of \emph{Model Counting} (\#SAT) \citep{DBLP:series/faia/GomesSS09}  is the problem of computing $M$.

The problem of \emph{Constrained Sampling} (CS) \citep{DBLP:conf/aaai/MeelVCFSFIM16} consists of sampling configurations $x$ with a uniform distribution given
that $\phi(x)=1$. In other words, we want to sample following this distribution
$$P(x)=\left\{\begin{array}{ll}
\frac{1}{M}&\mbox{if $\phi(x)=1$}\\
0&\mbox{if $\phi(x)=0$}
\end{array}\right.$$
In some case we want to sample a configuration $q$ of a subset $Q$ of the variables $X$ with a probability proportional to the number of models in which $q$ can be extended.
Supposing, without loss of generality, that $Q$ is equal to the first $l$ bits, then we want to sample configurations $q$ with probability
$$P(q)=\sum_{y:qy\models \phi}\frac{1}{M}$$
where $qy$ is a world where variables in $Q$ take value $q$ and variables in $Y=X\setminus Q$ take value $y$ and the sum is over all values $y$ of $Y$ such that $qy\models \phi$.

In this article we are concerned with weighted Boolean formulas which are pairs $(\phi,W)$ where
$\phi$ is a Boolean formula over variables $X$ and $W :\mathbb{B}^n \to R^{\geq 0 }$ is a weight function over the 
configurations of  $X$, i.e., it  assigns a weight $W(x)$ to a configuration $x$, that we abbreviate with $W_x$.
Then the \emph{Weighted
Model Count} (WMC) is defined as the sum of the weights of all satisfying
assignments:
$$\WMC(\phi,W) =\sum_{x:x\models \phi} W_x.$$
When $\phi$ and $W$ are clear from the context, we simply indicate $\WMC(\phi,W)$ with $\WMC$.
\emph{Weighted Model Counting} (WMC) \citep{DBLP:journals/ai/ChaviraD08} is the problem of computing $\WMC$.

We restrict our analysis to factorized weight functions, i.e., weight functions expressible as product of weights over
the literals built on $X$ (Boolean variables or their negation).
Seeing a configuration $x$ as a set of literals (e.g.
$[0,1,0,1]=[\neg X_1,X_2,\neg X_3, X_4]$, we can compute the weights $W_x$
with a function  $w : L \to R^{\geq 0 }$, where $L$ is the set of literals, such that
$$W_x =\prod_{l\in x}w(l).$$
An important special case is that in which $w(X_i)+w(\neg X_i)=1$, where the weights can be considered as the 
probabilities of the Boolean literals of being true and $\WMC$ is then the probability that $\phi$ takes value 1 assuming
that the Boolean variables are independent random variables.

\emph{Weighted Constrained Sampling} (WCS) or Distribution-Aware Sampling \citep{Chekraborty-Fremont}
is the problem of sampling a configuration $x$ with a probability proportional to its weight given that the formula is satisfied, i.e.:
\begin{equation}
P(x)=\left\{\begin{array}{ll}
\propto W_x&\mbox{if $\phi(x)=1$}\\
0&\mbox{if $\phi(x)=0$}
\end{array}\right.
=\left\{\begin{array}{ll}
\frac{W_x}{\sum_{x:\phi(x)=1}W_x}&\mbox{if $\phi(x)=1$}\\
0&\mbox{if $\phi(x)=0$}
\end{array}\right.
=\left\{\begin{array}{ll}
\frac{W_x}{\WMC}&\mbox{if $\phi(x)=1$}\\
0&\mbox{if $\phi(x)=0$}
\end{array}\right.
\label{wcs}
\end{equation}
It is a special case of the problem of sampling a set of query variables $Q$ with a probability proportional to the sum of the weights of the models in which $q$ can be extended:
\begin{equation}
P(q)=\frac{ \sum_{y:\phi(qy)=1}W_{qy}}{ \sum_{q,y:\phi(qy)=1}W_{qy}}=\frac{ \sum_{y:\phi(qy)=1}W_{qy}}{\WMC)}
\label{sampling-dist}
\end{equation}
WCS can play a role for the following problems. The \emph{Most Probable Explanation} (MPE) \citep{sang2007dynamic} problem involves finding the model that has the
maximum weight. The \emph{Maximum A Posteriori} (MAP) problem means finding an assignment of a subset of the variables such that 
the sum of the weights of the models that agree on the assignment is maximum.

Given a formula $\phi$  and a weight function $w : L \to R^{\geq 0 }$, 
the most probable state (Most Probable Explanation, MPE) of the variables  
is
$$\MPE(\phi,w)=\argmax_{x:x\models \phi} W_x$$
Let $\MPE^w(\phi,w)$ be the weight of $\MPE(\phi,w)$, i.e., 
$$\MPE^w(\phi,w)=\max_{x:x\models \phi} W_x.$$
Given formula $\phi$, a weight function $w : L \to R^{\geq 0 }$ and a set of query variables $Q$, the most probable state of the query
 variables (Maximum A Posteriori, MAP) is
$$\MAP_Q(\phi,w)=\argmax_q \sum_{y:qy\models \phi}W_{qy}$$
Let $\MAP_Q^w(\phi,w)$ be the  sum of the weights of the models that agree on the assignment  $\MAP_Q(\phi,w)$:
$$\MAP_Q^w(\phi,w)=\max_q \sum_{y:qy\models \phi}W_{qy}.$$
Clearly MPE is a special case of MAP when $Q=X$.
\begin{example}\label{sprinklerb}
Let us consider an example inspired by the sprinkler problem of \citep{pearl88}: we have three Boolean variables, $S$, $R$, $W$ representing propositions ``the sprinkler was on'', ``it rained last night'' and ``the grass is wet'', respectively. We know that: if the sprinkler was on, the grass is wet ($S\rightarrow W$); if it rained last night, the grass is wet ($R\rightarrow W$); and  the sprinkler being on and rain last night cannot be true at the same time ($S\wedge R\rightarrow$). The formula for this problem is:
$$\phi=(\neg S \vee W)\wedge(\neg R \vee W)\wedge(\neg S\vee \neg R).$$
Suppose the weights of literals are
$w(S)=0.55$, $w(\neg S)=0.45$, $w(R)=0.3$, $w(\neg R)=0.7$, $w(W)=0.7$ and $w(\neg W)=0.3$. Table \ref{sprinkler_table} shows the worlds together with the weight of each world.
The WMC of $\phi$ is thus
$\WMC(\phi,w)=0.0945+0.2205+0.0945+0.2695=0.679$.

The MPE is
$\MPE(\phi,w)=[1,0,1]$ with $\MPE^w(\phi,w)=0.2695$.
The MAP of query variables $S$ and $W$ is
$\MAP_{SW}(\phi,w)=[0,1]$ with $\MAP_{SW}^w(\phi,w)=0.2205+0.0945=0.315$.

So the most probable state of $S$ is 1 when we look for the overall MPE state and is 0 when we look for the most probable state of $S$ and $W$.
\end{example}
\begin{table}[tbp]
\centering
\begin{tabular}{|ccc|c|c|}
\hline
$S$&$R$&$W$&$\phi$&$weight$\\
\hline
0&0&0&1&$0.45\cdot 0.7\cdot 0.3=0.0945$\\
0&0&1&1&$0.45\cdot 0.7\cdot 0.7=0.2205$\\
0&1&0&0&$0.45\cdot 0.3\cdot 0.3=0.0405$\\
0&1&1&1&$0.45\cdot 0.3\cdot 0.7=0.0945$\\
1&0&0&0&$0.55\cdot 0.7\cdot 0.3=0.1155$\\
1&0&1&1&$0.55\cdot 0.7\cdot 0.7=0.2695$\\
1&1&0&0&$0.55\cdot 0.3\cdot 0.3=0.0495$\\
1&1&1&0&$0.55\cdot 0.3\cdot 0.7=0.1155$\\
\hline
\end{tabular}

\caption{Worlds for  Example \ref{sprinklerb}.\label{sprinkler_table}}
\end{table}
WCS  can be used to obtain an approximate algorithm
for performing MPE and MAP.
Given an algorithm for WCS, we can execute it for $o$ iterations.
At each iteration, we obtain a sample $q$.
After $o$ iterations, we  return the value that was found most frequently.

In fact $P(q)$ of Eq. (\ref{sampling-dist})  is a categorical distribution. Executing the algorithm $o$ times is 
equivalent to extracting a sample $d$ from a multinomial distribution $P(d;o,p)$ with $o$ trials, set of events $\{q|\exists y,\phi(qy)=1\}$ and event probabilities $P(q)$.
Let us rename the elements of the set of events as $q_1,\ldots,q_k$ and the event probabilities as $[p_1,\ldots,p_k]$ and let us reorder them so that 
the event probabilities are in ascending order: $p_1\leq p_1\leq \ldots\leq p_k$. So $q_k$ is $\MAP_Q(\phi,w)$. Let $d_i$ be the count associated to even $q_i$ in the
sample $d$.

We obtain a correct solution of the overall algorithm if $d_k>d_1\wedge\ldots\wedge d_k>d_{k-1}$. The probability that this event happens in a sample from a multinomial distribution
has been studied in \citep{multim-rank}. The authors consider a worst case scenario where the event probabilities are all equal to a common value $p$ except for the
last one which is equal to $Ap$ with $A\geq 1$ so the the event probabilities are $[p,\ldots,p,Ap]$. The authors do not present a closed formula for computing the probability of the event
$E=(d_k>d_1 \wedge \ldots \wedge d_k>d_{k-1})$ but tabulate the probability $P(E)$ for various values of $A$, $o$ and $k$.

\begin{figure}
\centering
\begin{subfigure}[b]{0.45\textwidth}
\centering
	\includegraphics[width=1\textwidth]{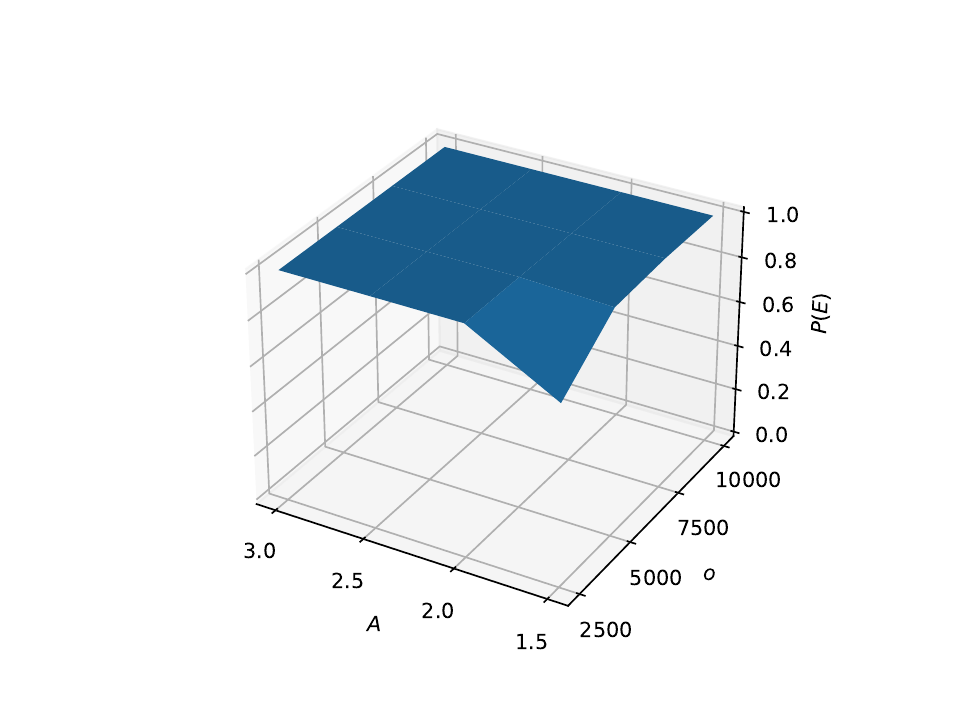}
   \caption{$k=50$}
         \label{fig:k50}
     \end{subfigure}
\begin{subfigure}[b]{0.45\textwidth}
\centering
	\includegraphics[width=1\textwidth]{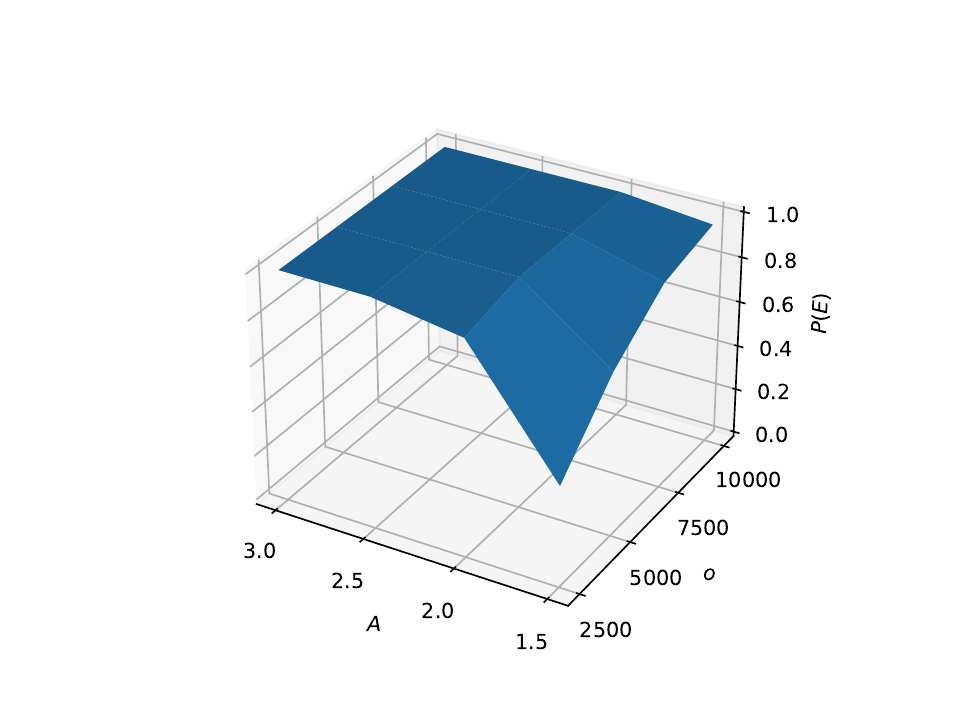}
   \caption{$k=100$}
         \label{fig:k100}
     \end{subfigure}

\begin{subfigure}[b]{0.45\textwidth}
\centering
	\includegraphics[width=1\textwidth]{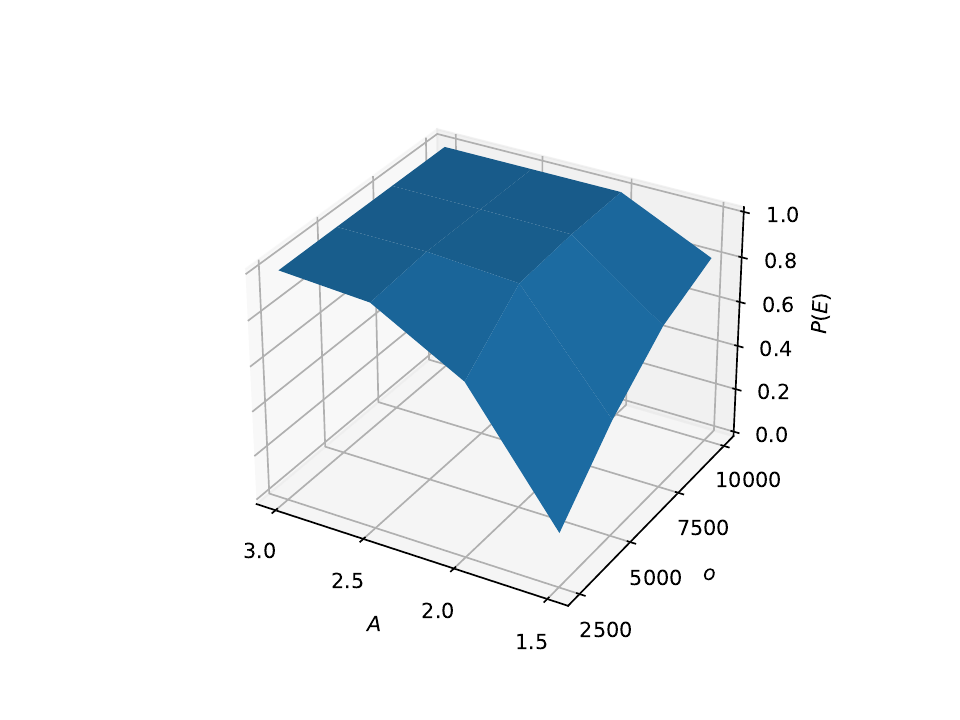}
   \caption{$k=150$}
         \label{fig:k150}
     \end{subfigure}
\begin{subfigure}[b]{0.45\textwidth}
\centering
	\includegraphics[width=1\textwidth]{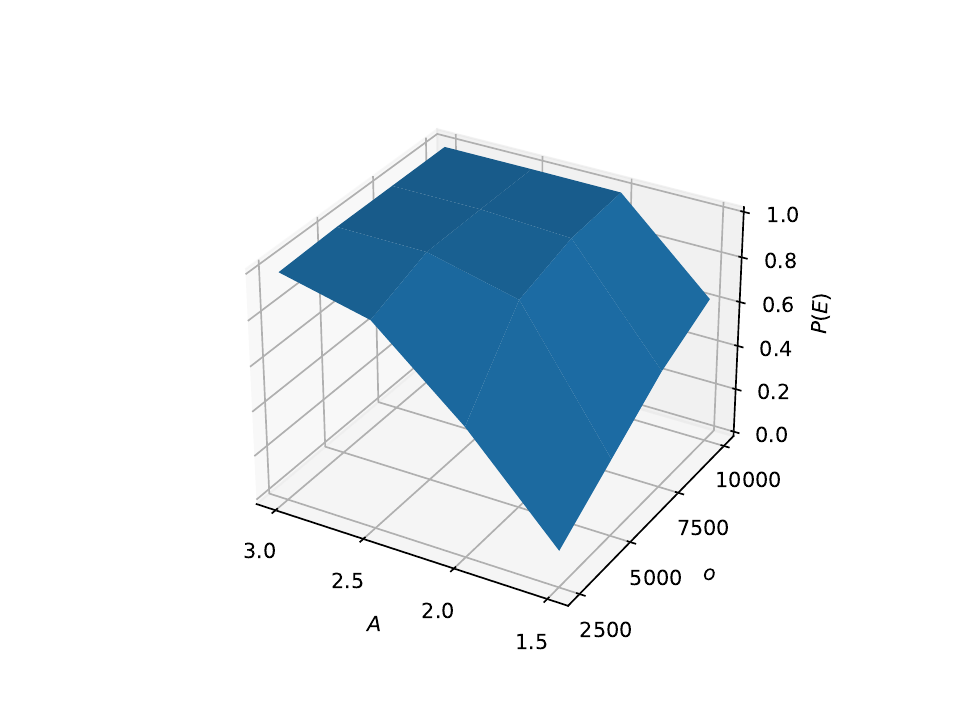}
   \caption{$k=200$}
         \label{fig:k200}
     \end{subfigure}

\caption{Probability of event $E$ in a set of samples from a multinomial random variable with distribution   $[p,\ldots, p,Ap]$. $k$ is the number of values of the random variable, $o$ is the number of trials and $A$ is the multiplier of the probability of the most probabile value.}
\label{fig:mult-rank}
\end{figure}

Unfortunately, the tables in \citep{multim-rank} do not include values of interests for us so we performed a simulation using probabilistic logic programming \citep{Rig23-BK} and
the MCINTYRE system in particular \citep{Rig13-FI-IJ}, for the set of values: $k=50,100,150,200$, $o=2500,5000,7500,10000$ and $A=1.5,2,2.5,3$.
The results are shown in Figure \ref{fig:mult-rank}. We can see that the probability of the event $E$, and so of the success of the algorithm, goes rapidly to 1 for increasing
values of $o$ and $A$.

\section{Quantum Search}
\label{qs}
CS can  be seen as a search problem: find a satisfying assignment of bits.
A quantum algorithm for solving this problem was proposed in \citep{Grover:1996:FQM:237814.237866,grover1996fast,grover1997quantum}. 
Here we follow the exposition of \citep{nielsen2010quantum} and \citep{hirvensalo2013quantum}.

We assume we have a black box  quantum circuit that evaluates $\phi$, called an \emph{oracle} $O$, that is such that
$$\ket{x} \overset{O}{\to} (-1)^{\phi(x)}\ket{x}$$
i.e., the oracle marks solutions to the search problems by changing the sign of the state. The oracle may use extra ancilla bits to do so. 

\begin{figure}[t]
\centering
\begin{tikzpicture}[scale=1.000000,x=1pt,y=1pt]
\filldraw[color=white] (0.000000, -7.500000) rectangle (177.000000, 22.500000);
\draw[color=black] (0.000000,15.000000) -- (165.000000,15.000000);
\draw[color=black] (165.000000,14.500000) -- (177.000000,14.500000);
\draw[color=black] (165.000000,15.500000) -- (177.000000,15.500000);
\draw[color=black] (0.000000,15.000000) node[left] {${X=\ket{0}}$};
\draw[color=black] (0.000000,0.000000) -- (177.000000,0.000000);
\draw[color=black] (0.000000,0.000000) node[left] {$Ancilla$};
\draw (6.000000, 9.000000) -- (14.000000, 21.000000);
\draw (12.000000, 18.000000) node[right] {$\scriptstyle{n}$};
\draw (6.000000, -6.000000) -- (14.000000, 6.000000);
\draw (12.000000, 3.000000) node[right] {$\scriptstyle{q}$};
\begin{scope}
\draw[fill=white] (37.000000, 15.000000) +(-45.000000:15.556349pt and 8.485281pt) -- +(45.000000:15.556349pt and 8.485281pt) -- +(135.000000:15.556349pt and 8.485281pt) -- +(225.000000:15.556349pt and 8.485281pt) -- cycle;
\clip (37.000000, 15.000000) +(-45.000000:15.556349pt and 8.485281pt) -- +(45.000000:15.556349pt and 8.485281pt) -- +(135.000000:15.556349pt and 8.485281pt) -- +(225.000000:15.556349pt and 8.485281pt) -- cycle;
\draw (37.000000, 15.000000) node {$H^{\otimes n}$};
\end{scope}
\draw (66.000000,15.000000) -- (66.000000,0.000000);
\begin{scope}
\draw[fill=white] (66.000000, 7.500000) +(-45.000000:8.485281pt and 19.091883pt) -- +(45.000000:8.485281pt and 19.091883pt) -- +(135.000000:8.485281pt and 19.091883pt) -- +(225.000000:8.485281pt and 19.091883pt) -- cycle;
\clip (66.000000, 7.500000) +(-45.000000:8.485281pt and 19.091883pt) -- +(45.000000:8.485281pt and 19.091883pt) -- +(135.000000:8.485281pt and 19.091883pt) -- +(225.000000:8.485281pt and 19.091883pt) -- cycle;
\draw (66.000000, 7.500000) node {G};
\end{scope}
\draw (90.000000,15.000000) -- (90.000000,0.000000);
\begin{scope}
\draw[fill=white] (90.000000, 7.500000) +(-45.000000:8.485281pt and 19.091883pt) -- +(45.000000:8.485281pt and 19.091883pt) -- +(135.000000:8.485281pt and 19.091883pt) -- +(225.000000:8.485281pt and 19.091883pt) -- cycle;
\clip (90.000000, 7.500000) +(-45.000000:8.485281pt and 19.091883pt) -- +(45.000000:8.485281pt and 19.091883pt) -- +(135.000000:8.485281pt and 19.091883pt) -- +(225.000000:8.485281pt and 19.091883pt) -- cycle;
\draw (90.000000, 7.500000) node {G};
\end{scope}
\draw[fill=white,color=white] (108.000000, -6.000000) rectangle (123.000000, 21.000000);
\draw (115.500000, 7.500000) node {$\cdots$};
\draw (141.000000,15.000000) -- (141.000000,0.000000);
\begin{scope}
\draw[fill=white] (141.000000, 7.500000) +(-45.000000:8.485281pt and 19.091883pt) -- +(45.000000:8.485281pt and 19.091883pt) -- +(135.000000:8.485281pt and 19.091883pt) -- +(225.000000:8.485281pt and 19.091883pt) -- cycle;
\clip (141.000000, 7.500000) +(-45.000000:8.485281pt and 19.091883pt) -- +(45.000000:8.485281pt and 19.091883pt) -- +(135.000000:8.485281pt and 19.091883pt) -- +(225.000000:8.485281pt and 19.091883pt) -- cycle;
\draw (141.000000, 7.500000) node {G};
\end{scope}
\draw[fill=white] (159.000000, 9.000000) rectangle (171.000000, 21.000000);
\draw[very thin] (165.000000, 15.600000) arc (90:150:6.000000pt);
\draw[very thin] (165.000000, 15.600000) arc (90:30:6.000000pt);
\draw[->,>=stealth] (165.000000, 9.600000) -- +(80:10.392305pt);
\draw[decorate,decoration={brace,amplitude = 4.000000pt},very thick] (57.000000,22.500000) -- (150.000000,22.500000);
\draw (103.500000, 26.500000) node[text width=144pt,above,text centered] {$R$};
\end{tikzpicture}
\caption{Quantum search algorithm.}
\label{qsa}
\end{figure}
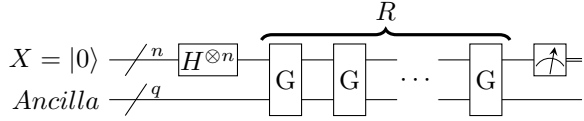

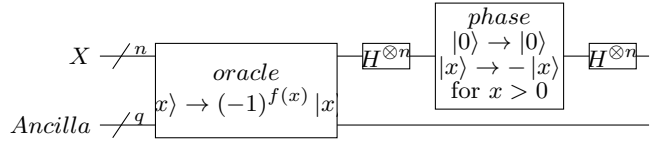
\begin{figure}[t]
\centering
\begin{scriptsize}
\begin{tikzpicture}[scale=0.800000,x=1pt,y=1pt]
\filldraw[color=white] (0.000000, -7.500000) rectangle (257.000000, 57.500000);
\draw[color=black] (0.000000,32.500000) -- (257.000000,32.500000);
\draw[color=black] (0.000000,32.500000) node[left] {$X$};
\draw[color=black] (0.000000,0.000000) -- (257.000000,0.000000);
\draw[color=black] (0.000000,0.000000) node[left] {$Ancilla$};
\draw (6.000000, 26.500000) -- (14.000000, 38.500000);
\draw (12.000000, 35.500000) node[right] {$\scriptstyle{n}$};
\draw (6.000000, -6.000000) -- (14.000000, 6.000000);
\draw (12.000000, 3.000000) node[right] {$\scriptstyle{q}$};
\draw (68.500000,32.500000) -- (68.500000,0.000000);
\begin{scope}
\draw[fill=white] (68.500000, 16.250000) +(-45.000000:60.104076pt and 31.466252pt) -- +(45.000000:60.104076pt and 31.466252pt) -- +(135.000000:60.104076pt and 31.466252pt) -- +(225.000000:60.104076pt and 31.466252pt) -- cycle;
\clip (68.500000, 16.250000) +(-45.000000:60.104076pt and 31.466252pt) -- +(45.000000:60.104076pt and 31.466252pt) -- +(135.000000:60.104076pt and 31.466252pt) -- +(225.000000:60.104076pt and 31.466252pt) -- cycle;
\draw (68.500000, 16.250000) node {$\begin{array}{c}oracle\\\ket{x}\to (-1)^{f(x)}\ket{x}\end{array}$};
\end{scope}
\begin{scope}
\draw[fill=white] (134.000000, 32.500000) +(-45.000000:15.556349pt and 8.485281pt) -- +(45.000000:15.556349pt and 8.485281pt) -- +(135.000000:15.556349pt and 8.485281pt) -- +(225.000000:15.556349pt and 8.485281pt) -- cycle;
\clip (134.000000, 32.500000) +(-45.000000:15.556349pt and 8.485281pt) -- +(45.000000:15.556349pt and 8.485281pt) -- +(135.000000:15.556349pt and 8.485281pt) -- +(225.000000:15.556349pt and 8.485281pt) -- cycle;
\draw (134.000000, 32.500000) node {$H^{\otimes n}$};
\end{scope}
\begin{scope}
\draw[fill=white] (187.000000, 32.500000) +(-45.000000:42.426407pt and 35.355339pt) -- +(45.000000:42.426407pt and 35.355339pt) -- +(135.000000:42.426407pt and 35.355339pt) -- +(225.000000:42.426407pt and 35.355339pt) -- cycle;
\clip (187.000000, 32.500000) +(-45.000000:42.426407pt and 35.355339pt) -- +(45.000000:42.426407pt and 35.355339pt) -- +(135.000000:42.426407pt and 35.355339pt) -- +(225.000000:42.426407pt and 35.355339pt) -- cycle;
\draw (187.000000, 32.500000) node {$\begin{array}{c}phase\\\ket{0}\to\ket{0}\\\ket{x}\to -\ket{x}\\\mbox{for }x>0\end{array}$};
\end{scope}
\begin{scope}
\draw[fill=white] (240.000000, 32.500000) +(-45.000000:15.556349pt and 8.485281pt) -- +(45.000000:15.556349pt and 8.485281pt) -- +(135.000000:15.556349pt and 8.485281pt) -- +(225.000000:15.556349pt and 8.485281pt) -- cycle;
\clip (240.000000, 32.500000) +(-45.000000:15.556349pt and 8.485281pt) -- +(45.000000:15.556349pt and 8.485281pt) -- +(135.000000:15.556349pt and 8.485281pt) -- +(225.000000:15.556349pt and 8.485281pt) -- cycle;
\draw (240.000000, 32.500000) node {$H^{\otimes n}$};
\end{scope}
\end{tikzpicture}
\end{scriptsize}
\caption{Grover operator $G$.}
\label{grover}
\end{figure}

Figure \ref{qsa} shows the circuit performing quantum search operating on an $n$-qubit register $X$ and the oracle ancilla  qubits $Ancilla$. 
All qubits of register $X$ start in state $\ket{0}$.
The circuit includes a gate $G$ that is called the \emph{Grover operator }and is implemented as shown in Figure \ref{grover}.

The first gate of the search circuit applies the $H$ gate to each qubit in   register $X$. Since all qubits in register $X$ start as $\ket{0}$ and the
effect of $H$ is to transform $\ket{0}$ to the state $\frac{\ket{0}+\ket{1}}{\sqrt{2}}$, then register $X$ is transformed to 
$$\begin{array}{l}
\ket{\psi}=\frac{\ket{0}+\ket{1}}{\sqrt{2}}\otimes\frac{\ket{0}+\ket{1}}{\sqrt{2}}\cdots\otimes\frac{\ket{0}+\ket{1}}{\sqrt{2}}\\
=\frac{\ket{00}+\ket{01}+\ket{10}+\ket{11}}{\sqrt{2^2}}\cdots\otimes\frac{\ket{0}+\ket{1}}{\sqrt{2}}\\
=\frac{\ket{000}+\ket{001}+\ket{010}+\ket{011}+\ket{100}+\ket{101}+\ket{110}+\ket{111}}{\sqrt{2^3}}\cdots\otimes\frac{\ket{0}+\ket{1}}{\sqrt{2}} \\
=\frac{1}{N^{1/2}}\sum_{x=0}^{N-1}\ket{x}
\end{array}
$$
where $N=2^n$.
This state is also called the 
\emph{uniform superposition state}.

The Grover operator can be written as 
$$G=H^{\otimes n}(2\ket{0}\bra{0}-I)H^{\otimes n}O$$
Since $H^\dagger=H$, $G$ can be rewritten as
\begin{eqnarray*}
G&=&H^{\otimes n}(2\ket{0}\bra{0}-I)(H^{\otimes n})^\dagger O\\
&=&\left(2H^{\otimes n}\ket{0}\bra{0}(H^{\otimes n})^\dagger -H^{\otimes n}I(H^{\otimes n})^\dagger \right)O\\
&=&(2\ket{\psi}\bra{\psi} -I)O
\end{eqnarray*}
We now show that the Grover operator is a rotation.
\begin{lemma}
\label{grov-is-rotation}
The Grover operation applied to the uniform superposition state  $\ket{\psi}$ rotates it by angle $2\arcsin\sqrt{M/N}$ where $M$ is the number of solutions of $\phi(x)=1$.
\end{lemma}
\begin{proof}
Consider the two states
$$\ket{\alpha}=\frac{1}{\sqrt{N-M}}\sum_{x:\phi(x)=0}\ket{x}$$
$$\ket{\beta}=\frac{1}{\sqrt{M}}\sum_{x:\phi(x)=1}\ket{x}.$$
These two states are orthonormal because they do not share any computational basis state.
The uniform superposition state $\ket{\psi}$ can be written as a linear combination of $\ket{\alpha}$ and
$\ket{\beta}$:
$$\ket{\psi}=\sqrt{\frac{N-M}{N}}\ket{\alpha}+\sqrt{\frac{M}{N}}\ket{\beta}$$
so $\ket{\psi}$ belongs to the plane defined by $\ket{\alpha}$ and $\ket{\beta}$. In this plane, the effect of
the oracle
operation $O$ is to perform a reflection about the vector $\alpha$ because $O(\ket{\alpha} + \ket{\beta}) = \ket{\alpha} - \ket{\beta}$, see Figure \ref{plane}.
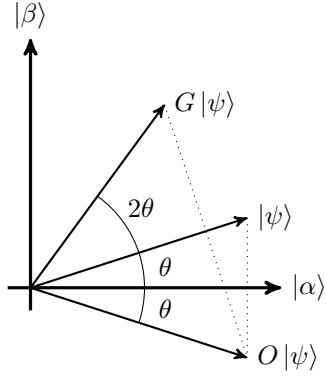
\begin{figure}[t]
\centering
\begin{tikzpicture}[
    scale=3,
    axis/.style={very thick, ->, >=stealth'},
    important line/.style={thick,->,>=stealth'},
    every node/.style={color=black}
    ]
    \draw[axis] (-0.1,0)  -- (1.1,0) coordinate(alpha) node(xline)[right]
        {$\ket{\alpha}$};
    \draw[axis] (0,-0.1) -- (0,1.1) node(yline)[above] {$\ket{\beta}$};
    \draw[important line] (0,0) coordinate (A) -- ({cos(\psiangle)},{sin(\psiangle)})
        coordinate (B) node[right, text width=5em] {$\ket{\psi}$};
    \draw[important line] (0,0) coordinate (C) -- ({cos(-\psiangle},{sin(-\psiangle)})
        coordinate (D) node[right, text width=5em] {$O\ket{\psi}$};
    \draw[important line] (0,0) coordinate (E) -- ({cos(3*\psiangle)},{sin(3*\psiangle)})
        coordinate (F) node[right, text width=5em] {$G\ket{\psi}$};
    \draw [dotted] (B) -- (D);
    \draw [dotted] (F) -- (D);
    \draw
    pic["$\theta$", draw=black, -, angle eccentricity=1.2, angle radius=1.5cm]
    {angle=D--A--alpha};
    \draw
    pic["$\theta$", draw=black, -, angle eccentricity=1.2, angle radius=1.5cm]
    {angle=alpha--A--B};
    \draw
    pic["$2\theta$", draw=black, -, angle eccentricity=1.2, angle radius=1.5cm]
    {angle=B--A--F};
\end{tikzpicture}
\caption{Visualization of the effect of Grover operator (Figure 6.3 from \citep{nielsen2010quantum}).}
\label{plane}
\end{figure}

The other component of the Grover operator, $2\ket{\psi}\bra{\psi}-I$, also performs a
reflection in the plane defined by $\ket{\alpha}$ and $\ket{\beta}$, about the vector $\ket{\psi}$.
The overall effect is that of a rotation \citep{aharonov1999quantum}. If we define $\cos \theta=\sqrt{(N-M)/N}$ and $\sin\theta=\sqrt{M/N}$, then $\ket{\psi}=\cos \theta\ket{\alpha}+\sin\theta\ket{\beta}$.


From Figure \ref{plane} we can see that the rotation applied by $G$ is by angle $2\theta$, so 
$$G\ket{\psi}=\cos 3\theta\ket{\alpha}+\sin 3\theta\ket{\beta}$$
\end{proof}
Repeated applications of $G$ take the state
to
$$G^k\ket{\psi}=\cos (2k+1)\theta\ket{\alpha}+\sin(2k+1)\theta\ket{\beta}.$$
These rotations bring the state of the system closer  to $\ket{\beta}$. If we perform the right number of rotations, a measurement in the
computational basis will produce one of the outcomes superposed in
$\ket{\beta}$, i.e., a solution to the search problem, with a non-zero probability. Moreover, since the weights of the computational basis states superimposed in $\ket{\beta}$ are all equal, 
then the measurement produces one of solutions with equal probability, solving CS when $Q=X$.
\begin{theorem}[Theorem 5.2.1 in \citep{hirvensalo2013quantum}]
\label{number-of-rotations}
Let $\phi:\mathbb{B}^n \rightarrow \mathbb{B}$ be such that there are $M$ elements $x\in \mathbb{B}^n$ satisfying $\phi(x)=1$. Assume that $0<M\leq \frac{3}{4}N$. and let $\theta \in [0,\pi/3]$ be chosen
such that $sin^2 \theta=\frac{M}{N}\leq \frac{3}{4}$. After $\left\lfloor \frac{\pi}{4\theta}\right\rfloor$ iterations of $G$ on an initial superposition
$$\frac{1}{N^{1/2}}\sum_{x=0}^{N-1}\ket{x}$$
the probability of seeing a solution is at least $\frac{1}{4}$.
\end{theorem}
\begin{proof}
The probability $P$ of seeing a solution is given by
$$P=\sin^2((2k+1)\theta)$$
To maximize $P$, we must find the least positive integer $k$  so that $P$ is as close to 1 as possible. This implies that 
$$(2k+1)\theta=\frac{\pi}{2}$$
so
$$k=\frac{\pi}{4\theta}-\frac{1}{2}$$.
Using the approximation $\theta^2\approx \sin^2\theta=\frac{M}{N}$
we have that
$$\theta\approx\sqrt{\frac{M}{N}}$$
and after
$$R=\left\lfloor \frac{\pi}{4}\sqrt{\frac{N}{M}}\right\rfloor$$
rotations, $P$ is close to 1.
Let us now compute $P$.
We can observe that
$$R=\frac{\pi}{4\theta}-\frac{1}{2}+\delta$$
with $|\delta|\leq \frac{1}{2}$. Therefore
$$(2R+1)\theta=\frac{\pi}{2}+2\delta\theta$$
The distance between $(2R+1)\theta$ and $\frac{\pi}{2}$ is thus $|2\delta\theta|$ and, since $\theta \in [0,\frac{\pi}{3}]$, $|2\delta\theta|\leq \frac{\pi}{3}$.
$P$ then becomes
$$P=\sin^2((2R+1)\theta)\geq\sin^2\left(\frac{\pi}{2}-\frac{\pi}{3}\right)=\frac{1}{4}$$
\end{proof}
Consider now the cases $M>\frac{3}{4}N$  and $M=0$. In the first case, we can guess a solution and the probability that it is correct is $\frac{3}{4}$. In the latter case $G$ does not alter the initial superposition.

This leads to Algorithm \ref{grover-alg}, where lines \ref{gstart}-\ref{gend} are implemented by the quantum circuit of Figure \ref{qsa}.

\begin{algorithm}[ht]
\begin{algorithmic}[1]
\Require A blackbox function $\phi: \mathbb{B}^n\rightarrow \mathbb{B}$ and $M=|\{x\in \mathbb{B}^n|\phi(x)=1\}|$
\Ensure $\phi(x)=1$ 
\If{$M>\frac{3}{4}N$}
\State Choose $x \in \mathbb{B}^n$ with uniform probability \label{sample}
\State\Return $x$
\Else
\State $R\gets\left\lfloor \frac{\pi}{4}\sqrt{\frac{M}{N}}\right\rfloor$
\State Prepare the initial superposition $\frac{1}{N^{1/2}}\sum_{x=0}^{N-1}\ket{x}$\label{gstart}
\State Apply operator $G$ $R$ times
\State Measure to get $x\in \mathbb{B}^n$\label{gend}
\State\Return $x$
\EndIf
\end{algorithmic}
\caption{Grover's algorithm.\label{grover-alg}}
\end{algorithm}

\begin{theorem}
\label{grover-compl}
Algorithm \ref{grover-alg} solves FSAT with probability at least $\frac{1}{4}$ and $O(\sqrt{N})$ queries to $\phi$ when $M$ is considered constant.
\end{theorem}
\begin{proof}
If $M\geq \frac{3}{4}N$, line \ref{sample} finds a solution with probability at least $\frac{3}{4}$.

Otherwise, in each application of $G$ we query $\phi$ and the number of applications is $R\leq  \frac{\pi}{4}\sqrt{\frac{N}{M}}$,
so $R\in O(\sqrt{N})$
\end{proof}
A notable value of $M$ is $N/4$ for which $k=\frac{\pi/3}{2\pi/6}=1$ and $R=1$, so one rotation is enough to find a solution with certainty.
For $N/4\leq M< N/2$ we have that $R=1$ and, for decreasing $M$, $R$  grows.

%
We can use this  as a probabilistic algorithm: if we have an algorithm $A$ that returns the required $x$ with probability $p$, with $0<p\leq 1$, and returns ''no answer`` with probability $1-p$, we can obtain an algorithm that 
fails to find the solution with a probability smaller than a given constant $\epsilon>0$. This is achieved by running $A$ multiple times: after
$o$ executions of $A$, the probability of not finding a solution is $(1-p)^o$. Since $(1-p)^o\leq e^{-op}$ and we want to achieve  $(1-p)^o\leq \epsilon$, picking
$o$ such that $o=-\frac{\log\epsilon}{p}$ guarantees that the algorithm fails to find a solution with probability smaller than $\epsilon$.

In the case of the Grover algorithm, this means running it for $-4\log\epsilon$ times to obtain a solution with probability smaller than $\epsilon$ while still requiring $O(\sqrt{N})$ calls to the oracle, when $M$ and $\epsilon$ are held constant.

When the number of solutions $M$ is not known, Algorithm \ref{grover-alg-unknownM} can be used \citep{hirvensalo2013quantum}.
\begin{algorithm}[ht]
\begin{algorithmic}[1]
\Require A blackbox function $\phi: \mathbb{B}^n\rightarrow \mathbb{B}$
\Ensure $\phi(x)=1$ 
\State Pick an element $x\in \mathbb{B}$
\If{$\phi(x)=1$}
\State\Return $x$
\Else
\State $m=\left\lfloor \sqrt{N}\right\rfloor+1$
\State Choose an integer $R$ uniformly in $[0,m-1]$
\State Prepare the initial superposition $\frac{1}{N^{1/2}}\sum_{x=0}^{N-1}\ket{x}$\label{gstart-u}
\State Apply operator $G$ $R$ times
\State Measure to get $x\in \mathbb{B}^n$\label{gend-u}
\State\Return $x$
\EndIf
\end{algorithmic}
\caption{Quantum search when the number of solutions is not known.\label{grover-alg-unknownM}}
\end{algorithm}
Let us first present two lemmas.
\begin{lemma}[Lemma 5.3.1 in \citep{hirvensalo2013quantum}]
\label{lemma-531hirvensalo}
For any real $\alpha$ and any positive integer $m$
$$\sum_{R=0}^{m-1}\cos((2R+1)\alpha)=\frac{\sin(2m\alpha)}{2\sin\alpha}$$
\end{lemma}
\begin{lemma}[\citep{boyer1998tight}]
\label{lemma-hirvensalo}
Let $\phi:\mathbb{B}^n \rightarrow \mathbb{B}$ be such that there are $M$ elements $x\in \mathbb{B}^n$ satisfying $\phi(x)=1$. Assume that $M\leq \frac{3}{4}N$. and let $\theta \in [0,\pi/3]$ be be defined by $\sin^2 \theta=\frac{M}{N}$. Let $m$ be any positive integer and $R\in [0,m-1]$ chosen with uniform distribution. If $G$ is applied to
 initial superposition
$$\frac{1}{N^{1/2}}\sum_{x=0}^{N-1}\ket{x}$$
$R$ times, then the probability of seeing a solution is
$$P_m=\frac{1}{2}-\frac{\sin(4m\theta)}{4m\sin(2\theta)}$$
\end{lemma}
\begin{proof}
After $R$ iterations of $G$, the probability of seeing a solution is $\sin^2((2R+1)\theta)$. So if $R$ is chosen uniformly in $[0,m-1]$, the probability of seeing 
a solution is
\begin{eqnarray*}
P_m&=&\frac{1}{m}\sum_{R=0}^{m-1}\sin^2((2R+1)\theta)\\
&=&\frac{1}{2m}\sum_{R=0}^{m-1}\left(1-\cos((2R+1)2\theta)\right)\\
&=&\frac{1}{2}-\frac{\sin(4m\theta)}{4m\sin(2\theta)}
\end{eqnarray*}
due to Lemma \ref{lemma-531hirvensalo}
\end{proof}
Now we can present the main theorem.
\begin{theorem}[\citep{hirvensalo2013quantum}]
\label{ext-grover-compl}
Algorithm \ref{grover-alg-unknownM} solves CS with  probability at least $\frac{1}{4}$ and $O(\sqrt{N})$ queries to $\phi$.
\end{theorem}
\begin{proof}
If $m\geq \frac{1}{\sin(2\theta)}$, then
$$\sin(4m\theta)\leq1=\frac{1}{\sin(2\theta)}\sin(2\theta)\leq m\sin(2\theta)$$
so
$$\frac{\sin(4m\theta)}{4m\sin(2\theta)}\leq\frac{1}{4}.$$
By Lemma \ref{lemma-hirvensalo}, then $P_m\geq\frac{1}{4}$.
Given that $0<M\leq \frac{3}{4}N$ then
$$\frac{1}{\sin(2\theta)}=\frac{1}{2\sin\theta\cos\theta}=\frac{N}{2\sqrt{M(N-M)}}\leq\sqrt{\frac{N}{M}}\leq\sqrt{N}.$$
So, if we pick $m\geq\sqrt{N}$, we ensure that $m\geq \frac{1}{\sin(2\theta)}$. Choosing  $m=\left\lfloor \sqrt{N}\right\rfloor+1$ we have that $P_m\geq \frac{1}{4}$ and 
the number of applications of $G$ is $O(\sqrt{N})$.
\end{proof}
Again, the probability of success can be made arbitrary close to 1 by repeating the algorithm.

An alternativa approach to make sure that $\frac{M}{N}\leq\frac{3}{4}$,  it is  to consider an extra qubit $X_{n+1}$, with 
$X'=[X_{1},\ldots,X_{n+1}]$ and $x'=[x_{1},\ldots,x_{n+1}]$ being two vectors of Boolean variables and values respectively, and defining a new formula $\phi'$ that is true only if both $\phi$ and $X_{n+1}$ are true, i.e., $\phi'=\phi\wedge X_{n+1}$. This leaves $M$ unchanged but multiplies $N$ by 2 so that $\sin^2\theta=\frac{M}{2N}$.
We thus obtain Algorithm \ref{grover-alg-unknownM-alt}.
\begin{algorithm}[ht]
\begin{algorithmic}[1]
\Require A blackbox function $\phi: \mathbb{B}^n\rightarrow \mathbb{B}$
\Ensure $\phi(x)=1$ 
\State $m=\left\lfloor \sqrt{\frac{N}{2}}\right\rfloor+1$
\State Choose an integer $R$ uniformly in $[0,m-1]$
\State Prepare the initial superposition $\frac{1}{N^{1/2}}\sum_{x=0}^{N-1}\ket{x}$\label{gstart-alt}
\State Apply operator $G$ $R$ times
\State Measure to get $x\in \mathbb{B}^n$\label{gend-alt}
\State\Return $x$
\end{algorithmic}
\caption{Alternative quantum search when the number of solutions is not known.\label{grover-alg-unknownM-alt}}
\end{algorithm}

\begin{lemma}
\label{lemma-hirvensalo-ext}
Let $\phi:\mathbb{B}^{n} \rightarrow \mathbb{B}$, $\phi'$ be defined as $\phi'=\phi\wedge X_{n+1}$ and be such that there are $M$ elements $x\in \mathbb{B}^{n}$ satisfying $\phi(x)=1$. Let $\theta \in [0,\pi/4]$ be defined by $sin^2 \theta=\frac{M}{2N}$. Let $m$ be any positive integer and $R\in [0,m-1]$ chosen with uniform distribution. If $G$ is applied to
 initial superposition
$$\frac{1}{\sqrt{2N}}\sum_{x'=0}^{2N-1}\ket{x'}$$
$R$ times, then the probability of seeing a solution is
$$P_m=\frac{1}{2}-\frac{\sin(4m\theta)}{4m\sin(2\theta)}$$
\end{lemma}
\begin{proof}
The proof is the same as than of Lemma \ref{lemma-hirvensalo}.
\end{proof}
\begin{theorem}
\label{ext-grover-compl-alt}
Algorithm \ref{grover-alg-unknownM-alt} solves CS with  probability at least $\frac{1}{4}$ and $O(\sqrt{N})$ queries to $\phi$.
\end{theorem}
\begin{proof}
We can follow the proof of Theorem  \ref{ext-grover-compl} with $N$ replaced by $2N$.
By Lemma \ref{lemma-hirvensalo-ext}, if $m\geq \frac{1}{\sin(2\theta)}$, then $P_m\geq \frac{1}{4}$.
Since $\sin\theta=\sqrt{\frac{M}{2N}}$ 
and  $\frac{M}{2N}\leq \frac{1}{2}$, then $\theta\in[0,\frac{\pi}{4}]$ and we have that
$$\frac{1}{\sin(2\theta)}=\frac{1}{2\sin\theta\cos\theta}=\frac{2N}{2\sqrt{M(2N-M)}}\leq\sqrt{\frac{N}{2M}}\leq\sqrt{\frac{N}{2}}.$$
So choosing  $m=\left\lfloor \sqrt{\frac{N}{2}}\right\rfloor+1$ we have that $P_m\geq \frac{1}{2}$ and 
the number of applications of $G$ is $O(\sqrt{N})$.
\end{proof}
In general, adding other extra qubits reduces the number of applications of $G$.


\section{Comparison of Quantum Search with Classical Algorithms}
\label{comparison}
Let us discuss classical  algorithms for solving  FSAT under a black box model
of computation \citep{nielsen2010quantum}, where the only knowledge we have of the Boolean function $\phi$ is the ability to evaluate it given an assignment of the 
Boolean variables, i.e., we have an oracle that answers queries over $\phi$  of the form ``given assignment $x$, is $\phi(x)$ equal to 1?''

Consider first the case that $M=1$. A deterministic algorithm for finding the single configuration $x$ of $n$ bits such that $\phi(x)=1$ under the black box model
 clearly requires $N=2^n$ evaluations of $\phi$ in the worst case. 

Let us consider a probabilistic algorithm,  i.e. an algorithm that returns the solution of the problem with  probability $p$, with $0<p\leq 1$, and returns ''no answer`` with probability $1-p$.

A classical probabilistic algorithm for solving the search problem with $M=1$ is the following: take $s$ samples of configurations of $X$ with uniform probability.
This can be performed by sampling each Boolean variable uniformly and combining the bit samples obtaining a configuration. 
Then, for each configurations,
test whether it is a solution. If it is a solution, return it and stop.
The probability of finding the single solution $x$ is $\frac{s}{N}$ so we need at least $pN$ queries to find $x$ with a probability at least $p$.

We may think that using a sampling distribution different from  uniform we may do better, but the following lemma proves that this is not true.
\begin{lemma}
\label{compl-search-1}
Let $N=2^n$ and $\phi$ be a black box function. Assume that $A_\phi$ is a probabilistic algorithm that makes queries to $\phi$ and 
returns an element $x\in \mathbb{B}^n$. If, for any non-constant $\phi$, the probability that $\phi(x)=1$ is at least $p>0$, then there is a function $\phi'$ such
$A_{\phi'}$ makes at least $pN$ queries\footnote{This lemma differs from Lemma 5.1.1 in \citep{hirvensalo2013quantum} because the bound on the number of queries here is $pN$ rather than $pN-1$, 
thus providing a tighter bound.}.
\end{lemma}
\begin{proof}
Let $\phi_y$ be a Boolean function such that
$$\phi_y(z)=\left\{\begin{array}{ll}
1&\mbox{if }z=y\\
0&\mbox{otherwise}
\end{array}\right.$$
Let $P_y(s)$ be the probability that $A_{\phi_y}$ returns $y$ using $s$ queries.
Suppose the algorithm samples the configurations to query with a probability distribution that assigns probability $q_z$ to $z$. We have that
$\sum_{z\in\mathbb{B}^n}q_z=1$

We show by induction that
$$\sum_{y\in\mathbb{B}^n}P_y(s)\leq s$$
For $s=1$, we have
$$\sum_{y\in\mathbb{B}^n}P_y(1)=\sum_{y\in\mathbb{B}^n}q_y=1$$
Now suppose
$$\sum_{y\in\mathbb{B}^n}P_y(s-1)\leq s-1$$
and consider the $s$-th query. $A_{\phi_y}$ queries $\phi(y)$ with probability $q_y$ so
$P_y(s)\leq P_y(s-1)+q_y$. So
$$\sum_{y\in\mathbb{B}^n}P_y(s)\leq \sum_{y\in\mathbb{B}^n}P_y(s-1)+\sum_{y\in\mathbb{B}^n}q_y\leq s-1+1=s$$
There are $N$ different choices for $y$ so there must exist one with
$$P_y(s)\leq \frac{s}{N}$$
because otherwise $P_y(s)>\frac{s}{N}$ for all $y$ implies that $\sum_{y\in\mathbb{B}^n}P_y(s)>N\frac{s}{N}=s$.

By assumption $P_y(s)\geq p$, so $\frac{s}{N}\geq P_y(s)\geq p$ implies that
$s\geq pN$.
\end{proof}
When $M>1$, the probabilistic algorithm that samples uniformly has probability $\frac{sM}{N}$ of sampling a solution by taking $s$ samples so we would need at least
$\frac{pN}{M}$ samples to obtain a solution with probability at least $p$. Again, using a non uniform sampling distribution does not provide improvements, as the next lemma shows.

\begin{lemma}
\label{compl-search-m}
Let $N=2^n$ and $\phi$ be a black box function with $M$ configurations $x$ for which $\phi(x)=1$. Assume that $A_\phi$ is a probabilistic algorithm that makes queries to $\phi$ and 
returns an element $x\in \mathbb{B}^n$. If, for any $\phi$ with $M$ solutions, the probability that $\phi(x)=1$ is at least $p>0$, then there is a function $\phi'$ such
$A_{\phi'}$ makes at least $p\frac{N}{M}$ queries.
\end{lemma}
\begin{proof}
Let $\phi_S$ be a Boolean function such that
$$\phi_S(z)=\left\{\begin{array}{ll}
1&\mbox{if }z\in S\\
0&\mbox{otherwise}
\end{array}\right.$$
where $S$ is a subset of $\mathbb{B}^n$ such that $|S|=M$.
Let $P_S(s)$ be the probability that $A_{\phi_S}$ returns $x$ such that $x\in S$ using $s$ queries.
Suppose the algorithm samples the configurations to query with a probability distribution that assign probability $q_z$ to $z$. We have that
$\sum_{z\in\mathbb{B}^n}q_z=1$

We show by induction that
$$\sum_{S\subseteq\mathbb{B}^n,|S|=M}P_S(s)\leq s {N-1 \choose M-1}$$
For $s=1$, we have
$$\sum_{S\subseteq\mathbb{B}^n,|S|=M}P_S(1)=\sum_{S\subseteq\mathbb{B}^n,|S|=M}\sum_{z\in S}q_z$$
Each $z$ appears in a number of subsets that is ${N-1 \choose M-1}$ so
$$\sum_{S\subseteq\mathbb{B}^n,|S|=M}P_S(1)=\sum_{z\in \mathbb{B}^n}{N-1 \choose M-1}q_z={N-1 \choose M-1}\sum_{z\in \mathbb{B}^n}q_z={N-1 \choose M-1}$$
Now suppose
$$\sum_{S\subseteq\mathbb{B}^n,|S|=M}P_S(s -1)\leq (s -1){N-1 \choose M-1}$$
and consider the $s$-th query. $A_{\phi_S}$ queries $\phi(z)$ with probability $q_z$ so
$P_S(s)\leq P_S(s -1)+\sum_{z\in S}q_z$. So
$$\sum_{S\subseteq\mathbb{B}^n,|S|=M}P_S(s )\leq \sum_{S\subseteq\mathbb{B}^n,|S|=M}P_S(s-1)+\sum_{S\subseteq\mathbb{B}^n,|S|=M}\sum_{z\in S}q_z$$
Again, 
$$\sum_{S\subseteq\mathbb{B}^n,|S|=M}P_S(s )\leq \sum_{S\subseteq\mathbb{B}^n,|S|=M}P_S(s -1)+\sum_{z\in \mathbb{B}^n}{N-1 \choose M-1}q_z$$
$$\sum_{S\subseteq\mathbb{B}^n,|S|=M}P_S(s)\leq \sum_{S\subseteq\mathbb{B}^n,|S|=M}P_S(s -1)+{N-1 \choose M-1}\sum_{z\in \mathbb{B}^n}q_z$$
$$\sum_{S\subseteq\mathbb{B}^n,|S|=M}P_S(s)\leq\sum_{S\subseteq\mathbb{B}^n,|S|=M}P_S(s -1)+{N-1 \choose M-1}$$
$$\sum_{S\subseteq\mathbb{B}^n,|S|=M}P_S(s)\leq s {N-1 \choose M-1}$$
There are ${N \choose M}$ different choices for $S$ so there must exist one with
$$P_S(s )\leq \frac{s }{\frac{N}{M}}$$
because otherwise $P_S(s )>\frac{s }{\frac{N}{M}}$ for all $S$ implies that 
$$\sum_{S\subseteq\mathbb{B}^n,|S|=M}P_S(s )>{N\choose M}\frac{sM}{N}$$
$$\sum_{S\subseteq\mathbb{B}^n,|S|=M}P_S(s )>\frac{N!}{(N-M)!M!}\frac{s M}{N}=s \frac{(N-1)!}{(N-M)!(M-1)!}=s {N-1 \choose M-1}$$

By assumption $P_S(s)\geq p$, so $\frac{s }{\frac{N}{M}}\geq P_S(s )\geq p$ implies that
$s \geq p\frac{N}{M}$.
\end{proof}
We can now present the main result.
\begin{theorem}
Any classical algorithm for solving the FSAT  problem with probability at least $\frac{1}{4}$ requires  $\Omega(N)$ queries to the oracle, when considering $M$ fixed.
\end{theorem}
\begin{proof}
By Lemma \ref{compl-search-m} a classical algorithm that solves FSAT with probability $\frac{1}{4}$ makes at least $\frac{1}{4}\frac{N}{M}$ queries.
\end{proof}
So quantum search offers a quadratic improvement over classical search.

If we consider the CS problem, sampling non-uniformly is not an option. The result thus is the same.
\begin{theorem}
Any classical algorithm for solving the CS  problem with $Q=X$ and probability at least $\frac{1}{4}$ requires  $\Omega(N)$ queries to the oracle, when considering $M$ fixed.
\end{theorem}

\section{Quantum Weighted Constrained Sampling}
\label{qmap}

Suppose first that the literal weights sum to 1, i.e., that $w(X_i)+w(\neg X_i)=1$ for all
bits $X_i$.

Given a Boolean function $\phi:\mathbb{B}^n \rightarrow \mathbb{B}$, a weight function $w:L\rightarrow [0,1]$ and set of variables $Q$,
we want to sample values for the variables $Q$ so that the probability of sampling configuration $q$ is
\begin{equation*}
P(q)=\frac{ \sum_{y:\phi(qy)=1}W_{qy}}{\WMC(\phi,w)}
\end{equation*}
Suppose, without loss of generality, that the query bits $Q$ come first and there are $l$ of them, while there are  and there are $n-l$ bits in $Y$.
Assume also that we add an extra bit $X_{n+1}$ so  
$X'=[X_{1},\ldots,X_{n+1}]$, $x'=[x_{1},\ldots,x_{n+1}]$, $\phi'=\phi\wedge X_{n+1}$. 
Let $Y'$ be $Y$ with the extra bit $X_{n+1}$, so overall $Y'$ has $n-l+1$ bits.

\begin{figure}[t]
\centering
\begin{scriptsize}
\begin{tikzpicture}[scale=0.700000,x=1pt,y=1pt]
\filldraw[color=white] (0.000000, -7.500000) rectangle (52.000000, 67.500000);
\draw[color=black] (0.000000,60.000000) -- (52.000000,60.000000);
\draw[color=black] (0.000000,45.000000) -- (52.000000,45.000000);
\draw[color=black] (0.000000,15.000000) -- (52.000000,15.000000);
\draw[color=black] (0.000000,0.000000) -- (52.000000,0.000000);
\begin{scope}
\draw[fill=white] (26.000000, 60.000000) +(-45.000000:28.284271pt and 8.485281pt) -- +(45.000000:28.284271pt and 8.485281pt) -- +(135.000000:28.284271pt and 8.485281pt) -- +(225.000000:28.284271pt and 8.485281pt) -- cycle;
\clip (26.000000, 60.000000) +(-45.000000:28.284271pt and 8.485281pt) -- +(45.000000:28.284271pt and 8.485281pt) -- +(135.000000:28.284271pt and 8.485281pt) -- +(225.000000:28.284271pt and 8.485281pt) -- cycle;
\draw (26.000000, 60.000000) node {$R_y(\theta_1)$};
\end{scope}
\begin{scope}
\draw[fill=white] (26.000000, 45.000000) +(-45.000000:28.284271pt and 8.485281pt) -- +(45.000000:28.284271pt and 8.485281pt) -- +(135.000000:28.284271pt and 8.485281pt) -- +(225.000000:28.284271pt and 8.485281pt) -- cycle;
\clip (26.000000, 45.000000) +(-45.000000:28.284271pt and 8.485281pt) -- +(45.000000:28.284271pt and 8.485281pt) -- +(135.000000:28.284271pt and 8.485281pt) -- +(225.000000:28.284271pt and 8.485281pt) -- cycle;
\draw (26.000000, 45.000000) node {$R_y(\theta_2)$};
\end{scope}
\begin{scope}
\draw (26.000000, 30.000000) node {$\cdots$};
\end{scope}
\begin{scope}
\draw[fill=white] (26.000000, 15.000000) +(-45.000000:28.284271pt and 8.485281pt) -- +(45.000000:28.284271pt and 8.485281pt) -- +(135.000000:28.284271pt and 8.485281pt) -- +(225.000000:28.284271pt and 8.485281pt) -- cycle;
\clip (26.000000, 15.000000) +(-45.000000:28.284271pt and 8.485281pt) -- +(45.000000:28.284271pt and 8.485281pt) -- +(135.000000:28.284271pt and 8.485281pt) -- +(225.000000:28.284271pt and 8.485281pt) -- cycle;
\draw (26.000000, 15.000000) node {$R_y(\theta_n)$};
\end{scope}
\begin{scope}
\draw[fill=white] (26.000000, -0.000000) +(-45.000000:8.485281pt and 8.485281pt) -- +(45.000000:8.485281pt and 8.485281pt) -- +(135.000000:8.485281pt and 8.485281pt) -- +(225.000000:8.485281pt and 8.485281pt) -- cycle;
\clip (26.000000, -0.000000) +(-45.000000:8.485281pt and 8.485281pt) -- +(45.000000:8.485281pt and 8.485281pt) -- +(135.000000:8.485281pt and 8.485281pt) -- +(225.000000:8.485281pt and 8.485281pt) -- cycle;
\draw (26.000000, -0.000000) node {$H$};
\end{scope}
\end{tikzpicture}
\end{scriptsize}
\caption{Circuit for gate $Rot$.}
\label{qwmc_rot}
\end{figure}
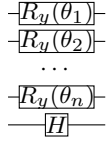

We perform quantum WCS by modifying the algorithm for quantum search, obtaining QWCS.
The circuit for performing QWCS  differs from the one in Figure \ref{qsa} because the Hadamard operations  applied to the lower register are replaced
by rotations $R_y(\theta_i)$ where $i$ is the qubit index except for the extra qubit for which the Hadamard operator is kept. Overall the gate $H^{\otimes n+1}$ is replaced by gate $Rot$ shown in Figure \ref{qwmc_rot}.
$\theta_i$ is computed as
$$\theta_i=2\arccos \sqrt{1-w_i}$$
where $w_i=w(X_i)$.

So
$$\cos \theta_i/2=\cos \arccos \sqrt{1-w_i}=\sqrt{1-w_i}$$
and
$$\sin \theta_i/2=\sqrt{1-(\cos\theta_i/2)^2}=\sqrt{w_i}$$
The effect of the rotation on the $i$th bit is
\begin{eqnarray*}
R_y(\theta_i)\ket{0}&=&\left[\begin{array}{cc}
\cos\frac{\theta_i}{2}&-\sin\frac{\theta_i}{2}\\
\sin\frac{\theta_i}{2}&\cos\frac{\theta_i}{2}
\end{array}\right] \left[\begin{array}{c}
1\\
0
\end{array}
\right]=\left[\begin{array}{c}
\cos\frac{\theta_i}{2}\\
\sin\frac{\theta_i}{2}\
\end{array}
\right]\\
&=&
\left[\begin{array}{c}
\sqrt{1-w_i}\\
\sqrt{w_i}
\end{array}
\right]=\sqrt{1-w_i}\ket{0}+\sqrt{w_i}\ket{1}
\end{eqnarray*}
Therefore the rotations prepare the state
\begin{eqnarray*}
\ket{\varphi}&=&\bigotimes_{i=1}^n( \sqrt{1-w_i}\ket{0}+\sqrt{w_i}\ket{1})\otimes \frac{1}{\sqrt{2}}(\ket{0}+\ket{1})\\
&=&\sum_{x_{1}\ldots x_{n+1}=0}^{2^{n+1}-1}\sqrt{w'_{{1}}\ldots w'_{n+1}}\ket{x_{1}\ldots x_{n+1}}
\end{eqnarray*}
where $w'_{i}$ is
$$w'_{i}=\left\{\begin{array}{ll}
w_i&\mbox{if }x_i=1\\
1-w_i&\mbox{if }x_i=0
\end{array}\right.
$$
for $i=1,\ldots,n$ and $w'_{n+1}=1/2$.
$\ket{\varphi}$ can be rewritten as
\begin{eqnarray*}
\ket{\varphi}&=&\sum_{x':\phi'(x')=0}\sqrt{\frac{w'_{{1}}\ldots w'_{n}}{2}}\ket{x'}+\sum_{x':\phi'(x')=1}\sqrt{\frac{w'_{{1}}\ldots w'_{n}}{2}}\ket{x'}
\end{eqnarray*}
where the first summation is over all configurations $x'$ of the $n+1$ bits such that $\phi'(x')=0$ and the latter over all confgurations $x'$ of the $n+1$ bits such that $\phi'(x')=1$. Then
\begin{eqnarray*}
\ket{\varphi}&=&\sum_{x,x_{n+1}:x_{n+1}=0\vee\phi(x)=0}\sqrt{\frac{w'_{{1}}\ldots w'_{n}}{2}}\ket{x'}+\sum_{x':\phi'(x')=1}\sqrt{\frac{w'_{{1}}\ldots w'_{n}}{2}}\ket{x'}
\end{eqnarray*}
because $\phi'(x')$ is 0 if $x_{n+1}$ is 0 or $\phi(x)=0$. Moreover
\begin{eqnarray*}
\ket{\varphi}&=&\sum_{x,x_{n+1}:x_{n+1}=0\vee(x_{n+1}=1\wedge\phi(x)=0)}\sqrt{\frac{w'_{{1}}\ldots w'_{n}}{2}}\ket{x'}\\
&&+\sum_{x':\phi'(x')=1}\sqrt{\frac{w'_{{1}}\ldots w'_{n}}{2}}\ket{x'}\\
&=&\sum_{x,x_{n+1}:x_{n+1}=0}\sqrt{\frac{w'_{{1}}\ldots w'_{n}}{2}}\ket{x'}+\sum_{x,x_{n+1}:x_{n+1}=1\wedge\phi(x)=0}\sqrt{\frac{w'_{{1}}\ldots w'_{n}}{2}}\ket{x'}+\\
&&+\sum_{x':\phi'(x')=1}\sqrt{\frac{w'_{{1}}\ldots w'_{n}}{2}}\ket{x'}\\
&=&\sum_{x}\sqrt{\frac{w'_{1}\ldots w'_{n}}{2}}\ket{x0}+\sum_{x:\phi(x)=0}\sqrt{\frac{w'_{{1}}\ldots w'_{n}}{2}}\ket{x1}\\
&&+\sum_{x:\phi(x)=1}\sqrt{\frac{w'_{{1}}\ldots w'_{n}}{2}}\ket{x1}
\end{eqnarray*}
where the last equation is obtained by setting the $x_{n+1}$ bit in the quantum state.

Define $W_{x}$ as $w'_{{1}}\cdot\ldots\cdot w'_{n}$
and normalized states
\begin{eqnarray*}
\ket{\gamma}&=&\frac{1}{\sqrt{\frac{1+\sum_{x:\phi(x)=0}W_x}{2}}}\left(\sum_{x}\sqrt{\frac{W_x}{2}}\ket{x0}+\sum_{x:\phi(x)=0}\sqrt{\frac{W_x}{2}}\ket{x1}\right)\\
\ket{\delta}&=&\frac{1}{\sqrt{\frac{\sum_{x:\phi(x)=1}W_x}{2}}}\sum_{x:\phi(x)=1}\sqrt{\frac{W_x}{2}}\ket{x1},
\end{eqnarray*}
then $\ket{\varphi}$ can be expressed as
\begin{equation*}
\ket{\varphi}=\left(\sqrt{\frac{1+\sum_{x:\phi(x)=0}W_x}{2}}\right)\ket{\gamma}+\left(\sqrt{\frac{\sum_{x:\phi(x)=1}W_x}{2}}\right)\ket{\delta}
\end{equation*}
so the initial state of the quantum computer is in the space spanned by $\ket{\gamma}$ and $\ket{\delta}$

Let $\cos \theta=\sqrt{\frac{1+\sum_{x:\phi(x)=0}W_x}{2}}$ and
 $\sin \theta=\sqrt{\frac{\sum_{x:\phi(x)=1}W_x}{2}}$
 so that 
$$\ket{\varphi}=\cos\theta/2\ket{\gamma}+\sin\theta/2\ket{\delta}.$$
Note that, correctly, $\cos^2 \theta + \sin^2\theta=1$ since $\sum_{x:\phi(x)=0}W_x+\sum_{x:\phi(x)=1}W_x=1$.
Gate $Rot$ replaces $H^{\otimes n+1}$ also in the Grover operator $G$ that becomes the weighted Grover operator $\WG$ shown in Figure \ref{grover_rot}:
\begin{eqnarray*}
\WG&=&Rot(2\ket{0}\bra{0}-I)Rot^\dagger O\\
&=&(2Rot\ket{0}\bra{0}Rot^\dagger- Rot I Rot^\dagger)O\\
&=&(2\ket{\varphi}\bra{\varphi}-I)O
\end{eqnarray*}
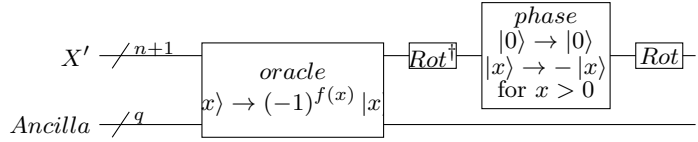
\begin{figure}[t]
\centering
\begin{scriptsize}
\begin{tikzpicture}[scale=0.800000,x=1pt,y=1pt]
\filldraw[color=white] (0.000000, -7.500000) rectangle (279.000000, 57.500000);
\draw[color=black] (0.000000,32.500000) -- (279.000000,32.500000);
\draw[color=black] (0.000000,32.500000) node[left] {$X'$};
\draw[color=black] (0.000000,0.000000) -- (279.000000,0.000000);
\draw[color=black] (0.000000,0.000000) node[left] {$Ancilla$};
\draw (6.000000, 26.500000) -- (14.000000, 38.500000);
\draw (12.000000, 35.500000) node[right] {$\scriptstyle{n+1}$};
\draw (6.000000, -6.000000) -- (14.000000, 6.000000);
\draw (12.000000, 3.000000) node[right] {$\scriptstyle{q}$};
\draw (90.500000,32.500000) -- (90.500000,0.000000);
\begin{scope}
\draw[fill=white] (90.500000, 16.250000) +(-45.000000:60.104076pt and 31.466252pt) -- +(45.000000:60.104076pt and 31.466252pt) -- +(135.000000:60.104076pt and 31.466252pt) -- +(225.000000:60.104076pt and 31.466252pt) -- cycle;
\clip (90.500000, 16.250000) +(-45.000000:60.104076pt and 31.466252pt) -- +(45.000000:60.104076pt and 31.466252pt) -- +(135.000000:60.104076pt and 31.466252pt) -- +(225.000000:60.104076pt and 31.466252pt) -- cycle;
\draw (90.500000, 16.250000) node {$\begin{array}{c}oracle\\\ket{x}\to (-1)^{f(x)}\ket{x}\end{array}$};
\end{scope}
\begin{scope}
\draw[fill=white] (156.000000, 32.500000) +(-45.000000:15.556349pt and 8.485281pt) -- +(45.000000:15.556349pt and 8.485281pt) -- +(135.000000:15.556349pt and 8.485281pt) -- +(225.000000:15.556349pt and 8.485281pt) -- cycle;
\clip (156.000000, 32.500000) +(-45.000000:15.556349pt and 8.485281pt) -- +(45.000000:15.556349pt and 8.485281pt) -- +(135.000000:15.556349pt and 8.485281pt) -- +(225.000000:15.556349pt and 8.485281pt) -- cycle;
\draw (156.000000, 32.500000) node {$Rot^\dagger$};
\end{scope}
\begin{scope}
\draw[fill=white] (209.000000, 32.500000) +(-45.000000:42.426407pt and 35.355339pt) -- +(45.000000:42.426407pt and 35.355339pt) -- +(135.000000:42.426407pt and 35.355339pt) -- +(225.000000:42.426407pt and 35.355339pt) -- cycle;
\clip (209.000000, 32.500000) +(-45.000000:42.426407pt and 35.355339pt) -- +(45.000000:42.426407pt and 35.355339pt) -- +(135.000000:42.426407pt and 35.355339pt) -- +(225.000000:42.426407pt and 35.355339pt) -- cycle;
\draw (209.000000, 32.500000) node {$\begin{array}{c}phase\\\ket{0}\to\ket{0}\\\ket{x}\to -\ket{x}\\\mbox{for }x>0\end{array}$};
\end{scope}
\begin{scope}
\draw[fill=white] (262.000000, 32.500000) +(-45.000000:15.556349pt and 8.485281pt) -- +(45.000000:15.556349pt and 8.485281pt) -- +(135.000000:15.556349pt and 8.485281pt) -- +(225.000000:15.556349pt and 8.485281pt) -- cycle;
\clip (262.000000, 32.500000) +(-45.000000:15.556349pt and 8.485281pt) -- +(45.000000:15.556349pt and 8.485281pt) -- +(135.000000:15.556349pt and 8.485281pt) -- +(225.000000:15.556349pt and 8.485281pt) -- cycle;
\draw (262.000000, 32.500000) node {$Rot$};
\end{scope}
\end{tikzpicture}
\end{scriptsize}
\caption{Weighted Grover operator $\WG$.}
\label{grover_rot}
\end{figure}
The overall circuit for performing QWCS is shown in Figure \ref{qmapcirc}, where the Grover operator of quantum search is replaced by the
Weighted Grover operator  $\WG$ shown in Figure \ref{grover_rot} and the initial $H^{\otimes n+1}$ gate is replaced by gate $Rot$ of Figure \ref{qwmc_rot}.

The initial state $\ket{\varphi}$ of the quantum computer is in the space spanned by $\ket{\gamma}$ and $\ket{\delta}$ which are orthonormal since they do not share computational basis states.
As for the Grover algorithm, the application of 
$\WG$  rotates $\ket{\varphi}$ in the space spanned by $\ket{\gamma}$ and $\ket{\delta}$ by angle
 $\theta$.

\begin{figure}[t]
\centering
\begin{tikzpicture}[scale=1.000000,x=1pt,y=1pt]
\filldraw[color=white] (0.000000, -3.500000) rectangle (219.000000, 122.500000);
\draw[color=black] (0.000000,115.000000) -- (207.000000,115.000000);
\draw[color=black] (207.000000,114.500000) -- (219.000000,114.500000);
\draw[color=black] (207.000000,115.500000) -- (219.000000,115.500000);
\draw[color=black] (0.000000,85.000000) -- (207.000000,85.000000);
\draw[color=black] (207.000000,84.500000) -- (219.000000,84.500000);
\draw[color=black] (207.000000,85.500000) -- (219.000000,85.500000);
\filldraw[color=white,fill=white] (0.000000,81.250000) rectangle (-4.000000,118.750000);
\draw[decorate,decoration={brace,amplitude = 4.000000pt},very thick] (0.000000,81.250000) -- (0.000000,118.750000);
\draw[color=black] (-4.000000,100.000000) node[left] {${\begin{array}{c}Q\\\ket{0}^{\otimes l}\end{array}}$};
\draw[color=black] (0.000000,70.000000) -- (219.000000,70.000000);
\draw[color=black] (0.000000,40.000000) -- (219.000000,40.000000);
\draw[color=black] (0.000000,25.000000) -- (219.000000,25.000000);
\filldraw[color=white,fill=white] (0.000000,21.250000) rectangle (-4.000000,73.750000);
\draw[decorate,decoration={brace,amplitude = 4.000000pt},very thick] (0.000000,21.250000) -- (0.000000,73.750000);
\draw[color=black] (-4.000000,47.500000) node[left] {${\begin{array}{c}Y'\\\ket{0}^{\otimes n-l+1}\end{array}}$};
\draw[color=black] (0.000000,14.000000) -- (219.000000,14.000000);
\draw[color=black] (0.000000,7.000000) -- (219.000000,7.000000);
\draw[color=black] (0.000000,0.000000) -- (219.000000,0.000000);
\filldraw[color=white,fill=white] (0.000000,-1.750000) rectangle (-4.000000,15.750000);
\draw[decorate,decoration={brace,amplitude = 4.000000pt},very thick] (0.000000,-1.750000) -- (0.000000,15.750000);
\draw[color=black] (-4.000000,7.000000) node[left] {${\begin{array}{c}\mbox{Ancilla}\\\ket{0}^{\otimes a}\end{array}}$};
\draw (21.000000,115.000000) -- (21.000000,25.000000);
\begin{scope}
\draw[fill=white] (21.000000, 70.000000) +(-45.000000:21.213203pt and 72.124892pt) -- +(45.000000:21.213203pt and 72.124892pt) -- +(135.000000:21.213203pt and 72.124892pt) -- +(225.000000:21.213203pt and 72.124892pt) -- cycle;
\clip (21.000000, 70.000000) +(-45.000000:21.213203pt and 72.124892pt) -- +(45.000000:21.213203pt and 72.124892pt) -- +(135.000000:21.213203pt and 72.124892pt) -- +(225.000000:21.213203pt and 72.124892pt) -- cycle;
\draw (21.000000, 70.000000) node {$Rot$};
\end{scope}
\draw (63.000000,115.000000) -- (63.000000,0.000000);
\begin{scope}
\draw[fill=white] (63.000000, 57.500000) +(-45.000000:21.213203pt and 89.802561pt) -- +(45.000000:21.213203pt and 89.802561pt) -- +(135.000000:21.213203pt and 89.802561pt) -- +(225.000000:21.213203pt and 89.802561pt) -- cycle;
\clip (63.000000, 57.500000) +(-45.000000:21.213203pt and 89.802561pt) -- +(45.000000:21.213203pt and 89.802561pt) -- +(135.000000:21.213203pt and 89.802561pt) -- +(225.000000:21.213203pt and 89.802561pt) -- cycle;
\draw (63.000000, 57.500000) node {$WG$};
\end{scope}
\draw (105.000000,115.000000) -- (105.000000,0.000000);
\begin{scope}
\draw[fill=white] (105.000000, 57.500000) +(-45.000000:21.213203pt and 89.802561pt) -- +(45.000000:21.213203pt and 89.802561pt) -- +(135.000000:21.213203pt and 89.802561pt) -- +(225.000000:21.213203pt and 89.802561pt) -- cycle;
\clip (105.000000, 57.500000) +(-45.000000:21.213203pt and 89.802561pt) -- +(45.000000:21.213203pt and 89.802561pt) -- +(135.000000:21.213203pt and 89.802561pt) -- +(225.000000:21.213203pt and 89.802561pt) -- cycle;
\draw (105.000000, 57.500000) node {$WG$};
\end{scope}
\draw[fill=white,color=white] (132.000000, -6.000000) rectangle (147.000000, 121.000000);
\draw (139.500000, 57.500000) node {$\cdots$};
\draw (174.000000,115.000000) -- (174.000000,0.000000);
\begin{scope}
\draw[fill=white] (174.000000, 57.500000) +(-45.000000:21.213203pt and 89.802561pt) -- +(45.000000:21.213203pt and 89.802561pt) -- +(135.000000:21.213203pt and 89.802561pt) -- +(225.000000:21.213203pt and 89.802561pt) -- cycle;
\clip (174.000000, 57.500000) +(-45.000000:21.213203pt and 89.802561pt) -- +(45.000000:21.213203pt and 89.802561pt) -- +(135.000000:21.213203pt and 89.802561pt) -- +(225.000000:21.213203pt and 89.802561pt) -- cycle;
\draw (174.000000, 57.500000) node {$WG$};
\end{scope}
\draw[fill=white] (201.000000, 109.000000) rectangle (213.000000, 121.000000);
\draw[very thin] (207.000000, 115.600000) arc (90:150:6.000000pt);
\draw[very thin] (207.000000, 115.600000) arc (90:30:6.000000pt);
\draw[->,>=stealth] (207.000000, 109.600000) -- +(80:10.392305pt);
\draw[fill=white] (201.000000, 79.000000) rectangle (213.000000, 91.000000);
\draw[very thin] (207.000000, 85.600000) arc (90:150:6.000000pt);
\draw[very thin] (207.000000, 85.600000) arc (90:30:6.000000pt);
\draw[->,>=stealth] (207.000000, 79.600000) -- +(80:10.392305pt);
\begin{scope}
\draw (207.000000, 100.000000) node {$\cdots$};
\end{scope}
\draw[decorate,decoration={brace,amplitude = 4.000000pt},very thick] (45.000000,122.500000) -- (192.000000,122.500000);
\draw (118.500000, 126.500000) node[text width=144pt,above,text centered] {$R$};
\end{tikzpicture}
\caption{The complete QWCS circuit.}
\label{qmapcirc}
\end{figure}
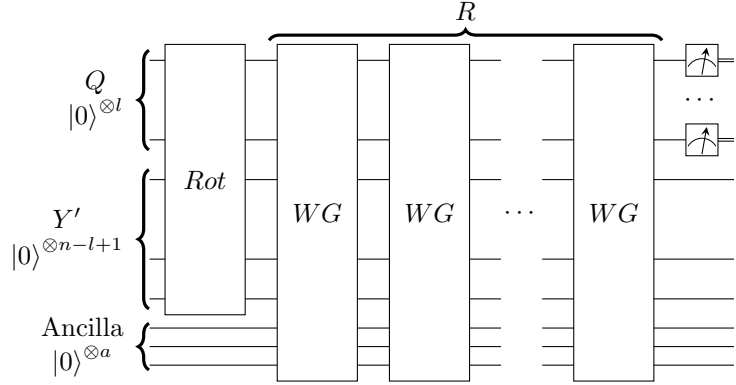

Repeated applications of $\WG$ take the state
to
$$\WG^k\ket{\varphi}=\cos ((2k+1)\theta)\ket{\gamma}+\sin((2k+1)\theta)\ket{\delta}.$$
These rotations bring the state of the system closer  to $\ket{\delta}$. 

Now we measure only the query qubits. Let us see what is the result of this measurement.
The density operator for the system is
\begin{equation*}
\rho=\left(\cos ((2k+1)\theta)\ket{\gamma}+\sin((2k+1)\theta)\ket{\delta}\right)\left(\cos ((2k+1)\theta)\bra{\gamma}+\sin((2k+1)\theta)\bra{\delta}\right)
\end{equation*}
For the distributivity of matrix multiplication, we can rewrite $\rho$ as
\begin{equation*}
\rho=\rho_1+\rho_2+\rho_3+\rho_4
\end{equation*}
with
\begin{eqnarray*}
\rho_1&=&\sin^2((2k+1)\theta)\ket{\delta}\bra{\delta}\\
\rho_2&=&\cos ((2k+1)\theta)\sin((2k+1)\theta)\ket{\gamma}\bra{\delta}\\
\rho_3&=&\sin((2k+1)\theta)\cos ((2k+1)\theta)\ket{\delta}\bra{\gamma}\\
\rho_4&=&\cos^2 ((2k+1)\theta)\ket{\gamma}\bra{\gamma}
\end{eqnarray*}
We now trace out bits $Y'$ obtaining the reduced density operator for $q$
\begin{eqnarray*}
\rho^Q&=&\tr_{Y'}(\rho_1+\rho_2+\rho_3+\rho_4)\nonumber\\
&=&\tr_{Y'}(\rho_1)+\tr_{Y'}(\rho_2)+\tr_{Y'}(\rho_3)+\tr_{Y'}(\rho_4)\\
&=&\rho^Q_1+\rho^Q_2+\rho^Q_3+\rho^Q_4
\end{eqnarray*}
for the linearity of the partial trace operator.

Let us consider first $\rho_1$
We  rewrite $\ket{\delta}$ as
\begin{eqnarray*}
\ket{\delta}&=&\frac{1}{\sqrt{\frac{\sum_{x:\phi(x)=1}W_x}{2}}}\sum_{x:\phi(x)=1}\sqrt{\frac{W_x}{2}}\ket{x1}\\
&=&\frac{1}{Z_1}\sum_{x:\phi(x)=1}\sqrt{\frac{W_x}{2}}\ket{x1}
\end{eqnarray*}
with $Z_1=\sqrt{\frac{\sum_{x:\phi(x)=1}W_x}{2}}$.

The density operator for $\rho_1$ is
\begin{equation*}
\rho_1=\sin^2((2k+1)\theta)\left(\frac{1}{Z_1}\sum_{x:\phi(x)=1}\sqrt{\frac{W_x}{2}}\ket{x1}\right)\left(\frac{1}{Z_1}\sum_{x:\phi(x)=1}\sqrt{\frac{W_x}{2}}\bra{x1}\right)
\end{equation*}
Writing a bit configuration $x$ as $qy$, $\rho_1$ becomes
\begin{small}
\begin{eqnarray}
\rho_1&=&\sin^2((2k+1)\theta)\left(\frac{1}{Z_1}\sum_{qy:\phi(qy)=1}\sqrt{\frac{W_{qy}}{2}}\ket{qy1}\right)\left(\frac{1}{Z_1}\sum_{qy:\phi(qy)=1}\sqrt{\frac{W_{qy}}{2}}\bra{qy1}\right)\nonumber\\
&=&\sin^2((2k+1)\theta)\left(\frac{1}{Z_1}\sum_{qy:\phi(qy)=1}\sqrt{\frac{W_{qy}}{2}}\ket{qy1}\right)\left(\frac{1}{Z_1}\sum_{rz:\phi(rz)=1}\sqrt{\frac{W_{rz}}{2}}\bra{rz1}\right)\label{renaming}\\
&=&\sin^2((2k+1)\theta)\frac{1}{Z_1^2}\left(\sum_{q}\sum_{y:\phi(qy)=1}\sqrt{\frac{W_{qy}}{2}}\ket{qy1}\right)\left(\sum_{r}\sum_{z:\phi(rz)=1}\sqrt{\frac{W_{rz}}{2}}\bra{rz1}\right)\label{split}\\
&=&\sin^2((2k+1)\theta)\frac{1}{2Z_1^2}\sum_{q,r}\sum_{y,z:\phi(qy)=1,\phi(rz)=1}\sqrt{W_{qy}W_{rz}}\ket{qy1}\bra{rz1}\label{join}
\end{eqnarray}
\end{small}
where we get Eq. (\ref{renaming}) by renaming $q$ and $y$ in the second factor as $r$ and $z$,  Eq. (\ref{split}) by splitting the sums and Eq. (\ref{join}) by multiplying the two factors and rearranging the terms.
We now trace out bits $Y'$ obtaining the reduced density operator for $Q$:
\begin{eqnarray}
\rho^Q_1&=&\tr_{Y'}(\rho_1)\nonumber\\
&=&\frac{\sin^2((2k+1)\theta)}{2Z_1^2}\sum_{q,r}\sum_{y,z:\phi(qy)=1,\phi(rz)=1}\sqrt{W_{qy}W_{rz}}\tr_{Y'}(\ket{qy1}\bra{rz1})\label{linearity}\\
&=&\frac{\sin^2((2k+1)\theta)}{2Z_1^2}\sum_{q,r}\sum_{y,z:\phi(qy)=1,\phi(rz)=1}\sqrt{W_{qy}W_{rz}}\ket{q}\bra{r}\tr(\ket{y1}\bra{z1})\label{trprop}\\
&=&\frac{\sin^2((2k+1)\theta)}{2Z_1^2}\sum_{q,r}\sum_{y,z:\phi(qy)=1,\phi(rz)=1}\sqrt{W_{qy}W_{rz}}\ket{q}\bra{r}\braket{z1|y1}\\
&=&\frac{\sin^2((2k+1)\theta)}{2Z_1^2}\sum_{q,r}\sum_{y:\phi(qy)=1,\phi(ry)=1}\sqrt{W_{qy}W_{ry}}\ket{q}\bra{r}\label{rhoq}
\end{eqnarray}
where we get Eq. (\ref{linearity}) for the linearity of the partial trace operator, Eq. (\ref{trprop}) for the following property of the partial trace operator 
\begin{equation}
\tr_B(\ket{a_1}\bra{a_2}\otimes\ket{b_1}\bra{b_2})=\ket{a_1}\bra{a_2}\tr (\ket{b_1}\bra{b_2}),\label{partial-trace}
\end{equation}
and  Eq. (\ref{rhoq})  because 
\begin{equation}
\label{dot-prod}
\braket{a|b}=\left\{\begin{array}{ll}
0& \mbox{if }a\neq b\\
1&\mbox{if }a= b
\end{array}\right.
\end{equation}
when $\ket{a}$ and $\ket{b}$ are computational basis states.

Let us consider $\rho_2$.
We  rewrite $\ket{\gamma}$ as
\begin{eqnarray*}
\ket{\gamma}&=&\frac{1}{\sqrt{\frac{1+\sum_{x:\phi(x)=0}W_x}{2}}}\left(\sum_{x}\sqrt{\frac{W_x}{2}}\ket{x0}+\sum_{x:\phi(x)=0}\sqrt{\frac{W_x}{2}}\ket{x1}\right)\\
&=&\frac{1}{Z_2}\left(\sum_{x}\sqrt{\frac{W_x}{2}}\ket{x0}+\sum_{x:\phi(x)=0}\sqrt{\frac{W_x}{2}}\ket{x1}\right)
\end{eqnarray*}
with $Z_2=\sqrt{\frac{1+\sum_{x:\phi(x)=0}W_x}{2}}$.

The density operator $\rho_2$ is
\begin{eqnarray*}
\rho_2&=&\cos ((2k+1)\theta)\sin((2k+1)\theta)\frac{1}{Z_2}\\
&&\left(\sum_{x}\sqrt{\frac{W_x}{2}}\ket{x0}+\sum_{x:\phi(x)=0}\sqrt{\frac{W_x}{2}}\ket{x1}\right)\left(\frac{1}{Z_1}\sum_{x:\phi(x)=1}\sqrt{\frac{W_x}{2}}\bra{x1}\right)
\end{eqnarray*}
Rewriting $x$ as $qy$ and $rz$, $\rho_2$ becomes
\begin{small}
\begin{eqnarray*}
\rho_2&=&\frac{\cos ((2k+1)\theta)\sin((2k+1)\theta)}{Z_1Z_2}\\
&&\left(\sum_{qy}\sqrt{\frac{W_{qy}}{2}}\ket{qy0}+\sum_{qy:\phi(qy)=0}\sqrt{\frac{W_{qy}}{2}}\ket{qy1}\right)\left(\sum_{rz:\phi(rz)=1}\sqrt{\frac{W_{rz}}{2}}\bra{rz1}\right)\nonumber\\
&=&\frac{\cos ((2k+1)\theta)\sin((2k+1)\theta)}{2Z_1Z_2}\sum_{q,r}\\
&&\left(\sum_{y,z:\phi(rz)=1}\sqrt{W_{qy}W_{rz}}\ket{qy0}\bra{rz1}+\sum_{y:\phi(qy)=0, z:\phi(rz)=1}\sqrt{W_{qy}W_{rz}}\ket{qy1}\bra{rz1}\right)
\end{eqnarray*}
\end{small}
We now trace out bits $Y'$ obtaining the reduced density operator for $Q$:
\begin{footnotesize}
\begin{eqnarray}
\rho^Q_2&=&\tr_{Y'}(\rho_2)\nonumber\\
&=&\frac{\cos ((2k+1)\theta)\sin((2k+1)\theta)}{2Z_1Z_2}\sum_{q,r}\nonumber\\
&&\left(\sum_{y,z:\phi(rz)=1}\sqrt{W_{qy}W_{rz}}\tr_{Y'}(\ket{qy0}\bra{rz1})+\sum_{y:\phi(qy)=0, z:\phi(rz)=1}\sqrt{W_{qy}W_{rz}}\tr_{Y'}(\ket{qy1}\bra{rz1})\right)\label{linearity2}\\
&=&\frac{\cos ((2k+1)\theta)\sin((2k+1)\theta)}{2Z_1Z_2}\sum_{q,r}\nonumber\\
&&\left(\sum_{y,z:\phi(rz)=1}\sqrt{W_{qy}W_{rz}}\ket{q}\bra{r}\braket{z1|y0})+\sum_{y:\phi(qy)=0, z:\phi(rz)=1}\sqrt{W_{qy}W_{rz}}\ket{q}\bra{r}\braket{z1|y1}\right)\label{tprop2}\\
&=&\frac{\cos ((2k+1)\theta)\sin((2k+1)\theta)}{2Z_1Z_2}\sum_{q,r}\left(\sum_{y:\phi(qy)=0, z:\phi(rz)=1}\sqrt{W_{qy}W_{rz}}\ket{q}\bra{r}\braket{z1|y1}\right)\label{first-comp}\\
&=&\frac{\cos ((2k+1)\theta)\sin((2k+1)\theta)}{2Z_1Z_2}\sum_{q,r}\left(\sum_{y:\phi(qy)=0,\phi(ry)=1}\sqrt{W_{qy}W_{ry}}\ket{q}\bra{r}\right)\label{second-comp}
\end{eqnarray}
\end{footnotesize}
where we get Eq. (\ref{linearity2}) for the linearity of the partial trace operator, Eq. (\ref{tprop2}) for Eq. (\ref{partial-trace}) and Eq. (\ref{first-comp}) and 
 Eq. (\ref{second-comp})  because of Eq. (\ref{dot-prod}).

For $\rho^Q_3$ we similarly get
\begin{eqnarray*}
\rho^Q_3&=&\tr_{Y'}(\rho_3)\nonumber\\
&=&\frac{\cos ((2k+1)\theta)\sin((2k+1)\theta)}{2Z_1Z_2}\sum_{q,r}\left(\sum_{y:\phi(qy)=0,\phi(ry)=1}\sqrt{W_{qy}W_{ry}}\ket{r}\bra{q}\right)
\end{eqnarray*}
The density operator $\rho_4$ is
\begin{eqnarray*}
\rho_4&=&\frac{\cos^2 ((2k+1)\theta)}{Z^2_2}\left(\sum_{x}\sqrt{\frac{W_x}{2}}\ket{x0}+\sum_{x:\phi(x)=0}\sqrt{\frac{W_x}{2}}\ket{x1}\right)\\
&&\left(\sum_{x}\sqrt{\frac{W_x}{2}}\bra{x0}+\sum_{x:\phi(x)=0}\sqrt{\frac{W_x}{2}}\bra{x1}\right)
\end{eqnarray*}
Rewriting $x$ as $qy$ and $rz$, $\rho_4$ becomes
\begin{small}
\begin{eqnarray*}
\rho_4&=&\frac{\cos^2 ((2k+1)\theta)}{2Z^2_2}\left(\sum_{qy}\sqrt{\frac{W_{qy}}{2}}\ket{qy0}+\sum_{qy:\phi(qy)=0}\sqrt{\frac{W_qy}{2}}\ket{qy1}\right)\\
&&\left(\sum_{rz}\sqrt{\frac{W_{rz}}{2}}\bra{rz0}+\sum_{rz:\phi(rz)=0}\sqrt{\frac{W_{rz}}{2}}\bra{rz1}\right)\\
&=&\frac{\cos^2 ((2k+1)\theta)}{2Z^2_2}\sum_{q,r}\left(\sum_{y,z}\sqrt{W_{qy}W_{rz}}\ket{qy0}\bra{rz0}\right.+\sum_{y,z:\phi(rz)=0}\sqrt{W_{qy}W_{rz}}\ket{qy0}\bra{rz1}\\
&&+\sum_{y:\phi(qy)=0,z}\sqrt{W_{qy}W_{rz}}\ket{qy1}\bra{rz0}\left.+\sum_{y:\phi(qy)=0,z:\phi(rz)=0}\sqrt{W_{qy}W_{rz}}\ket{qy1}\bra{rz1}\right)
\end{eqnarray*}
\end{small}
Tracing out bits $Y'$ obtaining the reduced density operator for $Q$ we get:
\begin{eqnarray*}
\rho^Q_4&=&\tr_{Y'}(\rho_4)\nonumber\\
&=&\frac{\cos^2 ((2k+1)\theta)}{2Z^2_2}\sum_{q,r}\left(\sum_{y,z}\sqrt{W_{qy}W_{rz}}\tr_{Y'}(\ket{qy0}\bra{rz0})\right.\\
&&+\sum_{y,z:\phi(rz)=0}\sqrt{W_{qy}W_{rz}}\tr_{Y'}(\ket{qy0}\bra{rz1}))\\
&&+\sum_{y:\phi(qy)=0,z}\sqrt{W_{qy}W_{rz}}\tr_{Y'}(\ket{qy1}\bra{rz0})\\
&&\left.+\sum_{y:\phi(qy)=0,z:\phi(rz)=0}\sqrt{W_{qy}W_{rz}}\tr_{Y'}(\ket{qy1}\bra{rz1})\right)\\
&=&\frac{\cos^2 ((2k+1)\theta)}{2Z^2_2}\sum_{q,r}\left(\sum_{y,z}\sqrt{W_{qy}W_{rz}}\ket{q}\bra{r}\braket{z0|y0}\right.\\
&&+\sum_{y,z:\phi(rz)=0}\sqrt{W_{qy}W_{rz}}\ket{q}\bra{r}\braket{z1|y0}\\
&&+\sum_{y:\phi(qy)=0,z}\sqrt{W_{qy}W_{rz}}\ket{q}\bra{r}\braket{z0|y1}\\
&&\left.+\sum_{y:\phi(qy)=0,z:\phi(rz)=0}\sqrt{W_{qy}W_{rz}}\ket{q}\bra{r}\braket{z1|y1}\right)\\
&=&\frac{\cos^2 ((2k+1)\theta)}{2Z^2_2}\sum_{q,r}\left(\sum_{y}\sqrt{W_{qy}W_{ry}}\ket{q}\bra{r}
+\sum_{y:\phi(qy)=0,\phi(ry)=0}\sqrt{W_{qy}W_{ry}}\ket{q}\bra{r}\right)
\end{eqnarray*}
Let's apply the measurement $\{M_m=\ket{q_m}\bra{q_m}\}$  to system $Q$ where $q_m$ is one of the computational basis state for  $Q$.
This is a measurement in the computational basis state of system $Q$ so $M_m^\dagger M_m=M_m$ and
\begin{eqnarray*}
P(m)&=&\tr(M_m\rho^Q)\\
&=&\tr\left(\ket{q_m}\bra{q_m}(\rho^Q_1+\rho^Q_2+\rho^Q_3+\rho^Q_4)\right)\\
&=&\tr(\ket{q_m}\bra{q_m}\rho^Q_1)+\tr(\ket{q_m}\bra{q_m}\rho^Q_2)+\tr(\ket{q_m}\bra{q_m}\rho^Q_3)+\tr(\ket{q_m}\bra{q_m}\rho^Q_4)
\end{eqnarray*}
Let us define $P_i(m)=\tr\left(\ket{q_m}\bra{q_m}\rho^Q_i\right)$ for $i=1,\ldots,4$.
Then
\begin{eqnarray*}
P_1(m)&=&\tr\left(\ket{q_m}\bra{q_m}\frac{\sin^2((2k+1)\theta)}{2Z_1^2}\sum_{q,r}\sum_{y:\phi(qy)=1:\phi(ry)=1}\sqrt{W_{qy}W_{ry}}\ket{q}\bra{r}\right)\\
&=&\tr\left(\frac{\sin^2((2k+1)\theta)}{2Z_1^2}\sum_{q,r}\sum_{y:\phi(qy)=1:\phi(ry)=1}\sqrt{W_{qy}W_{ry}}\ket{q_m}\braket{q_m|q}\bra{r}\right)
\end{eqnarray*}
Since $\braket{q_m|q}=0$ if $q_m\neq q$ and  $\braket{q_m|q}=1$ if $q_m= q$ then
\begin{eqnarray*}
P_1(m)&=&\tr\left(\frac{\sin^2((2k+1)\theta)}{2Z_1^2}\sum_{r}\sum_{y:\phi(ry)=1,\phi(q_m,y)=1}\sqrt{W_{q_my}W_{ry}}\ket{q_m}\bra{r}\right)\\
&=&\frac{\sin^2((2k+1)\theta)}{2Z_1^2}\sum_{r}\sum_{y:\phi(ry)=1,\phi(q_m,y)=1}\sqrt{W_{q_my}W_{ry}}\tr\left(\ket{q_m}\bra{r}\right)\\
&=&\frac{\sin^2((2k+1)\theta)}{2Z_1^2}\sum_{r}\sum_{y:\phi(ry)=1,\phi(q_m,y)=1}\sqrt{W_{q_my}W_{ry}}\braket{r|q_m}\\
&=&\frac{\sin^2((2k+1)\theta)}{2Z_1^2}\sum_{y:\phi(q_my)=1}\sqrt{W_{q_my}W_{q_my}}\\
&=&\frac{\sin^2((2k+1)\theta)}{2Z_1^2}\sum_{y:\phi(q_my)=1}W_{q_my}\\
&=&\frac{\sin^2((2k+1)\theta)}{\frac{2}{2}\sum_{x:\phi(x)=1}W_x}\sum_{y:\phi(q_my)=1}W_{q_my}\\
&=&\frac{\sin^2((2k+1)\theta)}{\sum_{x:\phi(x)=1}W_x}\sum_{y:\phi(q_my)=1}W_{q_my}\\
&=&\sin^2((2k+1)\theta)\frac{\sum_{y:\phi(q_my)=1}W_{q_my}}{\WMC}
\end{eqnarray*}
For $P_2(m)$ we have:
\begin{footnotesize}
\begin{eqnarray*}
P_2(m)&=&\tr\left(\ket{q_m}\bra{q_m}\frac{\cos ((2k+1)\theta)\sin((2k+1)\theta)}{2Z_1Z_2}\sum_{q,r}\sum_{y:\phi(qy)=0,\phi(ry)=1}\sqrt{W_{qy}W_{ry}}\ket{q}\bra{r}\right)\\
&=&\frac{\cos ((2k+1)\theta)\sin((2k+1)\theta)}{2Z_1Z_2}\sum_{q,r}\sum_{y:\phi(qy)=0,\phi(ry)=1}\sqrt{W_{qy}W_{ry}}\tr(\ket{q_m}\braket{q|q_m}\bra{r})\\
&=&\frac{\cos ((2k+1)\theta)\sin((2k+1)\theta)}{2Z_1Z_2}\sum_{r}\sum_{y:\phi(q_my)=0,\phi(ry)=1}\sqrt{W_{q_my}W_{ry}}\tr(\ket{q_m}\bra{r})\\
&=&\frac{\cos ((2k+1)\theta)\sin((2k+1)\theta)}{2Z_1Z_2}\sum_{r}\sum_{y:\phi(q_my)=0,\phi(ry)=1}\sqrt{W_{q_my}W_{ry}}\braket{r|q_m}\\
&=&\frac{\cos ((2k+1)\theta)\sin((2k+1)\theta)}{2Z_1Z_2}\sum_{y:\phi(q_my)=0,\phi(q_my)=1}\sqrt{W_{q_my}W_{q_my}}\\
&=&0
\end{eqnarray*}
\end{footnotesize}
because there is no $y$ such that $\phi(q_my)=0$ and $\phi(q_my)=1$.
Similarly $P_3(m)=0$. For $P_4(m)$ we have:
\begin{eqnarray*}
P_4(m)&=&\tr\left(\ket{q_m}\bra{q_m}\frac{\cos^2 ((2k+1)\theta)}{Z^2_2}\sum_{q,r}\left(\sum_{y}\sqrt{W_{qy}W_{ry}}\ket{q}\bra{r}\right.\right.\\
&&\left.\left.+\sum_{y:\phi(qy)=0,\phi(ry)=0}\sqrt{W_{qy}W_{ry}}\ket{q}\bra{r}\right)\right)\\
&=&\frac{\cos^2 ((2k+1)\theta)}{Z^2_2}\sum_{q,r}\left(\sum_{y}\sqrt{W_{qy}W_{ry}}\tr\left(\ket{q_m}\braket{q_m|q}\bra{r}\right)\right.\\
&&\left.+\sum_{y:\phi(qy)=0,\phi(ry)=0}\sqrt{W_{qy}W_{ry}}\tr\left(\ket{q_m}\braket{q_m|q}\bra{r}\right)\right)\\
&=&\frac{\cos^2 ((2k+1)\theta)}{Z^2_2}\sum_{r}\left(\sum_{y}\sqrt{W_{q_my}W_{ry}}\braket{r|q_m}\right.\\
&&\left.+\sum_{y:\phi(q_my)=0,\phi(ry)=0}\sqrt{W_{q_my}W_{ry}}\ket{q_m}\braket{r|q_m}\right)\\
&=&\frac{\cos^2 ((2k+1)\theta)}{1+\sum_{x:\phi(x)=0}W_x}\left(\sum_{y}W_{q_my}+\sum_{y:\phi(q_my)=0}W_{q_my}\right)\\
&=&\cos^2 ((2k+1)\theta)\frac{W_{q_m}+\sum_{y:\phi(q_my)=0}W_{q_my}}{2-\WMC}
\end{eqnarray*}
If $\sin((2k+1)\theta)=1$, this algorithm returns one of the  configurations of query bits $q_m$ that are superimposed in $\ket{\delta}$ with a  probability that is proportional to $\sum_{y:\phi(q_my)=1}W_{q_my}$.

This leads to Algorithm \ref{weighted-grover-alg}, where lines \ref{wgstart}-\ref{wgend} are implemented by the quantum circuit of Figure \ref{qmapcirc}.
\begin{algorithm}[ht]
\begin{algorithmic}[1]
\Require A blackbox function $\phi: \mathbb{B}^n\rightarrow \mathbb{B}$, normalized weight function $w$, query qbits $Q$ and $\WMC$
\Ensure $q\in \mathbb{B}^{l}$ sampled from  distribution (\ref{sampling-dist})
\State $\theta\gets \arcsin \sqrt{\frac{\WMC}{2}}$ with $\theta \in [0,\pi/4]$
\State $R\gets\left\lfloor \frac{\pi}{4\theta}\right\rfloor$
\State Prepare the initial superposition $Rot\ket{x'}$\label{wgstart}
\State Apply operator $\WG$ $R$ times
\State Measure $Q$ to get $q\in \mathbb{B}^{l}$\label{wgend}
\State\Return $q$
\end{algorithmic}
\caption{WCS when $\WMC$ is known.\label{weighted-grover-alg}}
\end{algorithm}

The properties of this algorithm are described by the theorem below.
\begin{theorem}
\label{number-of-weighted-rotations}
Let $\phi:\mathbb{B}^n \rightarrow \mathbb{B}$ Let $\theta \in [0,\pi/4]$ be chosen
such that $\sin^2 \theta=\frac{\WMC}{2}$. After $\left\lfloor \frac{\pi}{4\theta}\right\rfloor$  iterations of $\WG$ on an initial superposition
$$Rot\ket{x'}$$
the probability of sampling exactly from the distribution in Eq. (\ref{sampling-dist}) is at least $\sqrt{\frac{1}{2}}\approx 0.707$.
\end{theorem}
\begin{proof}
We can repeat the reasoning of Theorem \ref{number-of-rotations}:
the probability $P$ of sampling from the distribution in Eq. (\ref{sampling-dist}) is given by
$$P=\sin^2((2k+1)\theta)$$
Maximizing $P$ implies that 
$$(2k+1)\theta=\frac{\pi}{2}$$
so
$$k=\frac{\pi}{4\theta}-\frac{1}{2}$$.
From $\theta\approx\sqrt{\frac{\WMC}{2}}$ we obtain
$$R=\left\lfloor \frac{\pi}{4}\sqrt{\frac{2}{\WMC}}\right\rfloor$$
for the required number of applications of $\WG$.
Thus
$$(2R+1)\theta=\frac{\pi}{2}+2\delta\theta$$
with $|\delta|\leq \frac{1}{2}$. 
Since $\WMC\leq 1$, $\sin\theta\leq \sqrt{\frac{1}{2}}$, $\theta\leq\frac{\pi}{4}$, $|2\delta\theta|\leq \frac{\pi}{4}$.
So
$$P=\sin^2((2R+1)\theta)\geq\sin^2\left(\frac{\pi}{2}-\frac{\pi}{4}\right)=\sqrt{\frac{1}{2}}\approx 0.707$$
\end{proof}
We can now present the main result.
\begin{theorem}
\label{weighted-compl}
Algorithm \ref{weighted-grover-alg} samples from the distribution in Eq. (\ref{sampling-dist}) with probability at least $\sqrt{\frac{1}{2}}$ and $O(\frac{1}{\sqrt{\WMC}})$ queries to $\phi$.
\end{theorem}
\begin{proof}
By reasoning as in the proof of Theorem \ref{grover-compl},
in each application of $\WG$ we query $\phi$ and the number of applications $R$ is
$R\leq \frac{\pi}{4}\sqrt{\frac{2}{\WMC}}$,
so $R\in O(\frac{1}{\sqrt{\WMC}})$
\end{proof}

%
When $\WMC$ is not known, Algorithm \ref{weighted-grover-alg-unknownM} can be used, where  $W_{min{}}$ is the minimum weight of a configuration, i.e., $W_{min{}}=\prod_{i=1}^{n+1}w'_{i,min{}}$ and $w'_{i,min{}}=\min{(w'(X_i),w'(\neg X_i))}$ for $i=1,\ldots,n+1$.

.
\begin{algorithm}[ht]
\begin{algorithmic}[1]
\Require A blackbox function $\phi: \mathbb{B}^n\rightarrow \mathbb{B}$, normalized weight function $w$, query qubits $Q$
\Ensure $q\in \mathbb{B}^{l}$ sampled from  distribution (\ref{sampling-dist})
\State $m=\left\lfloor\frac{1}{\sqrt{W_{min}}}\right\rfloor+1$
\State Choose an integer $R$ uniformly in $[0,m-1]$
\State Prepare the initial superposition $\frac{1}{N^{1/2}}\sum_{x=0}^{N-1}\ket{x}$\label{gstart-wu}
\State Apply operator $G$ $R$ times
\State Measure to get $x\in \mathbb{B}^n$\label{gend-wu}
\State\Return $x$
\end{algorithmic}
\caption{WCS when  $\WMC$ is not known.\label{weighted-grover-alg-unknownM}}
\end{algorithm}

Lemma \ref{weighted-lemma-hirvensalo} and Theorem \ref{weigteh-ext-grover-compl} are analogous to Lemma \ref{lemma-hirvensalo-ext} and Theorem \ref{ext-grover-compl-alt}.
\begin{lemma}
\label{weighted-lemma-hirvensalo}
Let $\phi:\mathbb{B}^{n} \rightarrow \mathbb{B}$ and weight function $w$  be such the  weighed model count is $\WMC$. Let $\phi'$ be $\phi'=\phi\wedge X_{n+1}$. 
Let $\theta \in [0,\pi/4]$  be defined by $\sin^2 \theta=\frac{2}{WMC}$. Let $m$ be any positive integer and $R\in [0,m-1]$ chosen with uniform distribution. If $WG$ is applied to
 initial superposition
$$\left(\sum_{x}\sqrt{\frac{W_x}{2}}\ket{x0}+\sum_{x:\phi(x)=0}\sqrt{\frac{W_x}{2}}\ket{x1}\right)+
\sum_{x:\phi(x)=1}\sqrt{\frac{W_x}{2}}\ket{x1}$$
$R$ times, then the probability of seeing a solution is
$$P_m=\frac{1}{2}-\frac{\sin(4m\theta)}{4m\sin(2\theta)}$$
\end{lemma}
\begin{proof}
The proof of Lemma \ref{lemma-hirvensalo} still applies.
\end{proof}
\begin{theorem}
\label{weigteh-ext-grover-compl}
Algorithm \ref{weighted-grover-alg-unknownM} samples from Eq. (\ref{sampling-dist}) with  probability at least $\frac{1}{4}$ and $O(\frac{1}{\sqrt{W_{min}}})$ queries to $\phi'$.
\end{theorem}
\begin{proof}
If $m\geq \frac{1}{\sin(2\theta)}$, then
$$\frac{\sin(4m\theta)}{4m\sin(2\theta)}\leq\frac{1}{4}.$$
By Lemma \ref{weighted-lemma-hirvensalo}, then $P_m\geq\frac{1}{4}$.
Assuming $\phi$ is satisfiable,  then $0<W_{min}\leq \WMC \leq 1$. Since
 $\sin \theta=\sqrt{\frac{\WMC}{2}}$, we have:
 \begin{eqnarray*}
\frac{1}{\sin(2\theta)}&=&\frac{1}{2\sin\theta\cos\theta}=\frac{1}{2\sqrt{\frac{\WMC(2-\WMC)}{4}}}\\
&&=\frac{1}{\sqrt{\WMC(2-\WMC)}}\leq\frac{1}{\sqrt{\WMC}}\leq\frac{1}{\sqrt{W_{min}}}
\end{eqnarray*}
So, if $m\geq\frac{1}{\sqrt{W_{min}}}$, then $m\geq \frac{1}{\sin(2\theta)}$. Choosing  $m=\left\lfloor\frac{1}{\sqrt{W_{min}}}\right\rfloor+1$ we have that $P_m\geq \frac{1}{4}$ and 
the number of applications of $G$ is $O(\frac{1}{\sqrt{W_{min}}})$.
\end{proof}
If the literal weights do not sum to 1, i.e., $w(X_i)+w(\neg X_i)\neq 1$, we normalize them by considering the new weights
$\hat w(X_i)=\frac{w(X_i)}{w(X_i)+w(\neg X_i)}$ and $\hat w (\neg X_i)=\frac{w(\neg X_i)}{w(X_i)+w(\neg X_i)}$. Let $V_i$ be $w(X_i)+w(\neg X_i)$ for $i=1,\ldots,n$.
Then we perform the algorithms with $\hat w$ replacing $w$. The overall normalization factor $\prod_{i=1}^nV_i$ gets canceled out in $P(m)$.

Let the normalized weighted model count $\widehat{\WMC}$ be defined as
$$\widehat{\WMC}=\sum_{x:\phi(x)=1}\widehat W_x$$ 
where $\hat W_{x}$ is $\prod_{i=1}^n\hat w'_{i}$
and
$$\hat w'_{i}=\left\{\begin{array}{ll}
\hat w(X_i)&\mbox{if }x_i=1\\
1-\hat w(X_i)&\mbox{if }x_i=0
\end{array}\right.
$$
Then
\begin{eqnarray}
\widehat{\WMC}&=&\sum_{x:\phi(x)=1}\widehat W_x=\sum_{x_{1}\ldots x_n:\phi(x_{1}\ldots x_n)=1}\hat w'_{{1}}\ldots \hat w'_{n}\nonumber\\
&=&\sum_{x_{1}\ldots x_n:\phi(x_{1}\ldots x_n)=1}\frac{w_1}{V_{1}}\ldots \frac{w_n}{V_n}\nonumber\\
&=&\sum_{x_{1}\ldots x_n:\phi(x_{1}\ldots x_n)=1}\frac{1}{\prod_{i=1}^{n}V_i} w_{1} \ldots  w_n\nonumber\\
&=&\frac{1}{\prod_{i=1}^{n}V_i}\sum_{x_{1}\ldots x_n:\phi(x_{1}\ldots x_n)=1} w_{1}\ldots  w_n\nonumber\\
&=&\frac{1}{\prod_{i=1}^{n}V_i}\sum_{x:\phi(x)=1} W_{x}\nonumber\\
&=&\frac{1}{\prod_{i=1}^{n}V_i}\WMC\label{norm-wmc}
\end{eqnarray}
Theorems \ref{number-of-weighted-rotations} and \ref{weighted-compl} still hold for Algorithm \ref{weighted-grover-alg} provided that $\WMC$ is replaced by $\widehat\WMC$, while
Lemma \ref{weighted-lemma-hirvensalo} and Theorem \ref{weigteh-ext-grover-compl}  still hold for Algorithm \ref{weighted-grover-alg-unknownM} provided that $\WMC$ is replaced by $\widehat\WMC$, and $W_{min}$ is replaced by $\widehat W_{min}=\prod_{i=1}^{n}\min(\hat w(X_i),\hat w(\neg X_i))$.

In Section \ref{qwmc} we will see the QWMC algorithm that computes $\WMC$ in $\Theta(\sqrt{N})$ queries to $\phi$. Since $\min(\hat w(X_i),\hat w(\neg X_i))\leq \frac{1}{2}$, then
$\widehat W_{min}\leq \frac{1}{2^n}=\frac{1}{N}$, so
$\frac{1}{\sqrt{\widehat W_{min}}}\geq \sqrt{N}$. Moreover, $\frac{1}{\sqrt{\widehat W_{min}}}\geq \frac{1}{\sqrt{\widehat\WMC}}$, so it is more convenient to apply QWMC (Algorithm \ref{qwmc}) and then Algorithm \ref{weighted-grover-alg}, obtaining Algorithm \ref{qwcs} that we call QWCS.
\begin{algorithm}[ht]
\begin{algorithmic}[1]
\Require A blackbox function $\phi: \mathbb{B}^n\rightarrow \mathbb{B}$, normalized weight function $w$, query qbits $Q$
\Ensure $q\in \mathbb{B}^{l}$ sampled from  distribution (\ref{sampling-dist})
\State Call Algorithm \ref{qwmc} to obtain $\WMC$
\State Call Algorithm \ref{weighted-grover-alg} to obtain $q$
\State\Return $q$
\end{algorithmic}
\caption{Algorithm QWCS.\label{qwcs}}
\end{algorithm}

We are now ready to state our main result for WCS.
\begin{theorem}
\label{qwcs-compl}
QWCS samples the distribution from Eq. (\ref{sampling-dist}) with  probability at least $\sqrt{\frac{1}{2}}$ and $O(\sqrt{N}+\frac{1}{\sqrt{\widehat\WMC}})$ queries to $\phi'$.
\end{theorem}
\begin{proof}
Immediate from theorems  \ref{qwmc-compl} and \ref{weighted-compl}.
\end{proof}

\section{Comparison of QWCS with Classical Algorithms}
\label{qmap-class}

%
%

For classical probabilistic algorithms under a black box model of computation we have the following results.
\begin{theorem}
Any classical probabilistic algorithm for solving WCS under the black box model of computation takes $\Omega(\frac{1}{\widehat\WMC})$ oracle queries.
\end{theorem}
\begin{proof}
%
A classical algorithm for WCS needs a number of queries $s$ at least
$p\frac{1}{\widehat\WMC}$ to succeed with probability at least $p$. In fact, suppose $s< p\frac{1}{\widehat\WMC}$. If we set all 
weights to 0.5, then  $\widehat\WMC=\frac{M}{N}$ and we could solve FSAT by returning the solution found. By Lemma \ref{compl-search-m}, an algorithm for FSAT makes at least 
$pM/N$ queries to return a solution with probability $p$,  obtaining a contradiction. So a classical probabilistic algorithm takes $\Omega(\frac{1}{\WMC})$ queries.
\end{proof}

\section{Quantum Model Counting}
\label{qmc}
The algorithm for quantum model counting uses  the quantum Fourier transform and phase estimation, so we review those first.
\subsection{Quantum Fourier Transform}
\label{qft}
The discrete Fourier transform computes a vector of complex numbers $y_0,\ldots y_{N-1}$ given
a vector of complex numbers $x_0,\ldots, x_{N-1}$ as follows
$$y_k=\frac{1}{\sqrt{N}}\sum_{j=0}^{N-1}x_je^{2\pi ijk/N}$$
The \emph{quantum Fourier transform} \citep{coppersmith2002approximate} is similar, it takes an orthonormal basis $\ket{0},\ldots,\ket{N-1}$ 
and transforms it as:
$$\ket{j}\to \frac{1}{\sqrt{N}}\sum_{k=0}^{N-1}e^{2\pi i j k/N}\ket{k}$$
It is a Fourier transform because the action on an arbitrary state is
$$\sum_{j=0}^{N-1}x_j\ket{j}\to\sum_{k=0}^{N-1}y_k\ket{k}$$
with $y_k$ as in the discrete Fourier transform.

Assuming $N=2^n$, the quantum Fourier transform can be given a \emph{product representation} \citep{cleve1998quantum,griffiths1996semiclassical}:
\begin{equation}
\label{product_rep}
\begin{array}{l}
\ket{j_1\ldots j_n}\to\\
\frac{1}{2^{n/2}}\sum_{k=0}^{2^n-1}e^{2\pi ijk/2^n}\ket{k}\\
=\frac{1}{2^{n/2}}\sum_{k_1=0}^{1}\ldots\sum_{k_n=0}^1e^{2\pi ij\left(\sum_{l=1}^nk_l2^{-l}\right)}\ket{k_1\ldots k_n}\\
=\frac{1}{2^{n/2}}\sum_{k_1=0}^{1}\ldots\sum_{k_n=0}^1\bigotimes_{l=1}^ne^{2\pi ijk_l2^{-l}}\ket{k_l}\\
=\frac{1}{2^{n/2}}\bigotimes_{l=1}^n\left[\sum_{k_l=0}^{1}e^{2\pi ijk_l2^{-l}}\ket{k_l}\right]\\
=\frac{1}{2^{n/2}}\bigotimes_{l=1}^n\left[\ket{0}+e^{2\pi ij2^{-l}}\ket{1}\right]\\
 =\frac{\left(\ket{0}+e^{2\pi 0.j_n}\ket{1}\right)\otimes\left(\ket{0}+e^{2\pi 0.j_{n-1}j_n}\ket{1}\right)\otimes\cdots\left(\ket{0}+e^{2\pi 0.j_1j_2\cdots j_n}\ket{1}\right)}{2^{n/2}}
 \end{array}
 \end{equation}
where  the state $\ket{j}$ is written using the binary representation
$j=j_1j_2\ldots j_n$ and $0.j_lj_{l+1}\ldots j_m$ represents the number $j_l/2+j_{l+1}/4+\ldots+j_m/2^{m-l+1}$.
The quantum Fourier transform requires $\Theta(n^2)$ gates (see \citep{nielsen2010quantum} for the derivation of this formula).

\subsection{Quantum Phase Estimation}
\label{qpe}
In the problem of \emph{quantum phase estimation} \citep{cleve1998quantum}, we are given an operator $U$ and one of its eigenvectors
$\ket{u}$ with eigenvalue $e^{2\pi i\varphi}$ and we want to find the value of $\varphi$.
We assume that  we have black boxes that can prepare the state $\ket{u}$ and perform
controlled-$U^{2^j}$ operations for non negative integers $j$.

Phase estimation uses two registers, one with $t$ qubits initially in state $\ket{0}$ and the other with as many qubits as are necessary to store $\ket{u}$ that is also its initial state.

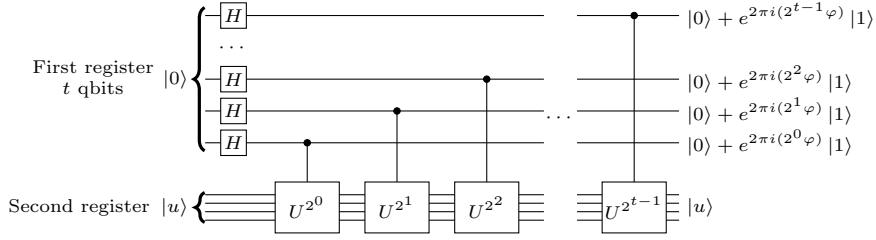
\begin{figure*}[t]
\centering
\begin{footnotesize}
\begin{tikzpicture}[scale=0.800000,x=1pt,y=1pt]
\filldraw[color=white] (0.000000, -2.000000) rectangle (222.000000, 104.000000);
\draw[color=black] (0.000000,96.500000) -- (222.000000,96.500000);
\draw[color=black] (0.000000,66.500000) -- (222.000000,66.500000);
\draw[color=black] (0.000000,51.500000) -- (222.000000,51.500000);
\draw[color=black] (0.000000,36.500000) -- (222.000000,36.500000);
\filldraw[color=white,fill=white] (0.000000,32.750000) rectangle (-4.000000,100.250000);
\draw[decorate,decoration={brace,amplitude = 4.000000pt},very thick] (0.000000,32.750000) -- (0.000000,100.250000);
\draw[color=black] (-4.000000,66.500000) node[left] {${\begin{array}{c}\mbox{First register}\\t\mbox{ qbits}\end{array}\ket{0}}$};
\draw[color=black] (0.000000,12.000000) -- (222.000000,12.000000);
\draw[color=black] (0.000000,8.000000) -- (222.000000,8.000000);
\draw[color=black] (0.000000,4.000000) -- (222.000000,4.000000);
\draw[color=black] (0.000000,0.000000) -- (222.000000,0.000000);
\filldraw[color=white,fill=white] (0.000000,-1.000000) rectangle (-4.000000,13.000000);
\draw[decorate,decoration={brace,amplitude = 3.500000pt},very thick] (0.000000,-1.000000) -- (0.000000,13.000000);
\draw[color=black] (-4.000000,6.000000) node[left] {${\mbox{Second register }\ket{u}}$};
\begin{scope}
\draw[fill=white] (13.500000, 96.500000) +(-45.000000:8.485281pt and 8.485281pt) -- +(45.000000:8.485281pt and 8.485281pt) -- +(135.000000:8.485281pt and 8.485281pt) -- +(225.000000:8.485281pt and 8.485281pt) -- cycle;
\clip (13.500000, 96.500000) +(-45.000000:8.485281pt and 8.485281pt) -- +(45.000000:8.485281pt and 8.485281pt) -- +(135.000000:8.485281pt and 8.485281pt) -- +(225.000000:8.485281pt and 8.485281pt) -- cycle;
\draw (13.500000, 96.500000) node {$H$};
\end{scope}
\begin{scope}
\draw[fill=white] (13.500000, 66.500000) +(-45.000000:8.485281pt and 8.485281pt) -- +(45.000000:8.485281pt and 8.485281pt) -- +(135.000000:8.485281pt and 8.485281pt) -- +(225.000000:8.485281pt and 8.485281pt) -- cycle;
\clip (13.500000, 66.500000) +(-45.000000:8.485281pt and 8.485281pt) -- +(45.000000:8.485281pt and 8.485281pt) -- +(135.000000:8.485281pt and 8.485281pt) -- +(225.000000:8.485281pt and 8.485281pt) -- cycle;
\draw (13.500000, 66.500000) node {$H$};
\end{scope}
\begin{scope}
\draw[fill=white] (13.500000, 51.500000) +(-45.000000:8.485281pt and 8.485281pt) -- +(45.000000:8.485281pt and 8.485281pt) -- +(135.000000:8.485281pt and 8.485281pt) -- +(225.000000:8.485281pt and 8.485281pt) -- cycle;
\clip (13.500000, 51.500000) +(-45.000000:8.485281pt and 8.485281pt) -- +(45.000000:8.485281pt and 8.485281pt) -- +(135.000000:8.485281pt and 8.485281pt) -- +(225.000000:8.485281pt and 8.485281pt) -- cycle;
\draw (13.500000, 51.500000) node {$H$};
\end{scope}
\begin{scope}
\draw[fill=white] (13.500000, 36.500000) +(-45.000000:8.485281pt and 8.485281pt) -- +(45.000000:8.485281pt and 8.485281pt) -- +(135.000000:8.485281pt and 8.485281pt) -- +(225.000000:8.485281pt and 8.485281pt) -- cycle;
\clip (13.500000, 36.500000) +(-45.000000:8.485281pt and 8.485281pt) -- +(45.000000:8.485281pt and 8.485281pt) -- +(135.000000:8.485281pt and 8.485281pt) -- +(225.000000:8.485281pt and 8.485281pt) -- cycle;
\draw (13.500000, 36.500000) node {$H$};
\end{scope}
\draw[fill=white,color=white] (6.000000, 75.500000) rectangle (21.000000, 87.500000);
\draw (13.500000, 81.500000) node {$\cdots$};
\draw (48.000000,36.500000) -- (48.000000,0.000000);
\begin{scope}
\draw[fill=white] (48.000000, 6.000000) +(-45.000000:21.213203pt and 16.970563pt) -- +(45.000000:21.213203pt and 16.970563pt) -- +(135.000000:21.213203pt and 16.970563pt) -- +(225.000000:21.213203pt and 16.970563pt) -- cycle;
\clip (48.000000, 6.000000) +(-45.000000:21.213203pt and 16.970563pt) -- +(45.000000:21.213203pt and 16.970563pt) -- +(135.000000:21.213203pt and 16.970563pt) -- +(225.000000:21.213203pt and 16.970563pt) -- cycle;
\draw (48.000000, 6.000000) node {$U^{2^0}$};
\end{scope}
\filldraw (48.000000, 36.500000) circle(1.500000pt);
\draw (90.000000,51.500000) -- (90.000000,0.000000);
\begin{scope}
\draw[fill=white] (90.000000, 6.000000) +(-45.000000:21.213203pt and 16.970563pt) -- +(45.000000:21.213203pt and 16.970563pt) -- +(135.000000:21.213203pt and 16.970563pt) -- +(225.000000:21.213203pt and 16.970563pt) -- cycle;
\clip (90.000000, 6.000000) +(-45.000000:21.213203pt and 16.970563pt) -- +(45.000000:21.213203pt and 16.970563pt) -- +(135.000000:21.213203pt and 16.970563pt) -- +(225.000000:21.213203pt and 16.970563pt) -- cycle;
\draw (90.000000, 6.000000) node {$U^{2^1}$};
\end{scope}
\filldraw (90.000000, 51.500000) circle(1.500000pt);
\draw (132.000000,66.500000) -- (132.000000,0.000000);
\begin{scope}
\draw[fill=white] (132.000000, 6.000000) +(-45.000000:21.213203pt and 16.970563pt) -- +(45.000000:21.213203pt and 16.970563pt) -- +(135.000000:21.213203pt and 16.970563pt) -- +(225.000000:21.213203pt and 16.970563pt) -- cycle;
\clip (132.000000, 6.000000) +(-45.000000:21.213203pt and 16.970563pt) -- +(45.000000:21.213203pt and 16.970563pt) -- +(135.000000:21.213203pt and 16.970563pt) -- +(225.000000:21.213203pt and 16.970563pt) -- cycle;
\draw (132.000000, 6.000000) node {$U^{2^2}$};
\end{scope}
\filldraw (132.000000, 66.500000) circle(1.500000pt);
\draw[fill=white,color=white] (159.000000, -6.000000) rectangle (174.000000, 102.500000);
\draw (166.500000, 48.250000) node {$\cdots$};
\draw (201.000000,96.500000) -- (201.000000,0.000000);
\begin{scope}
\draw[fill=white] (201.000000, 6.000000) +(-45.000000:21.213203pt and 16.970563pt) -- +(45.000000:21.213203pt and 16.970563pt) -- +(135.000000:21.213203pt and 16.970563pt) -- +(225.000000:21.213203pt and 16.970563pt) -- cycle;
\clip (201.000000, 6.000000) +(-45.000000:21.213203pt and 16.970563pt) -- +(45.000000:21.213203pt and 16.970563pt) -- +(135.000000:21.213203pt and 16.970563pt) -- +(225.000000:21.213203pt and 16.970563pt) -- cycle;
\draw (201.000000, 6.000000) node {$U^{2^{t-1}}$};
\end{scope}
\filldraw (201.000000, 96.500000) circle(1.500000pt);
\draw[color=black] (222.000000,96.500000) node[right] {${\ket{0}+e^{2\pi i(2^{t-1}\varphi)}\ket{1}}$};
\draw[color=black] (222.000000,66.500000) node[right] {${\ket{0}+e^{2\pi i(2^2\varphi)}\ket{1}}$};
\draw[color=black] (222.000000,51.500000) node[right] {${\ket{0}+e^{2\pi i(2^1\varphi)}\ket{1}}$};
\draw[color=black] (222.000000,36.500000) node[right] {${\ket{0}+e^{2\pi i(2^0\varphi)}\ket{1}}$};
\draw[color=black] (222.000000,6.000000) node[right] {${\ket{u}}$};
\end{tikzpicture}
\end{footnotesize}
\caption{First stage of phase estimation. On the right we have omitted normalization factors of $\frac{1}{\sqrt{2}}$.}
\label{phase_est}
\end{figure*}

The first stage of phase estimation is shown in Figure \ref{phase_est}. 
A controlled-$U^{2^j}$ operation on control qubit $b$ and target register in an eigenvector state $\ket{u}$   of $U$ acts as follows. If $b$ is $\ket{0}$, $U^{2^j}$ is not applied and the output is $\ket{0}\ket{u}$. If $b$ is $\ket{1}$, then $U^{2^j}$ is applied to $\ket{u}$.
Since $\ket{u}$ is an eigenvector of $U$, $\ket{u}$ is brought to $e^{2\pi i 2^j\varphi}\ket{u}$ and $\ket{1}\ket{u}$ becomes  $e^{2\pi i 2^j\varphi}\ket{1}\ket{u}$.

The result of the  controlled-$U^{2^j}$ operation on $(H\ket{0})\ket{u}=\frac{\ket{0}+\ket{1}}{\sqrt{2}}\ket{u}$ is
$$\left(\frac{\ket{0}+e^{2\pi i 2^j\varphi}\ket{1}}{\sqrt{2}}\right)\ket{u}$$

\begin{figure}[t]
\centering
\begin{tikzpicture}[scale=1.000000,x=1pt,y=1pt]
\filldraw[color=white] (0.000000, -7.500000) rectangle (150.000000, 37.500000);
\draw[color=black] (0.000000,22.500000) -- (138.000000,22.500000);
\draw[color=black] (138.000000,22.000000) -- (150.000000,22.000000);
\draw[color=black] (138.000000,23.000000) -- (150.000000,23.000000);
\draw[color=black] (0.000000,22.500000) node[left] {${\ket{0}}$};
\draw[color=black] (0.000000,0.000000) -- (150.000000,0.000000);
\draw[color=black] (0.000000,0.000000) node[left] {${\ket{u}}$};
\draw (6.000000, 16.500000) -- (14.000000, 28.500000);
\draw (12.000000, 25.500000) node[right] {$\scriptstyle{t}$};
\draw (6.000000, -6.000000) -- (14.000000, 6.000000);
\draw (12.000000, 3.000000) node[right] {$\scriptstyle{n}$};
\draw (56.000000,22.500000) -- (56.000000,0.000000);
\begin{scope}
\draw[fill=white] (56.000000, 11.250000) +(-45.000000:42.426407pt and 24.395184pt) -- +(45.000000:42.426407pt and 24.395184pt) -- +(135.000000:42.426407pt and 24.395184pt) -- +(225.000000:42.426407pt and 24.395184pt) -- cycle;
\clip (56.000000, 11.250000) +(-45.000000:42.426407pt and 24.395184pt) -- +(45.000000:42.426407pt and 24.395184pt) -- +(135.000000:42.426407pt and 24.395184pt) -- +(225.000000:42.426407pt and 24.395184pt) -- cycle;
\draw (56.000000, 11.250000) node {$\begin{array}{cc}\mbox{First phase}\\\mbox{Fig. \ref{phase_est}}\end{array}$};
\end{scope}
\begin{scope}
\draw[fill=white] (109.000000, 22.500000) +(-45.000000:15.556349pt and 8.485281pt) -- +(45.000000:15.556349pt and 8.485281pt) -- +(135.000000:15.556349pt and 8.485281pt) -- +(225.000000:15.556349pt and 8.485281pt) -- cycle;
\clip (109.000000, 22.500000) +(-45.000000:15.556349pt and 8.485281pt) -- +(45.000000:15.556349pt and 8.485281pt) -- +(135.000000:15.556349pt and 8.485281pt) -- +(225.000000:15.556349pt and 8.485281pt) -- cycle;
\draw (109.000000, 22.500000) node {$FT^\dagger$};
\end{scope}
\draw[fill=white] (132.000000, 16.500000) rectangle (144.000000, 28.500000);
\draw[very thin] (138.000000, 23.100000) arc (90:150:6.000000pt);
\draw[very thin] (138.000000, 23.100000) arc (90:30:6.000000pt);
\draw[->,>=stealth] (138.000000, 17.100000) -- +(80:10.392305pt);
\draw[color=black] (150.000000,0.000000) node[right] {${\ket{u}}$};
\end{tikzpicture}
\caption{The complete phase estimation circuit.}
\label{phase_est_all}
\end{figure}
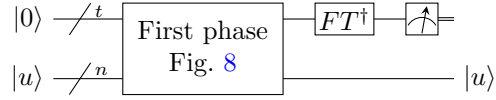

Thus the final state of the first register after the first phase of phase estimation is
\begin{equation}
\label{eq:phase}
\begin{array}{l}
\frac{1}{2^{t/2}}\left(\ket{0}+e^{2\pi i 2^{t-1}\varphi}\ket{1}\right)\otimes \left(\ket{0}+e^{2\pi i 2^{t-2}\varphi}\ket{1}\right)\ldots\left(\ket{0}+e^{2\pi i 2^{0}\varphi}\ket{1}\right)\\
\end{array}
\end{equation}
If the phase can be represented with exactly $t$ bits as $\varphi=0.\varphi_1\ldots\varphi_t$, Eq. (\ref{eq:phase}) can be rewritten as
\begin{equation}
 \frac{\left(\ket{0}+e^{2\pi i 0.\varphi_t}\ket{1}\right)\otimes\left(\ket{0}+e^{2\pi i0.\varphi_{t-1}\varphi_t}\ket{1}\right)\otimes\cdots\left(\ket{0}+e^{2\pi i0.\varphi_1\cdots \varphi_t}\ket{1}\right)}{2^{n/2}}
\end{equation}
This form is exactly the same as that of Eq. (\ref{product_rep}) so, if we apply the inverse of the Fourier transform, we  obtain $\ket{\varphi_1\ldots\varphi_t}$. The inverse of an operator is its adjoint so the overall phase estimation circuit is shown in Figure \ref{phase_est_all}.

If $\varphi$ cannot be represented exactly with $t$ bits, the algorithm provides approximation guarantees: if we want to approximate $\varphi$ to $m$ bits with probability of success at least $1-\epsilon$ we must choose $t=m+\lceil \log_2 \left(2+\frac{1}{2\epsilon}\right)\rceil$ (see \citep{nielsen2010quantum} for the derivation of this formula).

\subsection{Quantum Counting}
\label{sec:qc}
With quantum counting we want to count the number of solutions to the equation $\phi(x)=1$ where $\phi$ is a Boolean function as above. In the notation we are using, it means computing $M$.
An algorithm for quantum counting was proposed in \citep{boyer1998tight,DBLP:conf/icalp/BrassardHT98}.

Suppose $\ket{a}$ and $\ket{b}$ are the two eigenvectors of  the Grover operator $G$ in the space spanned by $\ket{\alpha}$ and $\ket{\beta}$. Since $G$ is a rotation of angle $2\theta$ in such a space, the eigenvalues of $\ket{a}$ and $\ket{b}$ are $e^{i2\theta}$ and $e^{i(2\pi-2\theta)}$.
If we know $\theta$, we can compute $M$ from $\sin^2(\theta) = M/N$. Since $\sin(\theta)=\sin(\pi-\theta)$, it does not matter which eigenvalue is estimated.

Let us consider an extra qubit $X_{n+1}$ as presented in Section \ref{qmap}, so that $\sin^2\theta=\frac{M}{2N}$.

\begin{figure}[t]
\centering
\begin{scriptsize}
\begin{tikzpicture}[scale=0.700000,x=1pt,y=1pt]
\filldraw[color=white] (0.000000, -6.500000) rectangle (277.000000, 123.500000);
\draw[color=black] (0.000000,117.000000) -- (265.000000,117.000000);
\draw[color=black] (265.000000,116.500000) -- (277.000000,116.500000);
\draw[color=black] (265.000000,117.500000) -- (277.000000,117.500000);
\draw[color=black] (0.000000,104.000000) -- (265.000000,104.000000);
\draw[color=black] (265.000000,103.500000) -- (277.000000,103.500000);
\draw[color=black] (265.000000,104.500000) -- (277.000000,104.500000);
\draw[color=black] (0.000000,91.000000) -- (265.000000,91.000000);
\draw[color=black] (265.000000,90.500000) -- (277.000000,90.500000);
\draw[color=black] (265.000000,91.500000) -- (277.000000,91.500000);
\filldraw[color=white,fill=white] (0.000000,87.750000) rectangle (-4.000000,120.250000);
\draw[decorate,decoration={brace,amplitude = 4.000000pt},very thick] (0.000000,87.750000) -- (0.000000,120.250000);
\draw[color=black] (-4.000000,104.000000) node[left] {${\ket{0}^{\otimes t}}$};
\draw[color=black] (0.000000,65.000000) -- (277.000000,65.000000);
\draw[color=black] (0.000000,52.000000) -- (277.000000,52.000000);
\draw[color=black] (0.000000,39.000000) -- (277.000000,39.000000);
\draw[color=black] (0.000000,26.000000) -- (277.000000,26.000000);
\draw[color=black] (0.000000,13.000000) -- (277.000000,13.000000);
\draw[color=black] (0.000000,0.000000) -- (277.000000,0.000000);
\filldraw[color=white,fill=white] (0.000000,-3.250000) rectangle (-4.000000,68.250000);
\draw[decorate,decoration={brace,amplitude = 4.000000pt},very thick] (0.000000,-3.250000) -- (0.000000,68.250000);
\draw[color=black] (-4.000000,32.500000) node[left] {${\ket{0}^{\otimes n+1}}$};
\draw (26.000000,117.000000) -- (26.000000,91.000000);
\begin{scope}
\draw[fill=white] (26.000000, 104.000000) +(-45.000000:28.284271pt and 26.870058pt) -- +(45.000000:28.284271pt and 26.870058pt) -- +(135.000000:28.284271pt and 26.870058pt) -- +(225.000000:28.284271pt and 26.870058pt) -- cycle;
\clip (26.000000, 104.000000) +(-45.000000:28.284271pt and 26.870058pt) -- +(45.000000:28.284271pt and 26.870058pt) -- +(135.000000:28.284271pt and 26.870058pt) -- +(225.000000:28.284271pt and 26.870058pt) -- cycle;
\draw (26.000000, 104.000000) node {$H^{\otimes t}$};
\end{scope}
\draw (26.000000,65.000000) -- (26.000000,0.000000);
\begin{scope}
\draw[fill=white] (26.000000, 32.500000) +(-45.000000:28.284271pt and 54.447222pt) -- +(45.000000:28.284271pt and 54.447222pt) -- +(135.000000:28.284271pt and 54.447222pt) -- +(225.000000:28.284271pt and 54.447222pt) -- cycle;
\clip (26.000000, 32.500000) +(-45.000000:28.284271pt and 54.447222pt) -- +(45.000000:28.284271pt and 54.447222pt) -- +(135.000000:28.284271pt and 54.447222pt) -- +(225.000000:28.284271pt and 54.447222pt) -- cycle;
\draw (26.000000, 32.500000) node {$H^{\otimes n+1}$};
\end{scope}
\draw (74.000000,91.000000) -- (74.000000,0.000000);
\begin{scope}
\draw[fill=white] (74.000000, 32.500000) +(-45.000000:22.627417pt and 54.447222pt) -- +(45.000000:22.627417pt and 54.447222pt) -- +(135.000000:22.627417pt and 54.447222pt) -- +(225.000000:22.627417pt and 54.447222pt) -- cycle;
\clip (74.000000, 32.500000) +(-45.000000:22.627417pt and 54.447222pt) -- +(45.000000:22.627417pt and 54.447222pt) -- +(135.000000:22.627417pt and 54.447222pt) -- +(225.000000:22.627417pt and 54.447222pt) -- cycle;
\draw (74.000000, 32.500000) node {$G^{2^0}$};
\end{scope}
\filldraw (74.000000, 91.000000) circle(1.500000pt);
\draw (118.000000,104.000000) -- (118.000000,0.000000);
\begin{scope}
\draw[fill=white] (118.000000, 32.500000) +(-45.000000:22.627417pt and 54.447222pt) -- +(45.000000:22.627417pt and 54.447222pt) -- +(135.000000:22.627417pt and 54.447222pt) -- +(225.000000:22.627417pt and 54.447222pt) -- cycle;
\clip (118.000000, 32.500000) +(-45.000000:22.627417pt and 54.447222pt) -- +(45.000000:22.627417pt and 54.447222pt) -- +(135.000000:22.627417pt and 54.447222pt) -- +(225.000000:22.627417pt and 54.447222pt) -- cycle;
\draw (118.000000, 32.500000) node {$G^{2^1}$};
\end{scope}
\filldraw (118.000000, 104.000000) circle(1.500000pt);
\draw[fill=white,color=white] (146.000000, -6.000000) rectangle (161.000000, 123.000000);
\draw (153.500000, 58.500000) node {$\cdots$};
\draw (189.000000,117.000000) -- (189.000000,0.000000);
\begin{scope}
\draw[fill=white] (189.000000, 32.500000) +(-45.000000:22.627417pt and 54.447222pt) -- +(45.000000:22.627417pt and 54.447222pt) -- +(135.000000:22.627417pt and 54.447222pt) -- +(225.000000:22.627417pt and 54.447222pt) -- cycle;
\clip (189.000000, 32.500000) +(-45.000000:22.627417pt and 54.447222pt) -- +(45.000000:22.627417pt and 54.447222pt) -- +(135.000000:22.627417pt and 54.447222pt) -- +(225.000000:22.627417pt and 54.447222pt) -- cycle;
\draw (189.000000, 32.500000) node {$G^{2^{t-1}}$};
\end{scope}
\filldraw (189.000000, 117.000000) circle(1.500000pt);
\draw (232.000000,117.000000) -- (232.000000,91.000000);
\begin{scope}
\draw[fill=white] (232.000000, 104.000000) +(-45.000000:21.213203pt and 26.870058pt) -- +(45.000000:21.213203pt and 26.870058pt) -- +(135.000000:21.213203pt and 26.870058pt) -- +(225.000000:21.213203pt and 26.870058pt) -- cycle;
\clip (232.000000, 104.000000) +(-45.000000:21.213203pt and 26.870058pt) -- +(45.000000:21.213203pt and 26.870058pt) -- +(135.000000:21.213203pt and 26.870058pt) -- +(225.000000:21.213203pt and 26.870058pt) -- cycle;
\draw (232.000000, 104.000000) node {$FT^\dagger$};
\end{scope}
\draw[fill=white] (259.000000, 111.000000) rectangle (271.000000, 123.000000);
\draw[very thin] (265.000000, 117.600000) arc (90:150:6.000000pt);
\draw[very thin] (265.000000, 117.600000) arc (90:30:6.000000pt);
\draw[->,>=stealth] (265.000000, 111.600000) -- +(80:10.392305pt);
\draw[fill=white] (259.000000, 98.000000) rectangle (271.000000, 110.000000);
\draw[very thin] (265.000000, 104.600000) arc (90:150:6.000000pt);
\draw[very thin] (265.000000, 104.600000) arc (90:30:6.000000pt);
\draw[->,>=stealth] (265.000000, 98.600000) -- +(80:10.392305pt);
\draw[fill=white] (259.000000, 85.000000) rectangle (271.000000, 97.000000);
\draw[very thin] (265.000000, 91.600000) arc (90:150:6.000000pt);
\draw[very thin] (265.000000, 91.600000) arc (90:30:6.000000pt);
\draw[->,>=stealth] (265.000000, 85.600000) -- +(80:10.392305pt);
\end{tikzpicture}
\end{scriptsize}
\caption{Circuit for quantum counting.}
\label{quantumcounting}
\end{figure}
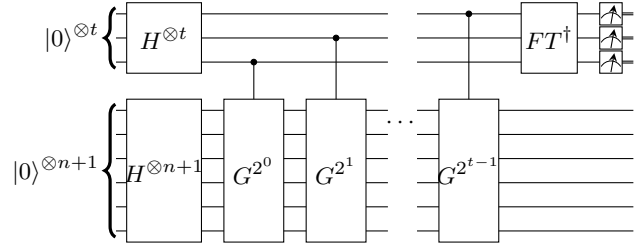

Thus quantum counting is performed by using quantum phase estimation to compute the eigenvalues of the 
Grover operator $G$. The circuit for quantum counting is shown in Figure \ref{quantumcounting}.

The upper register in Figure \ref{quantumcounting} has $t$ qubits while the lower register $n+1$. 
$\theta$ is estimated to $m$ bits of accuracy with probability at least $1-\epsilon$ if $t=m+\lceil \log_2 (2+1/2\epsilon)\rceil$ because of the use of quantum phase estimation. 
\begin{theorem}
The quantum counting algorithm of Figure \ref{quantumcounting}, using $t=\lceil n/2\rceil+5$ qubits in the upper register, performs $\Theta(\sqrt{N})$ applications of the Grover operator and returns a count $M$ estimated with $m=\lceil n/2\rceil+2$ bits of accuracy that, with probability 11/12, has error  $|\Delta M|=O(\sqrt{M})$.
\label{compl_q_counting}
\end{theorem}
\begin{proof}
The error on the estimate of the count $M$ is given by  
\citep{nielsen2010quantum}:
\begin{eqnarray*}
&&\frac{|\Delta M|}{2N}=\left|\sin^2\left(\theta+\Delta\theta\right)-\sin^2\theta\right|=\\
&&\left(\sin\left(\theta+\Delta\theta\right)+\sin\theta\right)\left|\sin\left(\theta+\Delta\theta\right)-\sin\theta\right|
\end{eqnarray*}
Since $\left|\sin(\theta+\Delta\theta)-\sin\theta\right|\leq|\Delta\theta|$ and $\left|\sin(\theta+\Delta\theta)\right|<\sin(\theta)+|\Delta\theta|$ from calculus and trigonometry respectively, we get
\begin{equation*}
\frac{|\Delta M|}{2N}<\left(2\sin\theta+|\Delta\theta|\right)|\Delta\theta|
\end{equation*}
Using $\sin^2(\theta)=M/2N$ and $|\Delta\theta|\leq 2^{-m}$ we obtain
\begin{eqnarray*}
|\Delta M|&<&2N\left(2\sqrt{\frac{M}{2N}}+\frac{1}{2^m}\right)2^{-m}\\
&=&2\left(\sqrt{2MN}+\frac{N}{2^m}\right)2^{-m}\\
&=&\left(\sqrt{8MN}+\frac{N}{2^{m-1}}\right)2^{-m}
\end{eqnarray*}
Consider this case: let $m=\lceil n/2\rceil+2$ and $\epsilon=1/12$. Then $t=\lceil n/2\rceil+5$. 
The number of applications of the Grover operator is $\Theta(\sqrt{N})$ and so is the number of oracle calls. The error, if $n$ is even,  is  
\begin{eqnarray*}
|\Delta M|&<&\left(\sqrt{8M2^n}+\frac{2^n}{2^{n/2+1}}\right)2^{-n/2-2}\\
&=&\left(\sqrt{2M}2^{n/2+1}+2^{n/2-1}\right)2^{-n/2-2}\\
&=&\sqrt{M/2}+1/8=O(\sqrt{M})
\end{eqnarray*}
If $n$ is odd:
\begin{eqnarray*}
|\Delta M|&<&\left(\sqrt{8M2^n}+\frac{2^n}{2^{n/2+1/2+1}}\right)2^{-n/2-1/2-2}\\
&=&\left(\sqrt{M}2^{n/2+3/2}+2^{n/2-3/2}\right)2^{-n/2-5/2}\\
&=&\sqrt{M}/2+1/16=O(\sqrt{M})
\end{eqnarray*}
\end{proof}

\section{Comparison of Quantum Counting with Classical Algorithms}
\label{compl}
Let us now discuss the advantages of quantum counting with respect to classical counting under a black box model of computation where
we only  have an oracle that answers queries over $\phi$. We want to know what is the minimum number 
of evaluations that are needed to solve counting problems.

A classical algorithm for probabilistically solving a model counting problem proceeds by taking $s$ samples uniformly from the search space. 
For each  sample $x$, we query the oracle and we obtain a value $F_i$ with $i=1,\ldots,s$, where $F_i$ is 
1 if $\phi(x)=1$ and $F_i$ is 0 if $\phi(x)=0$. Then we can estimate the count as
$$S=\frac{N}{s}\times\sum_{i=1}^s F_i=\frac{N\overline{F}}{s}$$
where $\overline{F}=\sum_{i=1}^sF_i$.
Variable $\overline{F}$ is binomially distributed with $s$ the number of trials and probability of success $M/N$ where $M$ is the model count of $\phi$. 
Therefore the mean of $\overline{F}$ is $sM/N$ and the mean of $S$ is $(N/s)s(M/N)=M$, so $S$ is an unbiased estimate of $M$. 

The following theorem appears as Exercise 6.13 in \citep{nielsen2010quantum}. Here we present it together with a proof that is absent in \citep{nielsen2010quantum}.
\begin{theorem}
\label{compl-class-count}
The complexity of the classical algorithm for estimating $M$ with a probability of at least $3/4$ within an accuracy of $\sqrt{M}$ is $\Omega(N)$ oracle calls.
\end{theorem}
\begin{proof}
We must prove that
\begin{equation}
P(s\frac{M-c\sqrt{M}}{N}\leq \overline{F}\leq s\frac{M+c\sqrt{M}}{N})\geq \frac{3}{4}\label{thesis-compl-cl-counting}
\end{equation}
We have
$$P(s\frac{M-c\sqrt{M}}{N}\leq \overline{F}\leq s\frac{M+c\sqrt{M}}{N})=P(\overline{F}\leq s\frac{M+c\sqrt{M}}{N})-P(\overline{F}< s\frac{M-c\sqrt{M}}{N})$$
For  a binomially distributed random variable $k$ with number of trials $s$ and probability of success $p$ we have that  \citep{feller1968introduction}:
\begin{equation}
P(k\geq r) \leq \frac{r(1-p)}{(r-sp)^2}\label{upper-interval}
\end{equation}
if $r\geq sp$. Moreover
$$P(k\leq r) \leq \frac{(s-r)p}{(sp-r)^2}$$
if $r\leq sp$.
Since $P(k \geq r)=1-P(k<r)$, from (\ref{upper-interval}) we have
\begin{eqnarray*}
1-P(k<r)&\leq& \frac{r(1-p)}{(r-sp)^2}\nonumber\\
P(k<r)&\geq& 1-\frac{r(1-p)}{(r-sp)^2}\label{lb-lower-interval}
\end{eqnarray*}
So
\begin{eqnarray*}
&&P(\overline{F}\leq s\frac{M+c\sqrt{M}}{N})-P(\overline{F}< s\frac{M-c\sqrt{M}}{N}) \\
&\geq&1-\frac{s\frac{M+c\sqrt{M}}{N}(1-p)}{\left(s\frac{M+c\sqrt{M}}{N}-sp\right)^2}-\frac{\left(s-s\frac{M-c\sqrt{M}}{N}\right)p}{\left(sp-s\frac{M-c\sqrt{M}}{N}\right)^2}=\\
&=&1-\frac{\frac{M+c\sqrt{M}}{N}(1-\frac{M}{N})}{s\left(\frac{M+c\sqrt{M}}{N}-\frac{M}{N}\right)^2}-\frac{\left(1-\frac{M-c\sqrt{M}}{N}\right)\frac{M}{N}}{s\left(\frac{M}{N}-\frac{M-c\sqrt{M}}{N}\right)^2}=\\
&=&1-\frac{\frac{M+c\sqrt{M}}{N}\frac{N-M}{N}}{s\left(\frac{M+c\sqrt{M}-M}{N}\right)^2}-\frac{\left(\frac{N-M+c\sqrt{M}}{N}\right)\frac{M}{N}}{s\left(\frac{M-M-c\sqrt{M}}{N}\right)^2}=\\
&=&1-\frac{\left(M+c\sqrt{M}\right)(N-M)}{s\left(c\sqrt{M}\right)^2}-\frac{\left(N-M+c\sqrt{M}\right)M}{s\left(-c\sqrt{M}\right)^2}=\\
&=&1-\frac{MN+cN\sqrt{M}-M^2-cM\sqrt{M}-NM+M^2-cM\sqrt{M}}{sc^2M}=\\
&=&1-\frac{cN\sqrt{M}-2cM\sqrt{M}}{sc^2M}=\\
&=&1-\frac{N-2M}{sc\sqrt{M}}=1-\frac{N}{sc\sqrt{M}}+\frac{2\sqrt{M}}{sc}
\end{eqnarray*}
and
\begin{eqnarray*}
&&1-\frac{N-2M}{sc\sqrt{M}}\geq \frac{3}{4}\\
&&-\frac{N-2M}{sc\sqrt{M}}\geq -\frac{1}{4}\\
&&\frac{1}{4}\geq \frac{N-2M}{sc\sqrt{M}}\\
&&s\geq \frac{4N-8M}{c\sqrt{M}}\\
&&s\geq \frac{4N}{c\sqrt{M}}- \frac{8\sqrt{M}}{c}
\end{eqnarray*}
So (\ref{thesis-compl-cl-counting}) is true if $s\in \Omega(N)$.
\end{proof}
One may think that using an algorithm that chooses non uniformly the next assignment to test could provide a better bound. However, the algorithm would not be correct, because the probability 
of sampling a solution would depend on the black box function and the estimate $\overline{F}$ would be biased, in a way that would not be possible to correct without more information on the function.
The fact that this is the best bound, in the sense that any classical counting algorithm with a probability at least
3/4 for estimating $M$ correctly to within an accuracy $c\sqrt{M}$ for some constant $c$
must make $\Omega(N)$ oracle calls, was previously stated without proof \citep[Exercise 6.14]{nielsen2010quantum}, \citep[Table 2.5]{mosca1999quantum}.
So quantum computing gives us a quadratic speedup.

\section{Quantum Weighted Model Counting}
\label{qwmc}
We now present the QWMC algorithm.
For the moment suppose that the literal weights sum to 1, i.e., that $w(X_i)+w(\neg X_i)=1$ for all
bits $X_i$.

\begin{figure}[t]
\centering
\begin{scriptsize}
\begin{tikzpicture}[scale=0.700000,x=1pt,y=1pt]
\filldraw[color=white] (0.000000, -3.500000) rectangle (336.000000, 143.500000);
\draw[color=black] (0.000000,140.000000) -- (324.000000,140.000000);
\draw[color=black] (324.000000,139.500000) -- (336.000000,139.500000);
\draw[color=black] (324.000000,140.500000) -- (336.000000,140.500000);
\draw[color=black] (0.000000,133.000000) -- (324.000000,133.000000);
\draw[color=black] (324.000000,132.500000) -- (336.000000,132.500000);
\draw[color=black] (324.000000,133.500000) -- (336.000000,133.500000);
\draw[color=black] (0.000000,126.000000) -- (324.000000,126.000000);
\draw[color=black] (324.000000,125.500000) -- (336.000000,125.500000);
\draw[color=black] (324.000000,126.500000) -- (336.000000,126.500000);
\filldraw[color=white,fill=white] (0.000000,124.250000) rectangle (-4.000000,141.750000);
\draw[decorate,decoration={brace,amplitude = 4.000000pt},very thick] (0.000000,124.250000) -- (0.000000,141.750000);
\draw[color=black] (-4.000000,133.000000) node[left] {${\begin{array}{c}\mbox{Register 1}\\\ket{0}^{\otimes t}\end{array}}$};
\draw[color=black] (0.000000,100.000000) -- (336.000000,100.000000);
\draw[color=black] (0.000000,85.000000) -- (336.000000,85.000000);
\draw[color=black] (0.000000,70.000000) -- (336.000000,70.000000);
\draw[color=black] (0.000000,40.000000) -- (336.000000,40.000000);
\draw[color=black] (0.000000,25.000000) -- (336.000000,25.000000);
\filldraw[color=white,fill=white] (0.000000,21.250000) rectangle (-4.000000,103.750000);
\draw[decorate,decoration={brace,amplitude = 4.000000pt},very thick] (0.000000,21.250000) -- (0.000000,103.750000);
\draw[color=black] (-4.000000,62.500000) node[left] {${\begin{array}{c}\mbox{Register 2}\\\ket{0}^{\otimes n+1}\end{array}}$};
\draw[color=black] (0.000000,14.000000) -- (336.000000,14.000000);
\draw[color=black] (0.000000,7.000000) -- (336.000000,7.000000);
\draw[color=black] (0.000000,0.000000) -- (336.000000,0.000000);
\filldraw[color=white,fill=white] (0.000000,-1.750000) rectangle (-4.000000,15.750000);
\draw[decorate,decoration={brace,amplitude = 4.000000pt},very thick] (0.000000,-1.750000) -- (0.000000,15.750000);
\draw[color=black] (-4.000000,7.000000) node[left] {${\begin{array}{c}\mbox{Ancilla}\\\ket{0}^{\otimes q}\end{array}}$};
\draw (28.500000,140.000000) -- (28.500000,126.000000);
\begin{scope}
\draw[fill=white] (28.500000, 133.000000) +(-45.000000:31.819805pt and 18.384776pt) -- +(45.000000:31.819805pt and 18.384776pt) -- +(135.000000:31.819805pt and 18.384776pt) -- +(225.000000:31.819805pt and 18.384776pt) -- cycle;
\clip (28.500000, 133.000000) +(-45.000000:31.819805pt and 18.384776pt) -- +(45.000000:31.819805pt and 18.384776pt) -- +(135.000000:31.819805pt and 18.384776pt) -- +(225.000000:31.819805pt and 18.384776pt) -- cycle;
\draw (28.500000, 133.000000) node {$H^{\otimes t}$};
\end{scope}
\draw (28.500000,100.000000) -- (28.500000,25.000000);
\begin{scope}
\draw[fill=white] (28.500000, 62.500000) +(-45.000000:31.819805pt and 61.518290pt) -- +(45.000000:31.819805pt and 61.518290pt) -- +(135.000000:31.819805pt and 61.518290pt) -- +(225.000000:31.819805pt and 61.518290pt) -- cycle;
\clip (28.500000, 62.500000) +(-45.000000:31.819805pt and 61.518290pt) -- +(45.000000:31.819805pt and 61.518290pt) -- +(135.000000:31.819805pt and 61.518290pt) -- +(225.000000:31.819805pt and 61.518290pt) -- cycle;
\draw (28.500000, 62.500000) node {$Rot$};
\end{scope}
\draw (85.500000,126.000000) -- (85.500000,0.000000);
\begin{scope}
\draw[fill=white] (85.500000, 50.000000) +(-45.000000:31.819805pt and 79.195959pt) -- +(45.000000:31.819805pt and 79.195959pt) -- +(135.000000:31.819805pt and 79.195959pt) -- +(225.000000:31.819805pt and 79.195959pt) -- cycle;
\clip (85.500000, 50.000000) +(-45.000000:31.819805pt and 79.195959pt) -- +(45.000000:31.819805pt and 79.195959pt) -- +(135.000000:31.819805pt and 79.195959pt) -- +(225.000000:31.819805pt and 79.195959pt) -- cycle;
\draw (85.500000, 50.000000) node {$WG^{2^0}$};
\end{scope}
\filldraw (85.500000, 126.000000) circle(1.500000pt);
\draw (142.500000,133.000000) -- (142.500000,0.000000);
\begin{scope}
\draw[fill=white] (142.500000, 50.000000) +(-45.000000:31.819805pt and 79.195959pt) -- +(45.000000:31.819805pt and 79.195959pt) -- +(135.000000:31.819805pt and 79.195959pt) -- +(225.000000:31.819805pt and 79.195959pt) -- cycle;
\clip (142.500000, 50.000000) +(-45.000000:31.819805pt and 79.195959pt) -- +(45.000000:31.819805pt and 79.195959pt) -- +(135.000000:31.819805pt and 79.195959pt) -- +(225.000000:31.819805pt and 79.195959pt) -- cycle;
\draw (142.500000, 50.000000) node {$WG^{2^1}$};
\end{scope}
\filldraw (142.500000, 133.000000) circle(1.500000pt);
\draw[fill=white,color=white] (177.000000, -6.000000) rectangle (192.000000, 146.000000);
\draw (184.500000, 70.000000) node {$\cdots$};
\draw (226.500000,140.000000) -- (226.500000,0.000000);
\begin{scope}
\draw[fill=white] (226.500000, 50.000000) +(-45.000000:31.819805pt and 79.195959pt) -- +(45.000000:31.819805pt and 79.195959pt) -- +(135.000000:31.819805pt and 79.195959pt) -- +(225.000000:31.819805pt and 79.195959pt) -- cycle;
\clip (226.500000, 50.000000) +(-45.000000:31.819805pt and 79.195959pt) -- +(45.000000:31.819805pt and 79.195959pt) -- +(135.000000:31.819805pt and 79.195959pt) -- +(225.000000:31.819805pt and 79.195959pt) -- cycle;
\draw (226.500000, 50.000000) node {$WG^{2^{t-1}}$};
\end{scope}
\filldraw (226.500000, 140.000000) circle(1.500000pt);
\draw (283.500000,140.000000) -- (283.500000,126.000000);
\begin{scope}
\draw[fill=white] (283.500000, 133.000000) +(-45.000000:31.819805pt and 18.384776pt) -- +(45.000000:31.819805pt and 18.384776pt) -- +(135.000000:31.819805pt and 18.384776pt) -- +(225.000000:31.819805pt and 18.384776pt) -- cycle;
\clip (283.500000, 133.000000) +(-45.000000:31.819805pt and 18.384776pt) -- +(45.000000:31.819805pt and 18.384776pt) -- +(135.000000:31.819805pt and 18.384776pt) -- +(225.000000:31.819805pt and 18.384776pt) -- cycle;
\draw (283.500000, 133.000000) node {$FT^\dagger$};
\end{scope}
\draw[fill=white] (318.000000, 134.000000) rectangle (330.000000, 146.000000);
\draw[very thin] (324.000000, 140.600000) arc (90:150:6.000000pt);
\draw[very thin] (324.000000, 140.600000) arc (90:30:6.000000pt);
\draw[->,>=stealth] (324.000000, 134.600000) -- +(80:10.392305pt);
\draw[fill=white] (318.000000, 127.000000) rectangle (330.000000, 139.000000);
\draw[very thin] (324.000000, 133.600000) arc (90:150:6.000000pt);
\draw[very thin] (324.000000, 133.600000) arc (90:30:6.000000pt);
\draw[->,>=stealth] (324.000000, 127.600000) -- +(80:10.392305pt);
\draw[fill=white] (318.000000, 120.000000) rectangle (330.000000, 132.000000);
\draw[very thin] (324.000000, 126.600000) arc (90:150:6.000000pt);
\draw[very thin] (324.000000, 126.600000) arc (90:30:6.000000pt);
\draw[->,>=stealth] (324.000000, 120.600000) -- +(80:10.392305pt);
\end{tikzpicture}
\end{scriptsize}
\caption{Circuit for quantum weighted model counting.}
\label{qwmc_fig}
\end{figure}
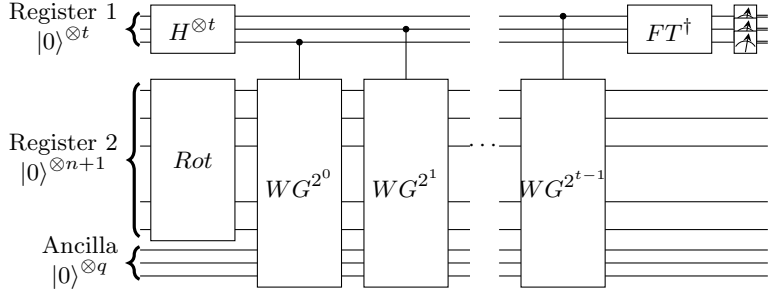

We can repeat the reasoning used for quantum counting: the application of the 
weighted Grover operator rotates $\ket{\varphi}$ in the space spanned by $\ket{\gamma}$ and $\ket{\delta}$ by angle
 $\theta$ and  $e^{i\theta}$ and $e^{i(2\pi-\theta)}$ are the eigenvalues of $\WG$. 
$\theta$ can be found by quantum phase estimation.
The overall circuit
 is shown in Figure \ref{qwmc_fig}.

From $\sin^2(\theta) =\frac{WMC}{2}$ we obtain
\begin{equation*}
\WMC=2\sin^2(\theta)
\end{equation*}
If the literal weights do not sum to 1, we consider the normalized weights as in Section \ref{qmap}.
Then we perform QWMC with $\hat w$ replacing $w$. We get a normalized count $\widehat{\WMC}$
from which we obtain $\WMC$ using Eq. \ref{norm-wmc}:
\begin{eqnarray*}
\WMC&=&\widehat{\WMC}\prod_{i=1}^{n}V_i\
\end{eqnarray*}
Let us consider the complexity of the algorithm. 
\begin{theorem}
\label{qwmc-compl}
QWMC on $n$ bits requires $\Theta(\sqrt{N})$ oracle calls to bound the error to $2^{-\frac{n}{2}-\frac{1}{2}}$ with probability $11/12$ using $t=\lceil n/2\rceil+5$ bits.
\end{theorem}
\begin{proof}
We can repeat the derivation of Theorem \ref{compl_q_counting} where $M$ is replaced by  $N\times \widehat{\WMC}$.
We get
$$\frac{|\Delta \widehat{\WMC}|}{2}<\left(2\sin\theta+|\Delta\theta|\right)|\Delta\theta|$$
Using $\sin^2(\theta)=\widehat{\WMC}/2$ and $|\Delta\theta|\leq 2^{-m}$ we obtain
$$|\Delta \widehat{\WMC}|<\left(\sqrt{2\widehat{\WMC}}+2^{-m+1}\right)2^{-m}$$
Since $\widehat{\WMC}\leq1$ we have
$$|\Delta \widehat{\WMC}|<\left(\sqrt{2}+2^{-m-1}\right)2^{-m}<2^{-m+\frac{1}{2}}+2^{-2m-1}
$$
If we choose $m=\lceil n/2\rceil+2$ and $\epsilon=1/12$, then $t=\lceil n/2\rceil+5$ and the algorithm 
requires $\Theta(\sqrt{N})$ oracle calls. For $n$ even, $m=n/2+2$ and the error becomes:
\begin{eqnarray*}
&&|\Delta \widehat{\WMC}|<2^{-\frac{n}{2}-2+\frac{1}{2}}+2^{-n-5}<\\
&&2^{-\frac{n}{2}-\frac{3}{2}}+2^{-\frac{n}{2}-\frac{3}{2}}<2^{-\frac{n}{2}-\frac{1}{2}}
\end{eqnarray*} 
For $n$ odd, $m=n/2+1/2+2=n/2+5/2$, so
\begin{eqnarray*}
&&|\Delta \widehat{\WMC}|<2^{-\frac{n}{2}-\frac{5}{2}+\frac{1}{2}}+2^{-n-3-1}<\\
&&2^{-\frac{n}{2}-2}+2^{-\frac{n}{2}-2}<2^{-\frac{n}{2}-1}<2^{-\frac{n}{2}-\frac{1}{2}}
\end{eqnarray*} 
Thus overall the error is bounded by $2^{-\frac{n}{2}-\frac{1}{2}}$.
\end{proof}
%
%
%
%
%

\section{Comparison of QWMC with Classical Algorithms}
\label{wcompl}
Let us now discuss the advantages of QWMC with respect to WMC under a black box model of computation.

For WMC, consider the following classical algorithm: take $s$ assignment samples by sampling each bit according
to its normalized weight. For each assignment sample, query the oracle obtaining value $F_i$ with $i=1,\ldots,s$ and 
estimate the WMC as for the unweighted case: $S=\frac{N}{s}\sum_{i=1}^sF_i.$
Variable $Ss/N$ is again binomially distributed with $s$ the number of trials and probability of success
$\widehat{\WMC}$.
In fact, the probability $P(F_i=1)$ is given by
$P(F_i=1)=\sum_{x}P(F_i=1,x)=\sum_{x}P(F_i=1|x)P(x)$
where $P(F_i=1|x)$ is 1 if $x$ is a model of $\phi$ and 0 otherwise. So
\begin{eqnarray*}
P(F_i=1)&=&\sum_{x:\phi(x)=1}P(x)=\\
&&\sum_{x_{n}\ldots x_n:\phi(x_{1}\ldots x_n)=1}P(x_{1}\ldots x_n)=\\
&&\sum_{x_{n}\ldots x_n:\phi(x_{1}\ldots x_n)=1}P(x_{1})\ldots P(x_n)=\\
&&\sum_{x_{n}\ldots x_n:\phi(x_{1}\ldots x_n)=1}\prod_{i=1}^{n}\hat w'_{i}=\\
&&\sum_{x_{n}\ldots x_n:\phi(x_{1}\ldots x_n)=1}\prod_{i=1}^{n}\frac{w_i}{V_i}=\\
&&\frac{\WMC}{\prod_{i=1}^{n}V_i}=\widehat{\WMC}
\end{eqnarray*}
\begin{theorem}
\label{compl-class-wmc}
The complexity of any classical algorithm for estimating $\widehat{\WMC}$ under a black box model 
with a probability of at least $3/4$ within an accuracy of $2^{-\lceil\frac{ n}{2}\rceil}$ is $\Omega(N)$ oracle calls and this is the best bound for a classical algorithm.
\end{theorem}
\begin{proof}
We can repeat the reasoning performed in the proof of Theorem \ref{compl-class-count} by considering
$\widehat{\WMC}$ in place of $\frac{M}{N}$ and $2^{-\lceil\frac{ n}{2}\rceil}N=2^{-\lceil\frac{ n}{2}\rceil+n}=2^{\lfloor\frac{n}{2}\rfloor}$ in place of $c\sqrt{M}$.
We obtain
$$s \geq \frac{4N}{2^{\lfloor\frac{ n}{2}\rfloor}}- \frac{8\sqrt{N\widehat\WMC}}{c}$$
Therefore $k=\Omega(N)$. 
This is also the best bound for a classical algorithm, as otherwise we could solve model counting with a better bound than $\Omega(N)$ by setting all weights to 0.5.
\end{proof}
Thus QWMC offers a quadratic speedup over classical computation in the black box model.

\section{Related Work}
\label{related}

\citet{YING2010162} presents a survey of the applications of quantum
computation in AI. The author says that quantum search was believed to be the one of the first
quantum computing techniques to play an important role in AI but few successful applications of
quantum search in AI have been reported in the '00s, the decade preceding the publication of 
\citep{YING2010162}. In this paper we try to remedy this and provide an application of quantum search 
to the important AI problems of WMC and WCS.

Recently the problem of computing the weighted count of eigenstates of Hamiltonians was tackled by
\citep{sundar2019quantum}. The authors proposed mixed quantum-Monte Carlo algorithms that repeat several times
quantum computations and then use statistics from the results to approximate the weighted count. In particular,
they propose applications of Adiabatic quantum optimization (AQO), quantum approximate optimization algorithm
(QAOA) and Grover's algorithm to find the eigenstates and provide individual samples.
The authors prove (numerically for QAOA) that the number of samples needed to achieve a certain relative
error is lower than that of classical optimal Monte Carlo simulation.
This work can be used to perform weighted model counting, we differ from it because we rely on quantum phase estimation rather than
sampling.

Knowledge compilation  \citep{DBLP:journals/jair/DarwicheM02,lagniez2017improved,huang2006solving} solves WMC classically by compiling the Boolean
formula to a representation such as deterministic Decomposable Negation Normal Form (d-DNNF) where counting is linear in the size of the representation.
Here the expensive part is the compilation one. 

WMC is an instance of a  ``sum of products'' problem, i.e. evaluating the sum of products of values from some semiring, 
a general framework that encompasses many problems in Artificial Intelligence and that has been much studied.
\citet{bacchus2009solving} use a modified Davis, Putamn, Logemann, Loveland (DPLL) procedure to solve these problems by means of backtracking.
\citet{DBLP:conf/icml/FriesenD16} propose conditions under which the problems become tractable and \citet{eiter2021complexity} study their complexity. 
\citet{ganian2022sum} present algorithms for Weighted Counting for Constraint Satisfaction with Default Values, a specialization of ``sum of products'' 
that still encompasses WMC. These algorithm are polynomial once the incidence treewidth, a more general version of treewidth, is bounded.

\citet{DBLP:conf/ijcai/ChakrabortyFMV15} present a transformation from a WMC problem to an unweighted model counting problem so that
unweighted model counters can be applied. Since these are usually better engineered, this approach leads to significant improvements on many problems.

A classical algorithm for  weighted model counting is proposed by
\citep{DBLP:conf/esa/FichteHWZ18} and exploits a dynamic programming algorithm on a tree decomposition and can be parallelized using GPUs.

All these approaches  require  knowledge of the structure of the formula that, furthermore, should have a small treewidth,  while our approach is targeted
to problem with high treewidth (more than half of the number of variables).

Approximate classical algorithms for counting are surveyed in \citep{chakraborty2021approximate}.  
\cite{DBLP:conf/ijcai/ChakrabortyMV16} present an approximate algorithm that uses a logarithmic number of calls to a SAT oracle. 
\cite{Chekraborty-Fremont} propose approximate algorithms WeigthMC and WeightGen for WMC and WCS respectively,
both using  a polynomial number of calls 
to a SAT oracle.
We differ from these work because in our case the oracle is the evaluation of the Boolean function which is much cheaper than a SAT call.

SampleSAT \citep{DBLP:conf/aaai/WeiES04} is an algorithm for sampling solutions to SAT problems nearly uniformly.
It combines WalkSAT with Simulated Annealing. QWCS generalize this algorithm to the case where
we also consider weights. 

Quantum computing has been applied to SAT and MaxSAT in
\citep{BIAN2020104609}, where the authors use a quantum annealer.

The complexity of the MAP problem for Bayesian networks has been studied in \citep{DBLP:journals/jair/ParkD04}.
The authors show that exact MAP is complete for $\mathrm{NP}^{\mathrm{PP}}$. Approximate 
algorithms for MAP in Bayesian networks are taken into acocunt in \citep{DBLP:journals/jair/Kwisthout15}.
The author considers various types of approximations: value-approximations, where the result has a value close to the optimal value;
structure-approximations, where the result has a small Hamming distance from the optimal solution; rank-approximations, where the result belongs
to the best $m$ solutions, and expectation-approximations, where the result is the optimal solution with high probability.
\citet{DBLP:journals/jair/Kwisthout15} proves that all these approximations are intractable unless $\mathrm{P}=\mathrm{NP}$ or
$\mathrm{NP}\subseteq\mathrm{BPP}$.

\citet{DBLP:journals/jair/Kwisthout15} then considers fixed-parameter tractability, where the problem becomes tractable if a limit
is imposed on parameters of the input data. For Bayesian networks an important parameter is treewidth: \citet{DBLP:journals/jair/Kwisthout15} shows that
limiting the treewidth is necessary to obtain tractable value-approximations, structure-approximations and rank-approximations, while it is not necessary
for expectation-approximations. A particular, constrained version of MAP can be efficiently approximated
under expectation-approximations if the probability of the MAP explanation
is high.

These results carry over to the weighted propositional problems we consider, as they can be encoded with Bayesian networks (and viceversa).
However, these results require knowledge of the formula (and Bayesian network), while we require only an oracle that returns the value of 
the propositional formula given the value of the variables: we use a black box model of computation. As such, our results are complementary to these.

\section{Discussion}
\label{discussion}
The idea of using rotation gates to represent weights was first proposed in  \citep{Rig20-ECAI-IC} but the QWMC algorithm there contained an error: it used the regular Grover operator instead of  the weighted Grover operator where $H$ gates are replaced by 
$Rot$. This article fixes this problem and proposes one more algorithm, QWCS, for solving the weighted constrained sampling problem. This algorithm exploits the same trick of using rotations to represent weights
and basically combines weighted searching together with projection
on the variables of interest.  

We have shown that QWMC  has  
a complexity of $\Theta(2^{\frac{n}{2}})$ evaluations of the Boolean formula, while QWCS solve its problem with a complexity of  $O(2^{\frac{n}{2}}+1/\sqrt{\WMC})$, where $\WMC$ is the
normalized  weighted model count of the formula.
We have also shown that if we consider the Boolean formula as a black box that we can only query asking for the value of the function given the inputs, QWMC  provides a quadratic 
speedup over classical algorithms with the same limitation. The black box setting
may be of interest when the Boolean formula is given by a quantum physical system of which we don’t know the internals.
In that case the quantum algorithms can plug in the system directly, improving over classical algorithms. 

In the majority of cases where we want to perform WMC, WCS,  we know the Boolean formula and, assuming the cost of implementing the circuit is linear in the number $n$ of variables, the complexity will be worse than classical algorithms for WMC, QWCS unless the treewidth of the model is larger than $n/2$.

However, QWMC can also be used as a subroutine in the junction tree algorithm \citep{DBLP:conf/uai/ShenoyS88,lauritzen1988local}:  after the probabilities are propagated in the tree, the nodes, whose maximum size minus 1 
is the treewidth, have to be processed to find the marginals of the individual variables. In this QWMC can help with a complexity of $\Theta(2^{\frac{n}{2}})$ with $n$ the number of variables in the node.

In general, the algorithms exploit quantum parallelism: all the models of the formula are  superimposed in the quantum state of the system. Unfortunately, however, the state is not directly accessible
and several applications of the Grover operator are required to extract a model or phase estimation is required to extract the count. 

\section{Conclusions}
\label{conc}
We have proposed quantum algorithms for performing WMC and WCS. 
The algorithms modify the quantum search and quantum 
counting algorithms  by taking into account weights.

Using the black box model of computation, QWMC makes $\Theta(\sqrt{N})$ oracle calls to return
a result whose errors is bounded by $ 2^{-\frac{ n+1}{2}}$ with probability 11/12.
By contrast, the best classical algorithm requires $\Theta(N)$ calls to the oracle.
Thus QWMC offers a quadratic speedup that may be useful, for example, for computing marginals for the variables of a tree node in
the junction tree algorithm.

Similarly, QWCS requires $O(2^{\frac{n}{2}}+1/\sqrt{\WMC})$ oracle queries, while  classical probabilistic algorithms
 take $\Omega(1/\WMC)$
oracle queries under the black box model of computation, again providing a quadratic improvement.

In the future, we plan to investigate the influence of noise on the quality of the results.


\section*{Acknowledgements}
The author would like to thank Mariia Mykhailova for interesting discussions on the topic of this paper and for her help in developing the Q\# code.

This work has been partially supported by  Spoke 1 ``FutureHPC \& BigData'' of the Italian Research Center on High-Performance Computing, Big Data and Quantum Computing (ICSC) funded by Ministero dell'Università e della Ricarca - Missione 4 - Next Generation EU (NGEU),  
by TAILOR, a project funded by EU Horizon 2020 research and innovation programme under GA No. 952215, and by the ``National Group of Computing Science (GNCS-INDAM)''.

\appendix
\section{Introduction to Quantum Computing}
\label{qc}
Here we provide a brief introduction to quantum computing following \citep{nielsen2010quantum}.
The bit is at the basis of classical computing and has a single value, either 0 or 1. The \emph{quantum bit} or \emph{qubit} is a generalization of the bit and is at the basis of quantum computing. A qubit represents a state defined by a pair of
complex numbers that can have various physical implementations.
From a mathematical point of view, a qubit is a unit vector in the $\C^2$ space,  where $\C$ is the set of complex numbers, i.e., 
$$\left[\begin{array}{l}
\alpha\\
\beta
\end{array}\right]
$$
where the normalization requirement is that $|\alpha|^2+|\beta|^2=1$.

Qubits are represented using the Dirac notation, where $\ket{\psi}$ (read ``ket psi'') is a two dimensional  column vector and  $\bra{\psi}$ (read ``bra psi'') is a two dimensional  complex conjugate row vector. The 
notation $\braket{\psi|\phi}$ (read ``braket'')  is the inner product of the $\C^2$ space, i.e., it is the dot product between $\ket{\phi}$ and the complex conjugate $\ket{\psi}^*$.
 The special states $\ket{0}$ and $\ket{1}$ are 
called \emph{computational basis states} and form the orthonormal basis 
$$\left[\begin{array}{c}1\\0\end{array}\right]\ \ \ \ \left[\begin{array}{c}0\\1\end{array}\right]$$
for $\C^2$.
Any qubit state $\ket{\psi}$ can  be expressed as a linear combination of the computational basis states:
$$\ket{\psi}=\alpha\ket{0}+\beta\ket{1}=\left[\begin{array}{c}\alpha\\\beta\end{array}\right]$$
In this case we say that $\ket{\psi}$ is in a \emph{superposition} of states $\ket{0}$ and $\ket{1}$.
Note that computational basis state are orthonormal: $\braket{0|1}=\braket{1|0}=0$.

\subsection{Composite Systems}
Whenever we have more than one qubit, we have a composite physical system {\color{blue} (also called a \emph{quantum register})} and the state space expands accordingly: for $n$ qubits, their state
 is a unit vector in the $\C^{2^n}$ space and
 there are $2^n$ computational basis states, 
e.g., if $n=2$, the basis states are $\ket{00}$, $\ket{01}$, $\ket{10}$ and
$\ket{11}$, and the state of the qubits can be written as
$$\ket{\psi}=\alpha_{00}\ket{00}+\alpha_{01}\ket{01}+\alpha_{10}\ket{10}+\alpha_{11}\ket{11}$$
where $\alpha_{00},\ldots,\alpha_{11}$ are complex numbers that form a unit length vector, i.e., such that $|\alpha_{00}|^2+|\alpha_{01}|^2+|\alpha_{10}|^2+|\alpha_{11}|^2=1$.
The state space of a composite physical system given  the component physical systems can be obtained using the tensor product of the states of the components.

The tensor product of two qubits 
$$\ket{a}=a_0\ket{0}+a_1\ket{1}=\left[\begin{array}{c}
a_0\\
a_1
\end{array}\right]$$
$$
\ket{b}=b_0\ket{0}+b_1\ket{1}=\left[\begin{array}{c}
b_0\\
b_1
\end{array}\right]
$$
is
$$\ket{a}\otimes \ket{b}=\left[\begin{array}{c}
a_0b_0\\
a_0b_1\\
a_1b_0\\
a_1b_1\\
\end{array}\right]=$$
$$a_0b_0\ket{00}+
a_0b_1\ket{01}+
a_1b_0\ket{10}+
a_1b_1\ket{11}
$$
The tensor product $\ket{a}\otimes\ket{b}$ is also written succinctly as $\ket{a}\ket{b}$ or also as $\ket{ab}$, so for example $\ket{0}\otimes\ket{0}=\ket{0}\ket{0}=\ket{00}$.

The Dirac notation is also useful for representing outer products $\ket{a}\bra{b}$: 
$$\ket{a}\bra{b}=\left[\begin{array}{l}
a_0\\
a_1
\end{array}\right]\left[\begin{array}{ll}b_0^*&b_1^*\end{array}\right]=\left[\begin{array}{ll}
a_0b_0^*&a_0b_1^*\\
a_1b_0^*&a_1b_1^*
\end{array}\right]$$
\subsection{Measurement}
One of the operations that can be performed on a quantum system is \emph{measurement}. There are various types of measurements, the simplest is measurement in the computational basis.
When we have a qubit in the state $\ket{\psi}=\alpha\ket{0}+\beta\ket{1}$ and we measure it in the computational basis, we obtain a classical bit as a result: $0$ with probability $|\alpha|^2$ and
$1$ with probability $|\beta|^2$. Since the states of qubits are unit vectors, then $|\alpha|^2+|\beta|^2=1$ and the probabilities of the outcomes are well-defined. We can also measure multi-qubit systems and in that 
case we obtain a vector of classical bits.

\subsection{Density Operators}
A qubit in a superposition state encodes uncertainty on the result of its measurement. In this case, the state is known with certainty, it is only the result of measurement
that is uncertain. Sometimes we want to represent uncertainty also on the state of the system.  For example, we may know that the system
is in one of several states $\ket{\psi_i}$, where $i$ is an index, with respective probabilities $p_i$.
In this case we can represent the state of the system using a \emph{density operator} $\rho$
 defined by the equation
$$\rho=\sum_ip_i\ket{\psi_i}\bra{\psi_i}$$
so density operators are matrices.
If the state $\ket{\psi}$ of a quantum system is known exactly, the system is said to be in a \emph{pure state} and the density operator is simply
$\rho=\ket{\psi}\bra{\psi}$. Otherwise, the system is said to be in a \emph{mixed state} and that it is  in a \emph{mixture} of different
pure states in the ensemble $\{(p_i,\ket{\psi_i})\}$ for $\rho$.

Quantum mechanics can be formulated in terms of density operators as well as in terms of state vectors. Each technique is more convenient in certain cases, for example, density operators are convenient when we 
want to identify the state of a subsystem of a quantum system.

 Consider a composite system $AB$ obtained by composing two systems $A$ and $B$ with density operators $\rho^A$ and $\rho^B$. The joint state is described by a density
operator $\rho^{AB}=\rho^A\otimes \rho^B$. The state of system $\rho^A$ is also called \emph{reduced density operator} and is obtained as
$$\rho^A=\tr_B(\rho^{AB})$$
where $\tr_B$ is a map called the \emph{partial trace over system} $B$ and is defined to apply to a tensor product as
\begin{equation}
\tr_B(\ket{a_1}\bra{a_2}\otimes\ket{b_1}\bra{b_2})=\ket{a_1}\bra{a_2}\tr (\ket{b_1}\bra{b_2})\label{partial-trace}
\end{equation}
where $\ket{a_1}$ and $\ket{a_2}$ are any two vectors in the state space of $A$,  $\ket{b_1}$ and $\ket{b_2}$ are any two vectors in the state space of $B$, and $\tr$ is the 
trace operator on matrices defined as  the sum of its diagonal elements
$$\tr(A)=\sum_{i}A_{ii}.$$
The trace operator has the following important property
\begin{eqnarray}
\tr(\ket{b_1}\bra{b_2})&=&\braket{b_2|b_1}\label{tr_outer}
\end{eqnarray}
When we extract the state of a subsystem of a system that is in a pure state, we can obtain a mixed state, meaning that we have uncertainty in the state of the subsystem.
Consider a system with two qubits in the Bell state $(\ket{00}+\ket{11})/\sqrt{2}$. Its density operator is
\begin{eqnarray*}
\rho^{AB }&=&\left(\frac{\ket{00}+\ket{11}}{\sqrt{2}}\right)\left(\frac{\bra{00}+\bra{11}}{\sqrt{2}}\right)\\
&=&\frac{\ket{00}\bra{00}+\ket{11}\bra{00}+\ket{00}\bra{11}+\ket{11}\bra{11}}{2}
\end{eqnarray*}
We want to compute the reduced density operator for the first qubit so
\begin{eqnarray*}
\rho^A&=&\tr_B(\rho)\\
&=&\frac{\tr_B(\ket{00}\bra{00})+\tr_B(\ket{11}\bra{00})+\tr_B(\ket{00}\bra{11})+\tr_B(\ket{11}\bra{11})}{2}\\
&=&\frac{\ket{0}\bra{0}\braket{0|0}+\ket{1}\bra{0}\braket{0|1}+\ket{0}\bra{1}\braket{1|0}+\ket{1}\bra{1}\braket{1|1}}{2}\\
&=&\frac{\ket{0}\bra{0}+\ket{1}\bra{1}}{2}
\end{eqnarray*}
As you can see, the system is in a mixture of the ensemble $\{(1/2,\ket{0}),(1/2,\ket{1})\}$. So, even if we have perfect knowledge of the system, we have uncertainty on the state of a subsystem.

\subsection{Quantum Circuits}
To present quantum algorithms, we use the quantum circuit model of computation, where each qubit corresponds to a wire and quantum gates are applied to sets of wires.

\begin{figure}
\centering
\begin{subfigure}[b]{0.3\textwidth}
\centering
\begin{tikzpicture}[scale=1.000000,x=1pt,y=1pt]
\filldraw[color=white] (0.000000, -7.500000) rectangle (24.000000, 7.500000);
\draw[color=black] (0.000000,0.000000) -- (24.000000,0.000000);
\begin{scope}
\draw[fill=white] (12.000000, -0.000000) +(-45.000000:8.485281pt and 8.485281pt) -- +(45.000000:8.485281pt and 8.485281pt) -- +(135.000000:8.485281pt and 8.485281pt) -- +(225.000000:8.485281pt and 8.485281pt) -- cycle;
\clip (12.000000, -0.000000) +(-45.000000:8.485281pt and 8.485281pt) -- +(45.000000:8.485281pt and 8.485281pt) -- +(135.000000:8.485281pt and 8.485281pt) -- +(225.000000:8.485281pt and 8.485281pt) -- cycle;
\draw (12.000000, -0.000000) node {$X$};
\end{scope}
\end{tikzpicture}
   \caption{$X$ gate.}
         \label{fig:xgate}
     \end{subfigure}
\begin{subfigure}[b]{0.3\textwidth}
\centering
\begin{tikzpicture}[scale=1.000000,x=1pt,y=1pt]
\filldraw[color=white] (0.000000, -7.500000) rectangle (24.000000, 7.500000);
\draw[color=black] (0.000000,0.000000) -- (24.000000,0.000000);
\begin{scope}
\draw[fill=white] (12.000000, -0.000000) +(-45.000000:8.485281pt and 8.485281pt) -- +(45.000000:8.485281pt and 8.485281pt) -- +(135.000000:8.485281pt and 8.485281pt) -- +(225.000000:8.485281pt and 8.485281pt) -- cycle;
\clip (12.000000, -0.000000) +(-45.000000:8.485281pt and 8.485281pt) -- +(45.000000:8.485281pt and 8.485281pt) -- +(135.000000:8.485281pt and 8.485281pt) -- +(225.000000:8.485281pt and 8.485281pt) -- cycle;
\draw (12.000000, -0.000000) node {$Z$};
\end{scope}
\end{tikzpicture}
   \caption{$Z$ gate.}
         \label{fig:zgate}
     \end{subfigure}
\begin{subfigure}[b]{0.3\textwidth}
\centering
\begin{tikzpicture}[scale=1.000000,x=1pt,y=1pt]
\filldraw[color=white] (0.000000, -7.500000) rectangle (24.000000, 7.500000);
\draw[color=black] (0.000000,0.000000) -- (24.000000,0.000000);
\begin{scope}
\draw[fill=white] (12.000000, -0.000000) +(-45.000000:8.485281pt and 8.485281pt) -- +(45.000000:8.485281pt and 8.485281pt) -- +(135.000000:8.485281pt and 8.485281pt) -- +(225.000000:8.485281pt and 8.485281pt) -- cycle;
\clip (12.000000, -0.000000) +(-45.000000:8.485281pt and 8.485281pt) -- +(45.000000:8.485281pt and 8.485281pt) -- +(135.000000:8.485281pt and 8.485281pt) -- +(225.000000:8.485281pt and 8.485281pt) -- cycle;
\draw (12.000000, -0.000000) node {$H$};
\end{scope}
\end{tikzpicture}
   \caption{$H$ gate.}
         \label{fig:hgate}
     \end{subfigure}

\begin{subfigure}[b]{0.3\textwidth}
\centering
\begin{tikzpicture}[scale=1.000000,x=1pt,y=1pt]
\filldraw[color=white] (0.000000, -7.500000) rectangle (42.000000, 7.500000);
\draw[color=black] (0.000000,0.000000) -- (42.000000,0.000000);
\begin{scope}
\draw[fill=white] (21.000000, -0.000000) +(-45.000000:21.213203pt and 8.485281pt) -- +(45.000000:21.213203pt and 8.485281pt) -- +(135.000000:21.213203pt and 8.485281pt) -- +(225.000000:21.213203pt and 8.485281pt) -- cycle;
\clip (21.000000, -0.000000) +(-45.000000:21.213203pt and 8.485281pt) -- +(45.000000:21.213203pt and 8.485281pt) -- +(135.000000:21.213203pt and 8.485281pt) -- +(225.000000:21.213203pt and 8.485281pt) -- cycle;
\draw (21.000000, -0.000000) node {$R_y(\theta)$};
\end{scope}
\end{tikzpicture}
   \caption{$R_y$ gate.}
         \label{fig:rygate}
     \end{subfigure}
\begin{subfigure}[b]{0.3\textwidth}
\centering
\begin{tikzpicture}[scale=1.000000,x=1pt,y=1pt]
\filldraw[color=white] (0.000000, -7.500000) rectangle (24.000000, 7.500000);
\draw[color=black] (0.000000,0.000000) -- (12.000000,0.000000);
\draw[color=black] (12.000000,-0.500000) -- (24.000000,-0.500000);
\draw[color=black] (12.000000,0.500000) -- (24.000000,0.500000);
\draw[fill=white] (6.000000, -6.000000) rectangle (18.000000, 6.000000);
\draw[very thin] (12.000000, 0.600000) arc (90:150:6.000000pt);
\draw[very thin] (12.000000, 0.600000) arc (90:30:6.000000pt);
\draw[->,>=stealth] (12.000000, -5.400000) -- +(80:10.392305pt);
\end{tikzpicture}
   \caption{Measurement gate.}
         \label{fig:mgate}
     \end{subfigure}
\begin{subfigure}[b]{0.3\textwidth}
\centering
\begin{tikzpicture}[scale=1.000000,x=1pt,y=1pt]
\filldraw[color=white] (0.000000, -7.500000) rectangle (18.000000, 22.500000);
\draw[color=black] (0.000000,15.000000) -- (18.000000,15.000000);
\draw[color=black] (0.000000,15.000000) node[left] {$\ket{a}$};
\draw[color=black] (0.000000,0.000000) -- (18.000000,0.000000);
\draw[color=black] (0.000000,0.000000) node[left] {$\ket{b}$};
\draw (9.000000,15.000000) -- (9.000000,0.000000);
\begin{scope}
\draw[fill=white] (9.000000, 0.000000) circle(3.000000pt);
\clip (9.000000, 0.000000) circle(3.000000pt);
\draw (6.000000, 0.000000) -- (12.000000, 0.000000);
\draw (9.000000, -3.000000) -- (9.000000, 3.000000);
\end{scope}
\filldraw (9.000000, 15.000000) circle(1.500000pt);
\draw[color=black] (18.000000,15.000000) node[right] {$\ket{a}$};
\draw[color=black] (18.000000,0.000000) node[right] {$\ket{b\oplus a}$};
\end{tikzpicture}
   \caption{CNOT gate.}
         \label{fig:cnotgate}
     \end{subfigure}
\caption{Examples of quantum gates and measurements.}
\label{xgate}
\end{figure}
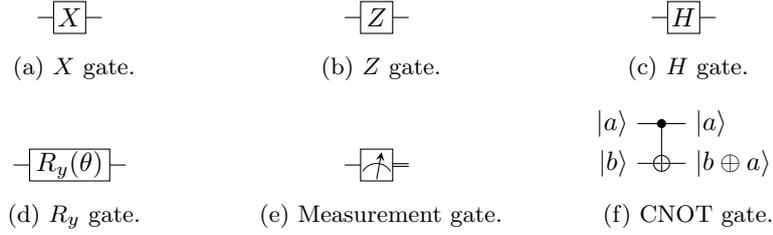

Quantum gates are linear operators represented by matrices with complex elements that must be unitary.
A matrix is \emph{unitary} if $M^\dagger M=I$, where $M^\dagger$
is the \emph{adjoint} or \emph{Hermitian conjugate} of a matrix $M$, i.e., the complex conjugate transpose matrix $M^\dagger=(M^*)^T$. 
We start from gates operating on single qubits that are described by matrices belonging to $\C^{2\times 2}$.
For  example, the quantum counterpart of the NOT Boolean gate for classical bits is the $X$ gate, defined as
$$X=\left[\begin{array}{cc}0&1\\
1&0\end{array}\right]$$
and drawn in circuits as in Figure \ref{fig:xgate}.
Quantum gates can also be defined by the effect they have on an orthonormal basis. For example, applying the $X$ gate to the computational basis produces:
$$X\ket{0}=\left[\begin{array}{cc}0&1\\
1&0\end{array}\right]\left[\begin{array}{c}1\\
0\end{array}\right]=\left[\begin{array}{c}0\\
1\end{array}\right]=\ket{1}$$
and
$$X\ket{1}=\left[\begin{array}{cc}0&1\\
1&0\end{array}\right]\left[\begin{array}{c}0\\
1\end{array}\right]=\left[\begin{array}{c}1\\
0\end{array}\right]=\ket{0}$$
so $X$ exchanges $\ket{0}$ and $\ket{1}$, justifying its role as the quantum counterpart of the NOT Boolean gate.

The $Z$ gate (see Figure \ref{fig:zgate}) is defined as
$$Z=\left[\begin{array}{cc}1&0\\
0&-1\end{array}\right]$$
i.e, it transforms $\ket{0}$ to $Z\ket{0}=\ket{0}$ and $\ket{1}$ to $Z\ket{1}=-\ket{1}$.

The \emph{Hadamard} gate (see Figure \ref{fig:hgate}) is 
$$H=\frac{1}{\sqrt{2}}\left[\begin{array}{cc}1&1\\
1&-1\end{array}\right]$$
and transforms $\ket{0}$ to $H\ket{0}=\frac{1}{\sqrt{2}}(\ket{0}+\ket{1})$ and $\ket{1}$ to $H\ket{1}=\frac{1}{\sqrt{2}}(\ket{0}-\ket{1})$.

Another useful gate is the parameterized $R_y$ \emph{rotation} gate (see Figure \ref{fig:rygate})
$$R_y(\theta)=\left[\begin{array}{cc}
\cos\frac{\theta}{2}&-\sin\frac{\theta}{2}\\
\sin\frac{\theta}{2}&\cos\frac{\theta}{2}
\end{array}\right]
$$
\emph{Measurement} can be seen as another type of operator applied to quantum states. 
In general, a measurement is described by a collection of
measurement operators $\{M_m\}$.  The index $m$ refers to the measurement outcomes that
may occur in the experiment. If the state of the quantum system is $\ket{\psi}$ immediately before the measurement, then the probability that result $m$ occurs is
$$P(m)=\bra{\psi}M_m^\dagger M_m\ket{\psi}$$
The measurement operators must satisfy the condition 
$$\sum_mM_m^\dagger M_m=I$$
This condition ensures that the probabilities add up to one. A measurement of a qubit in the computational basis  is given by the operators
$M_0=\ket{0}\bra{0}$ and $M_1=\ket{1}\bra{1}$. It's easy to verify that $M_0^\dagger M_0=M_0^2=M_0$, 
$M_1^\dagger M_1=M_1^2=M_1$, and $M_0^\dagger M_0+M_1^\dagger M_1=M_0+M_1=I$. This holds for all
measurements in the computational basis. 
 Measurement is drawn in circuits as in Figure \ref{fig:mgate}.

The results of measurement can also be computed with density operators. Suppose we have a measurement composed of measurement operators $\{M_m\}$ and that the system is described by density
operator $\rho$ given by  the ensemble $\{(p_i,\ket{\psi_i})\}$. If the system is in pure state $\ket{\psi_i}$,  the probability of obtaining result $m$ is
$$P(m|i)=\bra{\psi_i}M_m^\dagger M_m\ket{\psi_i}=\tr(M_m^\dagger M_m\ket{\psi_i}\bra{\psi_i})$$
where we applied
the following property of the trace operator
\begin{eqnarray*}
\tr(A\ket{\psi}\bra{\psi})&=&\bra{\psi}A\ket{\psi}
\end{eqnarray*}
Then the marginal probability of result $m$ is
\begin{eqnarray*}
P(m)&=&\sum_i p(m|i)p_i\\
&=&\sum_i p_i\tr(M_m^\dagger M_m\ket{\psi_i}\bra{\psi_i})\\
&=&\tr(M_m^\dagger M_m\rho)
\end{eqnarray*}
Turning to composite systems,
 the most important two-qubit gate  is  \emph{controlled-NOT} or \emph{CNOT}  that has two inputs, the 
control and the target qubits. The gate flips the target qubit if the control bit is set to 1 and does nothing if the control bit is set to 0. 
It can be seen as a gate that  operates as $\ket{ab}\to\ket{a,b\oplus a}$, where $\oplus$ is the XOR operation, see Figure \ref{fig:cnotgate}.

Any multi-qubit
logic gate can be composed from CNOT and single-qubit gates.

CNOT may be generalized to the case of more than two qubits: in this case, the extra qubits act as controls and the target is flipped if all controls are 1, see Figure \ref{ccnot}. 
Moreover, given an operator $U$, it is possible to define a controlled-$U$ operator defined as $\ket{ab}\to\ket{a,U^ab}$: if $a=0$ it does nothing, otherwise it applies operator $U$ to $b$.
\begin{figure}[t]
\centering
\begin{footnotesize}
\begin{tikzpicture}[scale=1.000000,x=1pt,y=1pt]
\filldraw[color=white] (0.000000, -7.500000) rectangle (18.000000, 37.500000);
\draw[color=black] (0.000000,30.000000) -- (18.000000,30.000000);
\draw[color=black] (0.000000,30.000000) node[left] {$\ket{a}$};
\draw[color=black] (0.000000,15.000000) -- (18.000000,15.000000);
\draw[color=black] (0.000000,15.000000) node[left] {$\ket{b}$};
\draw[color=black] (0.000000,0.000000) -- (18.000000,0.000000);
\draw[color=black] (0.000000,0.000000) node[left] {$\ket{c}$};
\draw (9.000000,30.000000) -- (9.000000,0.000000);
\begin{scope}
\draw[fill=white] (9.000000, 0.000000) circle(3.000000pt);
\clip (9.000000, 0.000000) circle(3.000000pt);
\draw (6.000000, 0.000000) -- (12.000000, 0.000000);
\draw (9.000000, -3.000000) -- (9.000000, 3.000000);
\end{scope}
\filldraw (9.000000, 30.000000) circle(1.500000pt);
\filldraw (9.000000, 15.000000) circle(1.500000pt);
\draw[color=black] (18.000000,30.000000) node[right] {$\ket{a}$};
\draw[color=black] (18.000000,15.000000) node[right] {$\ket{b}$};
\draw[color=black] (18.000000,0.000000) node[right] {$\ket{c\oplus ab}$};
\end{tikzpicture}
\end{footnotesize}
\caption{CNOT with two control qubits.}
\label{ccnot}
\end{figure}
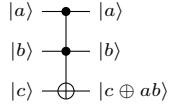
\begin{example}
\label{sprinklerq-ex}
Consider formula $\phi$ from Example 1:
$$\phi=(S\rightarrow W)\wedge (R\rightarrow W)\wedge (S\wedge R\rightarrow).$$ 
Transforming the formula into conjunctive normal form\footnote{A formula is in conjunctive normal form if it is a conjunction of 
clauses where  a clause is a disjunction of literals and a literal is a propositional variable or its negation.} we obtain
$$\phi=(\neg S \vee W)\wedge(\neg R \vee W)\wedge(\neg S\vee \neg R)$$
The quantum circuit for computing the value of formula $\phi$ from Example 1
is shown in Figure \ref{sprinklerq}. 
\end{example}

\begin{figure}[t]
\centering
\begin{footnotesize}
\begin{tikzpicture}[scale=0.800000,x=1pt,y=1pt]
\filldraw[color=white] (0.000000, -7.500000) rectangle (96.000000, 97.500000);
\draw[color=black] (0.000000,90.000000) -- (96.000000,90.000000);
\draw[color=black] (0.000000,90.000000) node[left] {$\ket{S}$};
\draw[color=black] (0.000000,75.000000) -- (96.000000,75.000000);
\draw[color=black] (0.000000,75.000000) node[left] {$\ket{R}$};
\draw[color=black] (0.000000,60.000000) -- (96.000000,60.000000);
\draw[color=black] (0.000000,60.000000) node[left] {$\ket{W}$};
\draw[color=black] (0.000000,45.000000) -- (96.000000,45.000000);
\draw[color=black] (0.000000,45.000000) node[left] {$\ket{1}$};
\draw[color=black] (0.000000,30.000000) -- (96.000000,30.000000);
\draw[color=black] (0.000000,30.000000) node[left] {$\ket{1}$};
\draw[color=black] (0.000000,15.000000) -- (96.000000,15.000000);
\draw[color=black] (0.000000,15.000000) node[left] {$\ket{1}$};
\draw[color=black] (0.000000,0.000000) -- (96.000000,0.000000);
\draw[color=black] (0.000000,0.000000) node[left] {$\ket{0}$};
\begin{scope}
\draw[fill=white] (12.000000, 60.000000) +(-45.000000:8.485281pt and 8.485281pt) -- +(45.000000:8.485281pt and 8.485281pt) -- +(135.000000:8.485281pt and 8.485281pt) -- +(225.000000:8.485281pt and 8.485281pt) -- cycle;
\clip (12.000000, 60.000000) +(-45.000000:8.485281pt and 8.485281pt) -- +(45.000000:8.485281pt and 8.485281pt) -- +(135.000000:8.485281pt and 8.485281pt) -- +(225.000000:8.485281pt and 8.485281pt) -- cycle;
\draw (12.000000, 60.000000) node {$X$};
\end{scope}
\draw (33.000000,90.000000) -- (33.000000,45.000000);
\begin{scope}
\draw[fill=white] (33.000000, 45.000000) circle(3.000000pt);
\clip (33.000000, 45.000000) circle(3.000000pt);
\draw (30.000000, 45.000000) -- (36.000000, 45.000000);
\draw (33.000000, 42.000000) -- (33.000000, 48.000000);
\end{scope}
\filldraw (33.000000, 90.000000) circle(1.500000pt);
\filldraw (33.000000, 75.000000) circle(1.500000pt);
\draw (51.000000,75.000000) -- (51.000000,30.000000);
\begin{scope}
\draw[fill=white] (51.000000, 30.000000) circle(3.000000pt);
\clip (51.000000, 30.000000) circle(3.000000pt);
\draw (48.000000, 30.000000) -- (54.000000, 30.000000);
\draw (51.000000, 27.000000) -- (51.000000, 33.000000);
\end{scope}
\filldraw (51.000000, 75.000000) circle(1.500000pt);
\filldraw (51.000000, 60.000000) circle(1.500000pt);
\draw (69.000000,90.000000) -- (69.000000,15.000000);
\begin{scope}
\draw[fill=white] (69.000000, 15.000000) circle(3.000000pt);
\clip (69.000000, 15.000000) circle(3.000000pt);
\draw (66.000000, 15.000000) -- (72.000000, 15.000000);
\draw (69.000000, 12.000000) -- (69.000000, 18.000000);
\end{scope}
\filldraw (69.000000, 90.000000) circle(1.500000pt);
\filldraw (69.000000, 60.000000) circle(1.500000pt);
\draw (87.000000,45.000000) -- (87.000000,0.000000);
\begin{scope}
\draw[fill=white] (87.000000, 0.000000) circle(3.000000pt);
\clip (87.000000, 0.000000) circle(3.000000pt);
\draw (84.000000, 0.000000) -- (90.000000, 0.000000);
\draw (87.000000, -3.000000) -- (87.000000, 3.000000);
\end{scope}
\filldraw (87.000000, 45.000000) circle(1.500000pt);
\filldraw (87.000000, 30.000000) circle(1.500000pt);
\filldraw (87.000000, 15.000000) circle(1.500000pt);
\draw[color=black] (96.000000,90.000000) node[right] {$\ket{S}$};
\draw[color=black] (96.000000,75.000000) node[right] {$\ket{R}$};
\draw[color=black] (96.000000,60.000000) node[right] {$\ket{\neg W}$};
\draw[color=black] (96.000000,45.000000) node[right] {$\ket{\neg (S \wedge R)}=\ket{\neg S \vee \neg R}$};
\draw[color=black] (96.000000,30.000000) node[right] {$\ket{\neg (R \wedge \neg W)}=\ket{\neg R\vee W}$};
\draw[color=black] (96.000000,15.000000) node[right] {$\ket{\neg (S \wedge \neg W)}=\ket{\neg S\vee W}$};
\draw[color=black] (96.000000,0.000000) node[right] {$\ket{(\neg S \vee \neg R)\wedge (\neg S\vee W)\wedge (\neg R\vee W)}$};
\end{tikzpicture}
\end{footnotesize}
\caption{Quantum circuit for computing $\phi$}
\label{sprinklerq}
\end{figure}
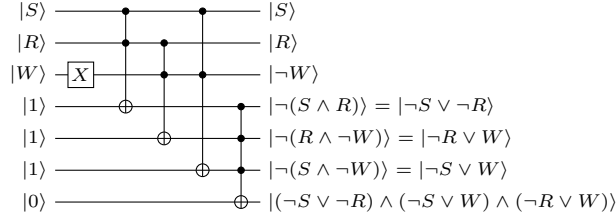
Quantum circuits are read from left to right. Each line or wire corresponds to a qubit and starts
in a computational basis state, usually $\ket{0}$ unless otherwise indicated.
The circuit in Figure \ref{sprinklerq} contains one wire for each Boolean variable of Example 1, three wires that represent the so-called \emph{ancilla qubits}, and one wire for the output qubit. 
Ancilla qubits are used in order to make the circuit reversible, a requirement of quantum circuits. The bottom qubit will contain the truth value of the function $\phi$ at the end of the computation. 

As can be seen, CNOT with more than one control qubit can be used to perform other Boolean functions. For example, the first CNOT with two control qubits computes the NAND of its control qubits because the
target qubits is initially set to $\ket{1}$: in fact $\ket{S,R,1}\to\ket{S,R,(S\wedge R)\oplus 1}=\ket{S,R,\neg(S\wedge R)}$.
On the other hand, the last CNOT with three control qubits and a target qubit set initially set to $\ket{0}$ computes the AND of the control qubits.

%

\section{Implementation}
\label{impl}
In this section we report the implementation of QWMC and QWCS applied to the problems of 
MPE and MAP for the sprinkler case of Example 1 using two languages for quantum computing:
Q\# by Microsoft \citep{qsharp} and Qiskit by IBM \citep{Qiskit-short}.
The code is available at
\url{https://github.com/friguzzi/qwmc}.
\subsection{Q\#}

Q\# is a domain-specific language for implementing quantum algorithms that is syntactically related to Python, C\# and F\#, see \url{https://docs.microsoft.com/en-us/azure/quantum/user-guide/} for a complete user guide,
here we will just present an introductory overview.

The main unit of Q\# code is an \emph{operation} which implements a sequence of quantum and classical operations.
For example, the implementation of the sprinkler circuit of Example \ref{sprinklerq-ex} is shown in Listing \ref{sprinkler-anc}.
It is applied to a query register and a target qubit and uses three ancilla qubits. With respect to Figure \ref{sprinklerq}, it includes an extra qubit in the query register to implement the doubling of $N$.
The operation flips the target qubit  if 
the sprinkler formula takes value 1. \mintinline{qsharp}{X} is the built-in $X$ gate, \mintinline{qsharp}{CCNOT} is the controlled $X$ gate with 
two control qubits, and \mintinline{qsharp}{Controlled X} is a multi-controlled $X$ gate, used in line 13 with four control qubits. The use of 
the \mintinline{qsharp}{within ... apply} construct allows to automate uncomputation - addition of the gates that revert the operations on the query register and ancilla bits in order to properly implement the classical computation as a quantum oracle.
\begin{listing}[h]
\begin{minted}
[
%frame=lines,
%framesep=2mm,
%baselinestretch=1.2,
%bgcolor=LightGray,
fontsize={\scriptsize},
linenos
]
{qsharp}
    operation Sprinkler(q : Qubit[],  target : Qubit) : Unit is Adj+Ctl 
    {
        use a = Qubit[3];
        within {
            X(q[2]);
            X(a[0]);
            X(a[1]);
            X(a[2]);
            CCNOT(q[0], q[1], a[0]);
            CCNOT(q[1], q[2], a[1]);
            CCNOT(q[0], q[2], a[2]);
        } apply {
            Controlled X(a + [q[3]], target);
        }
    }
\end{minted}

\caption{Sprinkler circuit in Q\#.}
\label{sprinkler-anc}
\end{listing}

Operation \verb|ApplyMarkingOracleAsPhaseOracle|, shown in Listing \ref{ApplyMarkingOracleAsPhaseOracle}, takes as input a marking oracle (an oracle that flips the state of the target qubit) and
a register of qubits, and transforms the marking oracle into a phase oracle (an oracle that flips the sign of the register).
The $X$ and $H$ gates applied to the target qubit put it into the state $\ket{-}=(\ket{0}-\ket{1})/\sqrt{2}$, then the marking oracle is applied. Since the target is in the $\ket{-}$ state,
flipping the target if the register satisfies the oracle condition will bring the target to $-\ket{-}=-(\ket{0}-\ket{1})\sqrt{2}$, applying a $-1$ factor to the whole register, effectively applying the phase oracle.
\begin{listing}[h]
\begin{minted}
[
%frame=lines,
%framesep=2mm,
%baselinestretch=1.2,
%bgcolor=LightGray,
fontsize={\scriptsize},
linenos
]
{qsharp}
operation ApplyMarkingOracleAsPhaseOracle(markingOracle : ((Qubit[], Qubit) => 
    Unit is Adj+Ctl), q : Qubit[]) :  Unit is Adj+Ctl 
{   
    use target = Qubit();
    within {
        X(target);
        H(target);
    } apply {
        markingOracle(q, target);
    }
}
\end{minted}
\caption{From a marking oracle to a phase oracle in Q\#.}
\label{ApplyMarkingOracleAsPhaseOracle}
\end{listing}

\begin{listing}[h]
\begin{minted}
[
%frame=lines,
%framesep=2mm,
%baselinestretch=1.2,
%bgcolor=LightGray,
fontsize={\scriptsize},
linenos
]
{qsharp}
operation Rot(q: Qubit[], weights : Double[]): Unit is Ctl+Adj 
{
    for i in 0 .. Length(weights) - 1 {
        let theta = 2.0 * ArcSin(Sqrt(weights[i]));
        Ry(theta, q[i]);
    }
    H(q[3]);
}
\end{minted}
\caption{$Rot$ operation in Q\#.}
\label{rot_op_code}
\end{listing}
Operation \verb|Rot|, shown in Listing \ref{rot_op_code}, implements the $Rot$ gate of Figure 7, which applies rotation $R_y(\theta_i)$ to the $i$th qubit, where  
$\theta_i=2\arcsin\sqrt{w_i}$, and $w_i$ the weight of the $i$th variable. The $H$ gate is applied to the $n+1$th bit.

Operation \verb|GroverIteration|, shown in Listing \ref{GroverIteration}, takes as input a register and a marking oracle and implements
the Weighted Grover operator $WG$ of Figure 8.
First it uses \verb|ApplyMarkingOracleAsPhaseOracle| to apply the phase oracle to the register.
Then it applies the adjoint (Hermitian conjugate) of gate  $Rot$ to all register qubits and multiplies by $-1$ all basis states except $\ket{00\ldots0}$.
This is performed by multiplying by $-1$ only configuration $\ket{00\ldots0}$ (lines 7-13) and then rotating the register by $\pi$ radians (line 14).

\begin{listing}[h]
\begin{minted}
[
%frame=lines,
%framesep=2mm,
%baselinestretch=1.2,
%bgcolor=LightGray,
fontsize={\scriptsize},
linenos
]
{qsharp}
operation GroverIteration(q : Qubit[], oracle : ((Qubit[],Qubit) => 
    Unit is Adj+Ctl), weights : Double[]) : Unit is Ctl+Adj
{
    ApplyMarkingOracleAsPhaseOracle(oracle, q);
    within {
        Adjoint Rot(q, weights);
        ApplyToEachCA(X, q);    // Brings |0..0> to |1..1>
    } apply {
        use a = Qubit();
        Controlled X(q, a);   // Bit flips the a to |1> if register is |1...1> 
        Z(a);                 // The phase of a (and therefore the whole 
                              // register phase)
                              // becomes -1 if above condition is satisfied
        Controlled X(q, a);   // Puts a back in |0>
        Ry(2.0 * PI(), q[0]);
    }
}   
\end{minted}

\caption{Grover iteration  in Q\#.}
\label{GroverIteration}
\end{listing}

Finally, operation \verb|QWMC|, shown in Listing \ref{QWMC}, implements the whole QWMC algorithm.
First it defines a constant \verb|t| -- the number of bits to use to represent the weighted model count.
It uses a four-qubit register for storing the state and a \verb|t|-qubit register for storing the phase.
It calls the \verb|OracleToDiscrete| library function to convert the Grover operator to a library type \verb|DiscreteOracle|, suitable for exponentiation.
Then it initializes the qubits of the register according to the weights using \verb|Rot|.
Next, the code calls a library operation to perform quantum phase estimation (line 12).
Then we measure the phase register to get an integer divided it by $2^t$ to obtain a number between 0 and 1.
Finally, lines 19 and 20 compute the weighted model count from the phase.
\begin{listing}[h]
\begin{minted}
[
%frame=lines,
%framesep=2mm,
%baselinestretch=1.2,
%bgcolor=LightGray,
fontsize={\scriptsize},
linenos
]
{qsharp}
operation QWMC() : Double 
{
    let t = 5;
    let weights = [0.55, 0.3, 0.7];
    let oracle = OracleToDiscrete(GroverIteration(_, Sprinkler(_, _), 
      weights));

    use (q, p) = (Qubit[4], Qubit[t]);
    Rot(q, weights);
    // Allocate qubits to hold the eigenstate of U and the phase in a big 
    // endian register 
    let pBE = BigEndian(p);
    // Call library
    QuantumPhaseEstimation(oracle, q, pBE);
    // Read out the phase
    let phase = IntAsDouble(MeasureInteger(BigEndianAsLittleEndian(pBE))) 
      / IntAsDouble(1 <<< (t));

    ResetAll(q + p);
    // The phase returned by QuantumPhaseEstimation is the value phi such 
    // that e^{2pi phi} is an eigenvalue
    let angle = 2.0 * PI() * phase; 
    let wmc = 2.0 * PowD(Sin(angle / 2.0), 2.0);
    // The 2.0 factor is added because there is an extra bit with weight 0.5
    // that is introduced to make the weighted count < 0.5

    return wmc;
}
\end{minted}
\caption{Quantum weighted model counting operator in Q\#.}
\label{QWMC}
\end{listing}

Operation \verb|QMPE|, shown in Listing \ref{QMPE}, implements the MPE algorithm using QWCS.
It uses a 4-qubit register for the state
and initializes it according to the weights (line 6).
The Grover operator is applied $\CI\left(\frac{\arccos\frac{\sum_{x;\phi(x)=1}W_x}{2}}{2\arcsin\frac{\sum_{x;\phi(x)=1}W_x}{2}}\right)=\CI(0.763)=1$ time (line 7)  and finally the register is measured (lines 8, 9).
\begin{listing}[h]
\begin{minted}
[
%frame=lines,
%framesep=2mm,
%baselinestretch=1.2,
%bgcolor=LightGray,
fontsize={\scriptsize},
linenos
]
{qsharp}
operation QMPE() : Result[] 
{
    let weights = [0.55, 0.3, 0.7];
    use reg = Qubit[4];
        
    Rot(reg, weights);
    GroverIteration(reg, Sprinkler, weights);
    let query= reg[0 .. 2];
    let state = MultiM(query);

    ResetAll(reg);
    return state;
}
\end{minted}

\caption{Quantum MPE operator in Q\#.}
\label{QMPE}
\end{listing}

QMAP (Listing \ref{QMAP}) differs from QMPE, because we measure only a subarray of the register, rather than the whole register (lines 8-9).
\begin{listing}[h]
\begin{minted}
[
%frame=lines,
%framesep=2mm,
%baselinestretch=1.2,
%bgcolor=LightGray,
fontsize={\scriptsize},
linenos
]
{qsharp}
operation QMAP() : Result[] 
{
    let weights = [0.55, 0.3, 0.7];
        
    use reg = Qubit[4];
    Rot(reg, weights);
    GroverIteration(reg, Sprinkler, weights);
    let query = Subarray([0, 2],reg);
    let state = MultiM(query);

    ResetAll(reg);
    return state;
}
\end{minted}

\caption{Quantum MAP operator. in Q\#}
\label{QMAP}
\end{listing}

\subsection{Qiskit}
Qiskit is a Python library for implementing quantum algorithms.
The code for performing QWMC presented in this section is based on the code
for quantum counting available in \citep{Qiskit-Textbook-etal}.

To use it, a number of import declaration are needed, see Listing \ref{qkit_import}.
\begin{listing}[h]
\begin{minted}
[
%frame=lines,
%framesep=2mm,
%baselinestretch=1.2,
%bgcolor=LightGray,
fontsize={\scriptsize},
linenos
]
{python}
from qiskit import Aer
from qiskit import QuantumCircuit, ClassicalRegister, QuantumRegister, execute
from qiskit.tools.visualization import plot_histogram
from math import pi, sqrt, sin, asin
\end{minted}
\caption{Importing components in Qiskit.}
\label{qkit_import}
\end{listing}

Listing \ref{qkit_sprinkler} shows the implementation of the sprinkler circuit by a function
that takes as input a quantum circuit \verb|qc| and two quantum registers, \verb|q|, for the
query qubits and \verb|a|, for the ancilla qubits. Methods \verb|x|, \verb|ccx| and \verb|mct|
of \verb|qc| implement respectively the $X$ gate, the $X$ gate with two control qubits and the
$X$ gate with a list of control qubits. 
\begin{listing}[h]
\begin{minted}
[
%frame=lines,
%framesep=2mm,
%baselinestretch=1.2,
%bgcolor=LightGray,
fontsize={\scriptsize},
linenos
]
{python}
def sprinkler(qc,q,a):
    qc.x(q[2])
    qc.x(a[0])
    qc.x(a[1])
    qc.x(a[2])
    qc.ccx(q[0],q[1],a[0])
    qc.ccx(q[1],q[2],a[1])
    qc.ccx(q[0],q[2],a[2])
    qc.mct([a[0],a[1],a[2],q[3]],a[3])
    qc.ccx(q[0],q[2],a[2])
    qc.ccx(q[1],q[2],a[1])
    qc.ccx(q[0],q[1],a[0])
    qc.x(a[2])
    qc.x(a[1])
    qc.x(a[0])
    qc.x(q[2])
\end{minted}
\caption{Sprinkler circuit in Qiskit.}
\label{qkit_sprinkler}
\end{listing}

Function \verb|rotations| shown in Listing \ref{qkit_rot} uses a different approach to implement
the $Rot$ operator: it builds internally the query register and the quantum circuit and returns it. Then the method
\verb|to_gate| is used to convert the circuit into a new gate, \verb|rot|, that implements $Rot$.
Method \verb|inverse| is used to obtain the inverse operation (the adjoint) \verb|invrot| of $Rot$. 
\begin{listing}[h]
\begin{minted}
[
%frame=lines,
%framesep=2mm,
%baselinestretch=1.2,
%bgcolor=LightGray,
fontsize={\scriptsize},
linenos
]
{python}
def rotations():
    q=QuantumRegister(4)
    qc=QuantumCircuit(q)
    weights=[0.55,0.3,0.7]
    for i in range(len(weights)):
       theta=2.0*asin(sqrt(weights[i]))
       qc.ry(theta,q[i])
    qc.h(q[3])
    return qc

rot=rotations().to_gate()
invrot=rot.inverse()
\end{minted}
\caption{$Rot$ circuit in Qiskit.}
\label{qkit_rot}
\end{listing}

The weighted Grover operator $WG$ is implemented by function \verb|grover_circ| of Listing \ref{qkit_wg} that applies the same gates as operator \verb|GroverIteration| of Listing \ref{GroverIteration}.
The circuit that is returned by \verb|grover_circ| is then turned into a gate and converted into a controlled gate by method \verb|control|.
\begin{listing}[h]
\begin{minted}
[
%frame=lines,
%framesep=2mm,
%baselinestretch=1.2,
%bgcolor=LightGray,
fontsize={\scriptsize},
linenos
]
{python}
def grover_circ():
    q=QuantumRegister(4)
    a=QuantumRegister(5)
    qc=QuantumCircuit(q,a)
    qc.x(a[3])
    qc.h(a[3])
    sprinkler(qc,q,a)
    qc.h(a[3])
    qc.x(a[3])
    qc.append(invrot,range(4))
    for i in range(q.size):
        qc.x(q[i])
    qc.mct([q[0],q[1],q[2],q[3]],a[4])
    qc.z(a[4])
    qc.mct([q[0],q[1],q[2],q[3]],a[4])
    for i in range(q.size):
        qc.x(q[i])
    qc.ry(2*pi,q[0]) 
    qc.append(rot,range(4))

    return qc
grover = grover_circ().to_gate()
cgrover = grover.control()
\end{minted}
\caption{Weighted Grover circuit in Qiskit.}
\label{qkit_wg}
\end{listing}    

Function \verb|qft| of Listing \ref{qkit_qft} builds an $n$-qubit QFT circuit  and is taken literally from \citep{Qiskit-Textbook-etal}.
The circuit is then transformed into a gate and inverted.

\begin{listing}[h]
\begin{minted}
[
%frame=lines,
%framesep=2mm,
%baselinestretch=1.2,
%bgcolor=LightGray,
fontsize={\scriptsize},
linenos
]
{python}
t = 5   # no. of counting qubits
n = 4   # no. of searching qubits

def qft(n):
    """Creates an n-qubit QFT circuit"""
    circuit = QuantumCircuit(n)
    def swap_registers(circuit, n):
        for qubit in range(n//2):
            circuit.swap(qubit, n-qubit-1)
        return circuit
    def qft_rotations(circuit, n):
        """Performs qft on the first n qubits in circuit (without swaps)"""
        if n == 0:
            return circuit
        n -= 1
        circuit.h(n)
        for qubit in range(n):
            circuit.cu1(pi/2**(n-qubit), qubit, n)
        qft_rotations(circuit, n)
    
    qft_rotations(circuit, n)
    swap_registers(circuit, n)
    return circuit

invqft = qft(t).to_gate().inverse()
\end{minted}
\caption{Quantum Fourier transform in Qiskit.}
\label{qkit_qft}
\end{listing}  

Listing \ref{qkit_qwmc} builds the overall circuit with \verb|t| counting qubits: it applies $H$ to all counting qubits,  it applies $Rot$ to the searching qubits,
it applies the series of controlled $WG$ operations and finally the inverse QFT.
\begin{listing}[h]
\begin{minted}
[
%frame=lines,
%framesep=2mm,
%baselinestretch=1.2,
%bgcolor=LightGray,
fontsize={\scriptsize},
linenos
]
{python}
q=QuantumRegister(n)
a=QuantumRegister(5)
s=QuantumRegister(t)
c=ClassicalRegister(t)

qc = QuantumCircuit(s,q,a,c) # Circuit with n+t qubits and t classical bits

for qubit in range(t):
    qc.h(qubit)

qc.append(rot, range(t,n+t))

# Begin controlled Grover iterations
iterations = 1
for qubit in range(t):
    for i in range(iterations):
        qc.append(cgrover, [qubit] + [*range(t, n+t+5)])
    iterations *= 2
    
# Do inverse QFT on counting qubits
qc.append(invqft, range(t))

# Measure counting qubits
qc.measure(range(t), range(t))
\end{minted}
\caption{Quantum weighted model counting operator in Qiskit.}
\label{qkit_qwmc}
\end{listing} 

To simulate the quantum circuit we use the code in Listing \ref{qkit_sim} that executes the simulation 1000 times and collects
the results in \verb|result_sim|.

\begin{listing}[h]
\begin{minted}
[
%frame=lines,
%framesep=2mm,
%baselinestretch=1.2,
%bgcolor=LightGray,
fontsize={\scriptsize},
linenos
]
{python}
backend = Aer.get_backend('qasm_simulator')
start = time.time()

job_sim = execute(qc, backend, shots=1000)
end = time.time()
print(end - start)

result_sim = job_sim.result()
\end{minted}
\caption{Simulation code in Qiskit.}
\label{qkit_sim}
\end{listing}     

The code for performing MPE with QWCS is shown in Listing \ref{qkit_qmpe}. The code for performing MAP differs because line 8 is replaced by \mintinline{python}|qc.measure([q[0],q[2]],c)|.
\begin{listing}[h]
\begin{minted}
[
%frame=lines,
%framesep=2mm,
%baselinestretch=1.2,
%bgcolor=LightGray,
fontsize={\scriptsize},
linenos
]
{python}
q=QuantumRegister(4)
a=QuantumRegister(5)
c=ClassicalRegister(3)

qc=QuantumCircuit(q,a,c)
qc.append(rot, range(4))
qc.append(grover,range(9))
qc.measure([q[0],q[1],q[2]],c)
\end{minted}
\caption{Quantum MPE code in Qiskit.}
\label{qkit_qmpe}
\end{listing}     

\section{Example of Execution}
\label{experiments}

In this section we present the results of the execution of QWMC, QWCS applied to MPE and MAP on the sprinkler case of Example \ref{sprinklerb}.

\begin{figure}
	\centering
	\includegraphics[width=.8\textwidth]{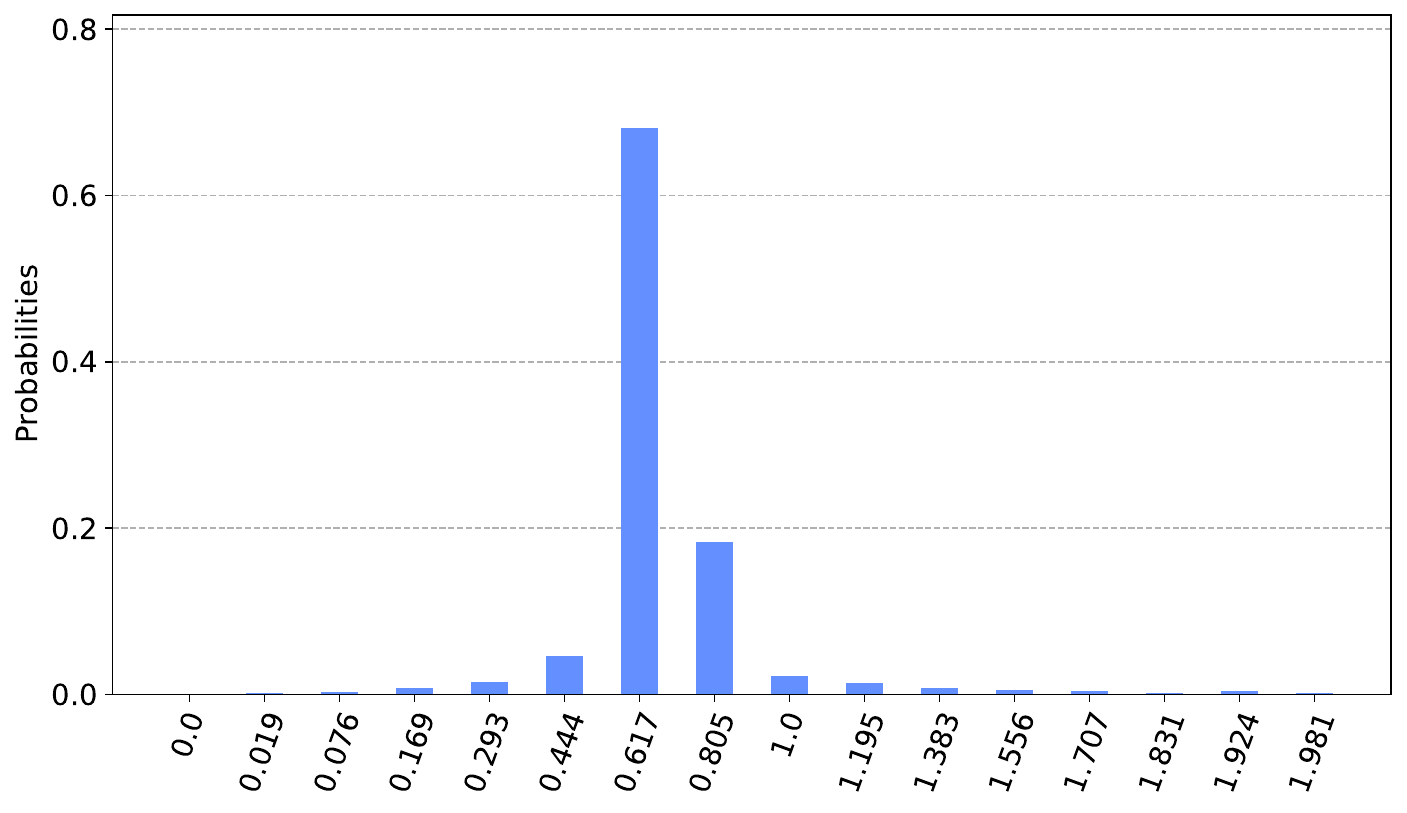}
	\caption{Distribution of results for QWMC.}
	\label{fig:qwmc}
\end{figure}

Here we report the results obtained with Q\#. Those for Qiskit are similar.
The circuit for QWMC was run 1000 times by simulation using ideal, noise free quantum circuits. The distribution of results is shown in Figure \ref{fig:qwmc}:  the peak at 0.617  shows that
most simulations came very close to the true value of 0.679.

\begin{figure}
	\centering
	\includegraphics[width=.8\textwidth]{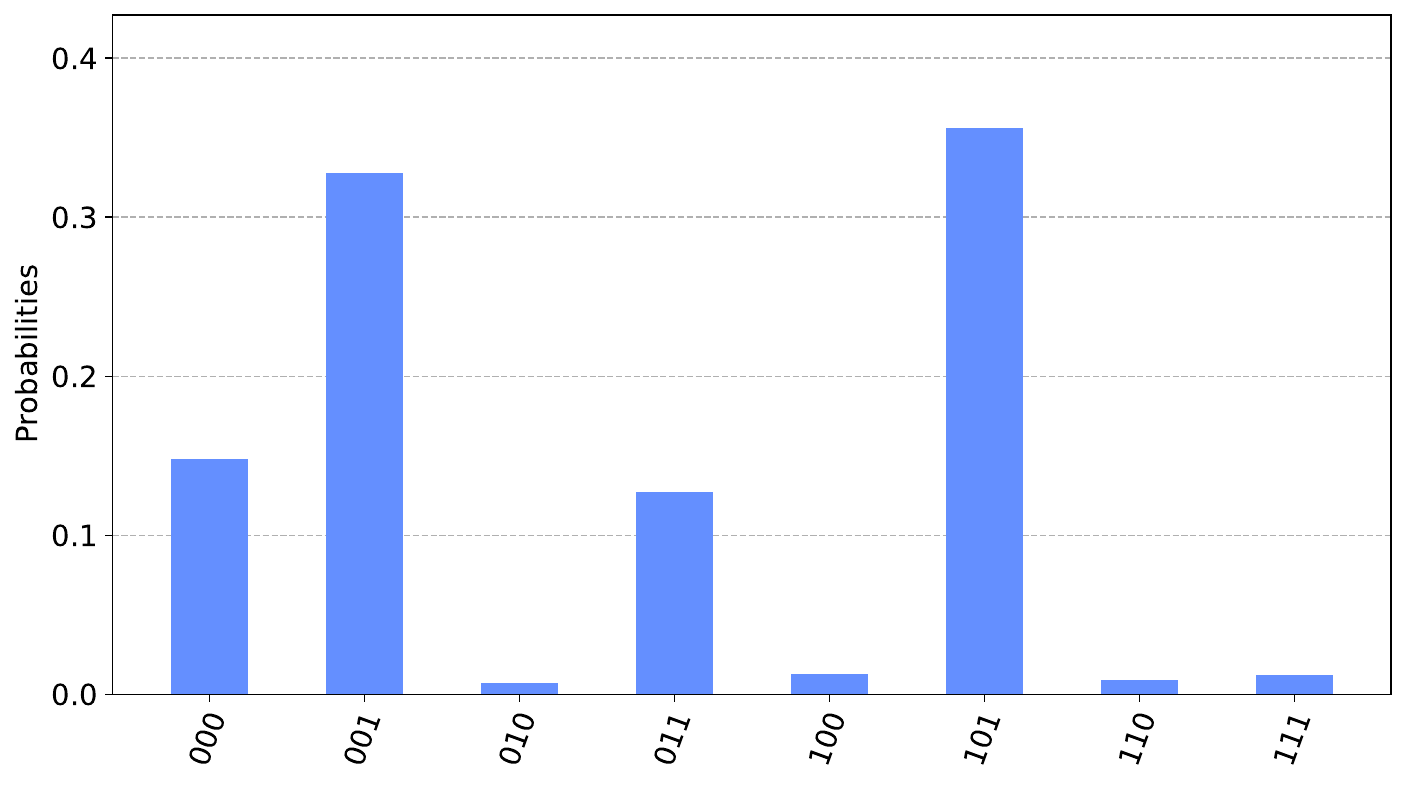}
	\caption{Distribution of results for MPE.}
	\label{fig:qmpe}
\end{figure}

\begin{figure}
	\centering
	\includegraphics[width=.8\textwidth]{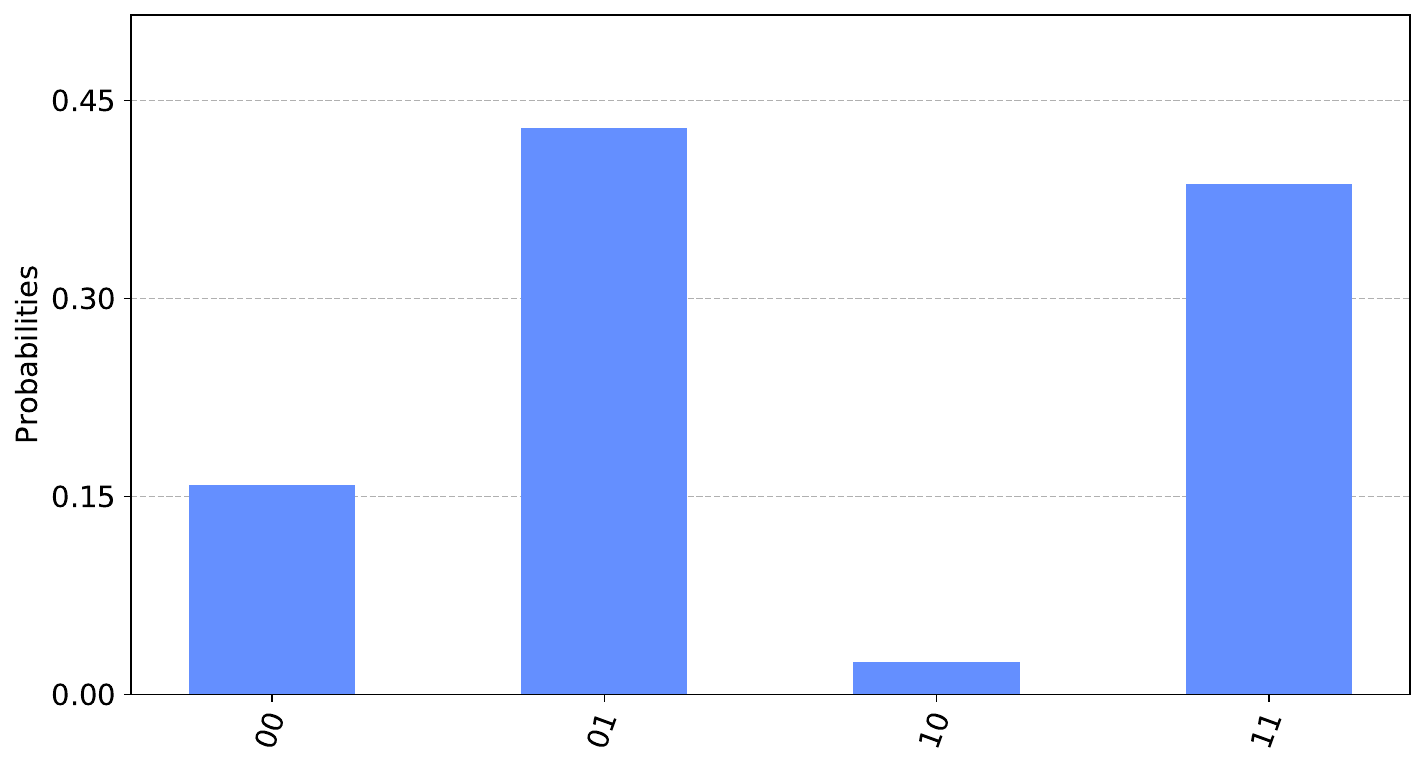}
	\caption{Distribution of results for MAP.}
	\label{fig:qmap}
\end{figure}

For MPE and MAP, we use the algorithm that repeatedly runs the circuits for QWCS and picks the most frequent result as the answer.
QWCS  was run by simulation 1000 times for MPE and MAP. Figures   \ref{fig:qmpe} and \ref{fig:qmap} show the distribution of results respectively for MPE and MAP: the highest peaks correspond to the
solutions of the problem, 101 for MPE and 01 for MAP (see Section \ref{wmc}).

\bibliography{bibtexrepository/journals_short,bibtexrepository/booktitles_long,bibtexrepository/series_springer,bibtexrepository/series_long,bibtexrepository/publishers_long,bibtexrepository/bibl}

\begin{thebibliography}{54}
\providecommand{\natexlab}[1]{#1}
\providecommand{\url}[1]{{#1}}
\providecommand{\urlprefix}{URL }
\providecommand{\doi}[1]{\url{https://doi.org/#1}}
\providecommand{\eprint}[2][]{\url{#2}}
 \bibcommenthead

\bibitem[{Abraham and et~al.(2019)}]{Qiskit-short}
Abraham H, et~al. (2019) Qiskit: An open-source framework for quantum computing. \doi{10.5281/zenodo.2562110}, 10.5281/zenodo.2562110

\bibitem[{Aharonov(1999)}]{aharonov1999quantum}
Aharonov D (1999) Quantum computation. In: Annual Reviews of Computational Physics VI. World Scientific, p 259--346

\bibitem[{Asfaw and et~al.(2020)}]{Qiskit-Textbook-etal}
Asfaw A, et~al. (2020) Learn quantum computation using qiskit. \url{http://community.qiskit.org/textbook}

\bibitem[{Bacchus et~al(2009)Bacchus, Dalmao, and Pitassi}]{bacchus2009solving}
Bacchus F, Dalmao S, Pitassi T (2009) Solving\# sat and bayesian inference with backtracking search. J Artif Intell Res 34:391--442

\bibitem[{Bian et~al(2020)Bian, Chudak, Macready, Roy, Sebastiani, and Varotti}]{BIAN2020104609}
Bian Z, Chudak F, Macready W, et~al (2020) Solving {SAT} (and {MaxSAT}) with a quantum annealer: Foundations, encodings, and preliminary results. Information and Computation 275:104609. \doi{https://doi.org/10.1016/j.ic.2020.104609}, \urlprefix\url{https://www.sciencedirect.com/science/article/pii/S0890540120300973}

\bibitem[{Bodlaender et~al(1993)}]{bodlaender1993tourist}
Bodlaender HL, et~al (1993) A tourist guide through treewidth. Acta Cybernetica 11(1-2):1--21

\bibitem[{Boyer et~al(1998)Boyer, Brassard, H{\o}yer, and Tapp}]{boyer1998tight}
Boyer M, Brassard G, H{\o}yer P, et~al (1998) Tight bounds on quantum searching. Fortschritte der Physik: Progress of Physics 46(4-5):493--505

\bibitem[{Brassard et~al(1998)Brassard, H{\o}yer, and Tapp}]{DBLP:conf/icalp/BrassardHT98}
Brassard G, H{\o}yer P, Tapp A (1998) Quantum counting. In: Larsen KG, Skyum S, Winskel G (eds) 25th International Colloquium on Automata, Languages and Programming (ICALP 1998), Lecture Notes in Computer Science, vol 1443. Springer, pp 820--831

\bibitem[{Chakraborty et~al(2014)Chakraborty, Fremont, Meel, Seshia, and Vardi}]{Chekraborty-Fremont}
Chakraborty S, Fremont DJ, Meel KS, et~al (2014) Distribution-aware sampling and weighted model counting for sat. In: Proceedings of the Twenty-Eighth AAAI Conference on Artificial Intelligence. AAAI Press, AAAI'14, pp 1722--1730

\bibitem[{Chakraborty et~al(2015)Chakraborty, Fried, Meel, and Vardi}]{DBLP:conf/ijcai/ChakrabortyFMV15}
Chakraborty S, Fried D, Meel KS, et~al (2015) From weighted to unweighted model counting. In: Yang Q, Wooldridge MJ (eds) 24th International Joint Conference on Artificial Intelligence (IJCAI 2015). AAAI Press, pp 689--695, \urlprefix\url{http://ijcai.org/Abstract/15/103}

\bibitem[{Chakraborty et~al(2016)Chakraborty, Meel, and Vardi}]{DBLP:conf/ijcai/ChakrabortyMV16}
Chakraborty S, Meel KS, Vardi MY (2016) Algorithmic improvements in approximate counting for probabilistic inference: From linear to logarithmic {SAT} calls. In: Kambhampati S (ed) 25th International Joint Conference on Artificial Intelligence (IJCAI 2016). AAAI Press/IJCAI, pp 3569--3576

\bibitem[{Chakraborty et~al(2021)Chakraborty, Meel, and Vardi}]{chakraborty2021approximate}
Chakraborty S, Meel KS, Vardi MY (2021) Approximate model counting. In: Handbook of Satisfiability. IOS Press, p 1015--1045

\bibitem[{Chavira and Darwiche(2008)}]{DBLP:journals/ai/ChaviraD08}
Chavira M, Darwiche A (2008) On probabilistic inference by weighted model counting. Artif Intell 172(6-7):772--799

\bibitem[{Cleve et~al(1998)Cleve, Ekert, Macchiavello, and Mosca}]{cleve1998quantum}
Cleve R, Ekert A, Macchiavello C, et~al (1998) Quantum algorithms revisited. Proceedings of the Royal Society of London Series A: Mathematical, Physical and Engineering Sciences 454(1969):339--354

\bibitem[{Coppersmith(2002)}]{coppersmith2002approximate}
Coppersmith D (2002) An approximate {Fourier} transform useful in quantum factoring. ArXiv preprint quant-ph/0201067

\bibitem[{Darwiche(2001)}]{darwiche2001recursive}
Darwiche A (2001) Recursive conditioning. Artif Intell 126(1-2):5--41

\bibitem[{Darwiche and Marquis(2002)}]{DBLP:journals/jair/DarwicheM02}
Darwiche A, Marquis P (2002) A knowledge compilation map. J Artif Intell Res 17:229--264. \doi{10.1613/jair.989}

\bibitem[{Dechter(1999)}]{dechter1999bucket}
Dechter R (1999) Bucket elimination: A unifying framework for reasoning. Artif Intell 113(1-2):41--85

\bibitem[{Eiter and Kiesel(2021)}]{eiter2021complexity}
Eiter T, Kiesel R (2021) On the complexity of sum-of-products problems over semirings. In: 35th {AAAI} Conference on Artificial Intelligence, {AAAI} 2021. {AAAI} Press, pp 6304--6311

\bibitem[{Feller(1968)}]{feller1968introduction}
Feller W (1968) An Introduction to Probability Theory and Its Applications: Volume I. Wiley series in probability and mathematical statistics, John Wiley \& sons.

\bibitem[{Fichte et~al(2018)Fichte, Hecher, Woltran, and Zisser}]{DBLP:conf/esa/FichteHWZ18}
Fichte JK, Hecher M, Woltran S, et~al (2018) Weighted model counting on the {GPU} by exploiting small treewidth. In: Azar Y, Bast H, Herman G (eds) 26th Annual European Symposium on Algorithms (ESA 2018). Schloss Dagstuhl - Leibniz-Zentrum fuer Informatik, pp 28:1--28:16, \doi{10.4230/LIPIcs.ESA.2018.28}

\bibitem[{Fichte et~al(2021)Fichte, Hecher, and Hamiti}]{fichte2021model}
Fichte JK, Hecher M, Hamiti F (2021) The model counting competition 2020. Journal of Experimental Algorithmics (JEA) 26:1--26

\bibitem[{Friesen and Domingos(2016)}]{DBLP:conf/icml/FriesenD16}
Friesen AL, Domingos PM (2016) The sum-product theorem: {A} foundation for learning tractable models. In: Balcan M, Weinberger KQ (eds) 33nd International Conference on Machine Learning, {ICML} 2016, {JMLR} Workshop and Conference Proceedings, vol~48. JMLR.org, pp 1909--1918

\bibitem[{Ganian et~al(2022)Ganian, Kim, Slivovsky, and Szeider}]{ganian2022sum}
Ganian R, Kim EJ, Slivovsky F, et~al (2022) Sum-of-products with default values: Algorithms and complexity results. J Artif Intell Res 73:535--552

\bibitem[{Gomes et~al(2009)Gomes, Sabharwal, and Selman}]{DBLP:series/faia/GomesSS09}
Gomes CP, Sabharwal A, Selman B (2009) Model counting. In: Biere A, Heule M, van Maaren H, et~al (eds) Handbook of Satisfiability, vol 185. {IOS} Press, p 633--654

\bibitem[{Griffiths and Niu(1996)}]{griffiths1996semiclassical}
Griffiths RB, Niu CS (1996) Semiclassical fourier transform for quantum computation. Physical Review Letters 76(17):3228

\bibitem[{Grover(1996{\natexlab{a}})}]{Grover:1996:FQM:237814.237866}
Grover LK (1996{\natexlab{a}}) A fast quantum mechanical algorithm for database search. In: 28th Annual ACM Symposium on Theory of Computing (STOC 1996). ACM Press, New York, NY, USA, pp 212--219

\bibitem[{Grover(1996{\natexlab{b}})}]{grover1996fast}
Grover LK (1996{\natexlab{b}}) A fast quantum mechanical algorithm for database search. ArXiv preprint quant-ph/9605043

\bibitem[{Grover(1997)}]{grover1997quantum}
Grover LK (1997) Quantum mechanics helps in searching for a needle in a haystack. Physical review letters 79(2):325

\bibitem[{Gupta and Nagel(1967)}]{multim-rank}
Gupta SS, Nagel K (1967) On selection and ranking procedures and order statistics from the multinomial distribution. Sankhyā: The Indian Journal of Statistics, Series B (1960-2002) 29(1/2):1--34. \urlprefix\url{http://www.jstor.org/stable/25051587}

\bibitem[{Hirvensalo(2013)}]{hirvensalo2013quantum}
Hirvensalo M (2013) Quantum Computing. Natural Computing Series, Springer

\bibitem[{Huang et~al(2006)Huang, Chavira, and Darwiche}]{huang2006solving}
Huang J, Chavira M, Darwiche A (2006) Solving {MAP} exactly by searching on compiled arithmetic circuits. In: Proceedings of the 21st National Conference on Artificial Intelligence - Volume 2. AAAI Press, AAAI'06, pp 1143--1148

\bibitem[{Jerrum and Sinclair(1996)}]{Jerrum1996TheMC}
Jerrum M, Sinclair A (1996) The markov chain monte carlo method: an approach to approximate counting and integration. In: Hochbaum D (ed) Approximation algorithms for NP-hard problems. PWS Publishing Company, p 482– 520

\bibitem[{Kwisthout(2015)}]{DBLP:journals/jair/Kwisthout15}
Kwisthout J (2015) Tree-width and the computational complexity of {MAP} approximations in bayesian networks. J Artif Intell Res 53:699--720. \doi{10.1613/jair.4794}, \urlprefix\url{https://doi.org/10.1613/jair.4794}

\bibitem[{Lagniez and Marquis(2017)}]{lagniez2017improved}
Lagniez JM, Marquis P (2017) An improved decision-dnnf compiler. In: 26th International Joint Conference on Artificial Intelligence (IJCAI 2017), pp 667--673

\bibitem[{Lauritzen and Spiegelhalter(1988)}]{lauritzen1988local}
Lauritzen SL, Spiegelhalter DJ (1988) Local computations with probabilities on graphical structures and their application to expert systems. Journal of the Royal Statistical Society: Series B (Methodological) 50(2):157--194

\bibitem[{Madras and Piccioni(1999)}]{Madras-Piccioni}
Madras N, Piccioni M (1999) Importance sampling for families of distributions. The Annals of Applied Probability 9(4):1202--1225. \urlprefix\url{http://www.jstor.org/stable/2667147}

\bibitem[{Meel et~al(2016)Meel, Vardi, Chakraborty, Fremont, Seshia, Fried, Ivrii, and Malik}]{DBLP:conf/aaai/MeelVCFSFIM16}
Meel K, Vardi M, Chakraborty S, et~al (2016) Constrained sampling and counting: Universal hashing meets {SAT} solving. In: Darwiche A (ed) Beyond NP, Papers from the 2016 {AAAI} Workshop, Phoenix, Arizona, USA, February 12, 2016, {AAAI} Technical Report, vol {WS-16-05}. {AAAI} Press, \urlprefix\url{http://www.aaai.org/ocs/index.php/WS/AAAIW16/paper/view/12618}

\bibitem[{Mosca(1999)}]{mosca1999quantum}
Mosca M (1999) Quantum computer algorithms. PhD thesis, University of Oxford. 1999.

\bibitem[{Naveh et~al(2006)Naveh, Rimon, Jaeger, Katz, Vinov, Marcus, and Shurek}]{Naveh}
Naveh Y, Rimon M, Jaeger I, et~al (2006) Constraint-based random stimuli generation for hardware verification. In: Proceedings of the 18th Conference on Innovative Applications of Artificial Intelligence - Volume 2. AAAI Press, IAAI'06, p 1720–1727

\bibitem[{Nielsen and Chuang(2010)}]{nielsen2010quantum}
Nielsen M, Chuang I (2010) Quantum Computation and Quantum Information: 10th Anniversary Edition. Cambridge University Press

\bibitem[{Park and Darwiche(2004)}]{DBLP:journals/jair/ParkD04}
Park JD, Darwiche A (2004) Complexity results and approximation strategies for {MAP} explanations. J Artif Intell Res 21:101--133. \doi{10.1613/jair.1236}

\bibitem[{Pearl(1988)}]{pearl88}
Pearl J (1988) Probabilistic Reasoning in Intelligent Systems: Networks of Plausible Inference. Morgan Kaufmann

\bibitem[{Riguzzi(2013)}]{Rig13-FI-IJ}
Riguzzi F (2013) {MCINTYRE}: A {Monte Carlo} system for probabilistic logic programming. Fund Inform 124(4):521--541. \doi{10.3233/FI-2013-847}

\bibitem[{Riguzzi(2020)}]{Rig20-ECAI-IC}
Riguzzi F (2020) Quantum weighted model counting. In: {De Giacomo} G, Catala A, Dilkina B, et~al (eds) 24th European Conference on Artificial Intelligence (ECAI 2020). IOS Press, Amsterdam, Berlin, Washington DC, pp 2640--2647, \doi{10.3233/FAIA200401}

\bibitem[{Riguzzi(2023)}]{Rig23-BK}
Riguzzi F (2023) Foundations of Probabilistic Logic Programming Languages, Semantics, Inference and Learning, Second Edition. River Publishers

\bibitem[{Sang et~al(2005)Sang, Beame, and Kautz}]{DBLP:conf/aaai/SangBK05}
Sang T, Beame P, Kautz HA (2005) Performing bayesian inference by weighted model counting. In: 20th National Conference on Artificial Intelligence (AAAI 2005). AAAI Press, Palo Alto, California USA, pp 475--482

\bibitem[{Sang et~al(2007)Sang, Beame, and Kautz}]{sang2007dynamic}
Sang T, Beame P, Kautz HA (2007) A dynamic approach for {MPE} and weighted {MAX-SAT}. In: 20th International Joint Conference on Artificial Intelligence (IJCAI 2007). AAAI Press/IJCAI, pp 173--179

\bibitem[{Shenoy and Shafer(1990)}]{DBLP:conf/uai/ShenoyS88}
Shenoy PP, Shafer G (1990) Axioms for probability and belief-function proagation. In: Shachter RD, Levitt TS, Kanal LN, et~al (eds) 4th Conference Conference on Uncertainty in Artificial Intelligence (UAI 1988). North-Holland, pp 169--198

\bibitem[{Sundar et~al(2019)Sundar, Paredes, Damanik, Duenas-Osorio, and Hazzard}]{sundar2019quantum}
Sundar B, Paredes R, Damanik DT, et~al (2019) A quantum algorithm to count weighted ground states of classical spin hamiltonians. ArXiv preprint arXiv:1908.01745

\bibitem[{Svore et~al(2018)Svore, Geller, Troyer, Azariah, Granade, Heim, Kliuchnikov, Mykhailova, Paz, and Roetteler}]{qsharp}
Svore K, Geller A, Troyer M, et~al (2018) Q\#: Enabling scalable quantum computing and development with a high-level dsl. In: Real World Domain Specific Languages Workshop (RWDSL 2018). ACM, New York, NY, USA, pp 7:1--7:10, \doi{10.1145/3183895.3183901}

\bibitem[{Wei et~al(2004)Wei, Erenrich, and Selman}]{DBLP:conf/aaai/WeiES04}
Wei W, Erenrich J, Selman B (2004) Towards efficient sampling: Exploiting random walk strategies. In: McGuinness DL, Ferguson G (eds) 19th National Conference on Artificial Intelligence, AAAI'04, San Jose, California, USA. {AAAI} Press / The {MIT} Press, pp 670--676

\bibitem[{Ying(2010)}]{YING2010162}
Ying M (2010) Quantum computation, quantum theory and {AI}. Artif Intell 174(2):162--176. \doi{10.1016/j.artint.2009.11.009}, special Review Issue

\bibitem[{Zhang and Poole(1996)}]{DBLP:journals/jair/ZhangP96}
Zhang NL, Poole DL (1996) Exploiting causal independence in {Bayesian} network inference. J Artif Intell Res 5:301--328

\end{thebibliography}

\end{document}